\documentclass[prd,aps,twocolumn,nofootinbib,showpacs,superscriptaddress]{revtex4-1}
\usepackage{amsfonts}
\usepackage{amsmath}
\usepackage{amssymb}
\usepackage{bm}
\usepackage{dcolumn}
\usepackage[dvips]{graphicx}
\usepackage{graphics}
\usepackage[latin1]{inputenc}
\usepackage{latexsym}
\usepackage{rotating}
\usepackage[colorlinks=true]{hyperref}
\usepackage{xspace} 
\usepackage[usenames]{color}
\usepackage{mathrsfs}

\definecolor {darkgreen}{rgb}{0.2,0.7,0.2}
\definecolor{purple}{rgb}{0.5,0,0.5}

\newcommand\be{\begin{equation}}
\newcommand\ba{\begin{eqnarray}}
\newcommand\ee{\end{equation}}
\newcommand\ea{\end{eqnarray}}

\def\a{\alpha}
\def\b{\beta}
\def\r{\rho}
\def\s{\sigma}

\def\m{\mu}
\def\n{\nu}

\def\d{\delta}

\def\kh{{kh}}

\newcommand\bw{\begin{widetext}}
\newcommand\ew{\end{widetext}}

\newcommand{\nn}{\nonumber}
\newcommand{\tot}{{\mbox{\tiny tot}}}

\newcommand{\wf}{{\mbox{\tiny wf}}}

\newcommand{\WD}{{\mbox{\tiny WD}}}
\newcommand{\NSS}{{\mbox{\tiny NS}}}

\newcommand{\GR}{{\mbox{\tiny GR}}}
\newcommand{\obs}{{\mbox{\tiny obs}}}
\newcommand{\inferred}{{\mbox{\tiny inf}}}

\newcommand{\MAT}{{\mbox{\tiny mat}}}

\newcommand{\NS}{*}

\newcommand{\mrm}{\mathrm}

\begin{document}

\title{Constraints on Einstein-\AE ther theory and  Ho\v rava gravity \\ from binary pulsar observations}

\author{Kent Yagi}
\affiliation{Department of Physics, Montana State University, Bozeman, MT 59717, USA.}

\author{Diego Blas}
\affiliation{CERN, Theory Division, 1211 Geneva, Switzerland.}

\author{Enrico Barausse}
\affiliation{CNRS, UMR 7095, Institut d'Astrophysique de Paris, 98bis Bd Arago, 75014 Paris, France}
\affiliation{Sorbonne Universit\'es, UPMC Univ Paris 06, UMR 7095, 98bis Bd Arago, 75014 Paris, France}

\author{Nicol\'as Yunes}
\affiliation{Department of Physics, Montana State University, Bozeman, MT 59717, USA.}

\date{\today}

\begin{abstract} 
Binary pulsars are ideal to test the foundations of General Relativity, such as Lorentz symmetry, which
requires that experiments produce the same results in all free-falling (i.e.~inertial) frames. 
We here break this symmetry in the gravitational sector by specifying a preferred time direction, and thus a preferred frame, at each spacetime point. 
We then examine the consequences of this gravitational Lorentz symmetry breaking in the orbital evolution of binary pulsars,  focusing
on the dissipative effects. 
We find that Lorentz symmetry breaking modifies these effects, and thus the orbital dynamics, in two different ways.
First, it generically causes the emission of dipolar radiation, which makes the orbital separation decrease faster than in General Relativity. Second, the quadrupole component of the emission is also modified.
The orbital evolution depends critically on the sensitivities of the stars, which measure how their binding energies depend on the motion relative to the preferred frame. 
We calculate the sensitivities numerically and compute the predicted orbital decay rate of binary pulsars in Lorentz-violating gravity.  
By testing these predictions against  observations, we place very stringent constraints on gravitational Lorentz violation.

\end{abstract}


\onecolumngrid
\begin{flushright}
\small  CERN-PH-TH/2013-262
\end{flushright}

\maketitle

\section{Introduction}

The social scientist and epistemologist Karl Popper believed that
scientific hypotheses could never be proved correct, but rather could
only be disproved~\cite{popper}. In his view, the role of a scientist
is to attempt to disprove the canonical set of hypotheses of the day,
the {\textit{status quo}}, by whatever means possible (experimental, observational or mathematical). 
These efforts are an important engine for scientific progress -- in particular to generate what Kuhn
would later call  {\textit{scientific revolutions}} that leap us
forward in our understanding of the Universe~\cite{khun}. 

General Relativity (GR) is today the {\it status quo} when it comes to
describing the gravitational interaction and the motion of large
bodies. As Popper suggested, scientists have spent decades putting
Einstein's theory to the test, essentially since its
conception. Einstein himself would repeatedly look for
observational confirmation of the predictions that his theory would
make. Today, we have an overwhelming amount of data that confirms GR
to incredible precision~\cite{will-living}, and one may wonder why
we should bother with alternative theories.  

Three reasons come immediately to mind. First, almost all the existing tests of GR involve systems that are governed by the
weakly-gravitating, mildly-relativistic regime of the Einstein
equations. In the Solar System, typical orbital velocities are much below $0.1\%$ of the speed of light, and curvatures
are barely measurable. Stronger tests of GR come from
binary pulsars, e.g.~from observations
of their orbital decay rate or time-dilation effects when photons from
the secondary graze the surface of the primary. The
orbital velocity in observed pulsar binaries, however, is not very different
from velocities in the Solar System. In fact, their shortest orbital periods
are on the order of hours, and thus orbital velocities are still $\lesssim 1$ \% the speed of light, 
although curvatures are much higher inside pulsars than in the Solar System. 
Thus, the {\textit{highly-relativistic and dynamical strong-field regime}} of Einstein's equations,
i.e.~the regime characterized by velocities comparable to
the speed of light, by large, dynamical curvatures,  
has not yet been tested and could in principle highlight large deviations 
from GR's predictions (see e.g.~Refs.~\cite{ST0,ST1,ST2,ST3} and~\cite{CSreview,Yagi:2011xp,Yagi:2012vf,Yagi:2013mbt} for two theories of gravity that are indistinguishable from GR with current Solar System and binary pulsar observations, but
deviate from it in strong-gravity systems).  Tests of this highly-relativistic strong-field regime will probably have to wait for the detection
of gravitational waves (GWs) from merging neutron stars (NSs) or black holes~\cite{ligo,virgo,Yunes:2013dva,ST1,ST2,ST3}. 
To understand how much will be learnt about gravitation from these observations, it is important to 
clarify the corresponding predictions from GR and alternative theories \cite{Yunes:2013dva}.

Another reason why GR cannot be the final word  for gravitation in Nature is 
its intrinsic incompatibility with quantum mechanics and the presence
of mathematical pathologies, like singularities in 
gravitational collapse. One may conjecture that the modifications to GR
induced by any approach addressing these issues will be suppressed by
a high-energy scale, e.g.~the Planck mass $M_P\sim 10^{19}\ \mathrm{GeV}$ 
where GR fails as an effective field theory. This would mean that no information about
quantum gravity could be obtained by any current or foreseen experiment.
 However, such arguments
hold only on the very thin thread of dimensional analysis. In fact, they are\
known to fail for existing candidates of quantum gravity, such as
models with large extra-dimensions, where the scale of suppression can
be much lower~\cite{Antoniadis:1998ig}, or in Ho\v rava
gravity~\cite{Horava:2009uw}, where GR modifications can enter at any
scale. 

Finally, cosmological observations
\cite{Ade:2013zuv,Spergel:2006hy} reveal a model of the Universe (the
$\Lambda$CDM model) that is not completely satisfactory from a
theoretical standpoint. The most famous puzzle is the conflict between
the value of the energy density of the source behind the current
cosmological acceleration and the order of magnitude that would be expected
based on na\"ive dimensional arguments. Any
explanation addressing this \textit{cosmological-constant} problem
typically requires a modification of gravitation at cosmological (and possibly smaller)
scales~\cite{DEBook,Clifton:2011jh}. 

\subsection{Lorentz-Violating Gravity}
\label{sec:LVG-intro}

Lorentz symmetry has been put to the test in a variety of circumstances. 
Particle physics experiments have stringently done so in the matter 
sector~\cite{Kostelecky:2003fs,Kostelecky:2008ts,Mattingly:2005re,Jacobson:2005bg}. 
A model-independent formalism, the \textit{standard model 
extension} (SME)~\cite{Colladay:1998fq,Kostelecky:1998id,Kostelecky:1999rh}, 
has been devised to  translate a wide variety of observations into tests of Lorentz invariance. 
This is very efficient for  bounds  on violations of Lorentz
symmetry primarily in the matter sector \cite{Kostelecky:2008ts} or in the sector coupling matter to gravity~\cite{Kostelecky:2010ze}. 

Given the previous constraints, one could question the possibility of having observable effects in gravitation from Lorentz violation. Indeed,
one could argue that any degree of gravitational Lorentz violation should percolate into particle physics. Therefore, stringent constraints from the matter sector 
would 
require that Lorentz violation in gravity also be very small (and thus undetectable with current observations). 
This argument, however, is not watertight. Different mechanisms have been put forward 
to justify the possibility of Lorentz violation in gravity to a degree that is orders of magnitude larger than any violation in the matter sector (see e.g. Ref.~\cite{Liberati:2013xla}).

A first possibility would be to assume that this happens due to the finely tuned (small) value of the operators that violate Lorentz invariance in the matter sector, as compared
to those in the gravity sector. More interestingly, Lorentz invariance in the matter sector could be an emergent feature at low energies \cite{Froggatt:1991ft}, either due to a renormalization group phenomenon \cite{Chadha:1982qq,Bednik:2013nxa} or to it being an accidental symmetry \cite{GrootNibbelink:2004za}. Finally, it has been pointed out recently  that two sectors with different degrees of Lorentz violation can easily coexist provided that the interaction between them is suppressed by a high energy-scale \cite{Pospelov:2010mp}. This could be the case for the matter and  gravity sectors, each with a very different scale of Lorentz violation.

Thus, it is reasonable to seek for independent tests of Lorentz invariance in gravity. These are neither as developed nor as stringent as
for the matter sector.  Gravitational Lorentz invariance bounds in the context of the SME can be found in Refs.~\cite{Bailey:2006fd,Bailey:2009me,Bailey:2013oda}. 
More relevant for this work are the bounds related to models where Lorentz invariance is broken by the presence of a preferred timelike vector.
Those include constraints placed with Solar System~\cite{will-living,Blas:2011zd,Jacobson:2008aj} and cosmological 
observations~\cite{Jacobson:2008aj,Zuntz:2008zz,Li:2007vz,Audren:2013dwa}, but these are
rather weak. In the Solar System, observations can only probe certain aspects of gravitational Lorentz violation, 
namely leading-order post-Newtonian (PN) effects, i.e.~leading in the expansion in the ratio of the orbital velocity to the speed of light. In cosmology, only the linear perturbative regime over an expanding background has been studied.
Binary pulsar observations have also been used to place constraints on gravitational Lorentz invariance~\cite{Jacobson:2008aj,Bell:1995jz,Shao:2012eg,Shao:2013wga}, but these concentrate only on preferred-frame corrections
to the conservative orbital dynamics, neglecting dissipative effects. 

In this paper, we will focus on theories that modify GR by assuming
the existence of a preferred time direction (frame) at every spacetime point. 
This directly implies the violation of boost symmetry and therefore of Lorentz invariance.
The preferred time direction will be described by a timelike
unit vector field $U^\mu$ (the \AE ther field), whose dynamics may be
described by two different theories: Einstein-\AE ther~\cite{Jacobson:2000xp} 
and khronometric theory~\cite{Blas:2009qj}.
As we will argue in Sec.~\ref{sec:MGT}, these are
 generic theories that may arise 
from more fundamental theories at low energies.

Einstein-\AE ther theory is a Lorentz-violating,
metric theory of gravity where the \AE ther field is generic. The theory  is
characterized by the \AE ther's couplings 
to gravity.  A direct coupling of the \AE ther  field to matter is absent, 
because that would have consequences that are not observed experimentally, e.g.~a 
fifth force would arise in the matter interactions and the weak equivalence principle (i.e.
the universality of free fall for weakly gravitating bodies) would not hold.
Einstein-\AE ther theory can be thought of as representing the
low-energy description of some high-energy unknown dynamics, i.e. it can be understood as an effective field theory \cite{ArmendarizPicon:2010mz}.

Einstein-\AE ther theory passes all theoretical and phenomenological
constraints in a certain region of the parameter space of the \AE ther's couplings. Of particular
importance are the bounds related to Solar System observations, which
force the theory to depend on only two combinations of coupling
constants, $c_+$ and $c_-$, up to corrections
of ${\cal O}(10^{-4})$~\cite{will-living,Foster:2005dk}. These two combinations, however, are currently 
weakly constrained, mostly by requiring the linear stability of Minkowski space (i.e.~absence of tachyonic instabilities) 
and the absence of gravitational Cherenkov radiation~\cite{Elliott:2005va}
(i.e.~requiring the speeds of the gravitational modes be larger than the speed of light,
so that photons and high-energy particles do not lose energy in a Cherenkov-like process).
An order of magnitude constraint (of about $\mathcal{O}(10^{-2})$) was also obtained using the orbital decay rate of binary 
pulsars~\cite{Foster:2007gr}, but this analysis is only valid in the small coupling region ($c_{i} \ll 1$).

Khronometric theory is defined through the same action as that of Einstein-\AE ther theory, but with
the additional requirement that the \AE ther field be hypersurface-orthogonal.
 This choice reduces the dynamical degrees of freedom of the theory, and more importantly 
it forces the theory to exactly coincide with the low-energy limit of  Ho\v rava gravity~\cite{Horava:2009uw}.
Ho\v rava gravity is a proposal for quantum gravity where
Lorentz invariance is broken by the existence of a preferred foliation
of spacetime into spacelike hypersurfaces, and which provides a power-counting renormalizable
completion of GR in the ultraviolet regime.
In this work, we will focus on the version of this theory introduced
in Ref.~\cite{Blas:2009qj}, which has Minkowski spacetime as a
ground-state, and which reduces exactly to khronometric theory
at low energies.

As in Einstein-\AE ther theory, any viable khronometric theory is
characterized by two coupling constants, $\beta$ and $\lambda$, after imposing constraints  
from Solar System observations.
As in Einstein-\AE ther theory,  $\beta$ and $\lambda$ are weakly constrained by requiring linear
stability of Minkowski space and the absence of gravitational Cherenkov radiation~\cite{Blas:2010hb,Barausse:2011pu}. 
However, unlike in Einstein-\AE ther theory~\cite{Carroll:2004ai,Zuntz:2008zz,Li:2007vz,Jacobson:2008aj}, cosmological observations can place stringent bounds on the couplings of khronometric theory 
[roughly $|\beta, \lambda|\lesssim {\cal{O}}(10^{-1})$]. 
In particular, strong constraints can be obtained by requiring that big bang nucleosynthesis
(BBN) in khronometric theory produces element abundances in agreement with observations~\cite{Audren:2013dwa}.
Cosmological constraints are instead much weaker in Einstein-\AE ther theory 
because, unlike in khronometric theory, the Newtonian constant regulating the cosmological evolution
coincides with the locally measured constant to within ${\cal O}(10^{-4})$,
once Solar System constraints are imposed.

\subsection{Executive Summary}

In this paper, we explain in detail and extend the analysis of Ref.~\cite{Yagi:2013qpa} to place very stringent constraints on Lorentz-violating gravity, 
focusing on Einstein-\AE ther and khronometric theory. These constraints are obtained by comparing the evolution of binary systems in these 
theories to binary pulsar observations. 

Let us first consider modifications to the {\it dissipative sector}, which regulates how fast binary systems lose energy and shrink,
forcing the orbital period to decay. A generic property of 
 Einstein-\AE ther and khronometric theory is the excitation of propagating modes that are absent in GR, and which carry
energy away from the system at dipole order~\cite{Foster:2006az,Foster:2007gr,Blas:2011zd}. 
Since dipolar radiation is generally stronger than GR's quadrupole radiation, 
the binary's separation decreases much faster in Lorentz-violating gravity, leading to a strong modification to the predicted evolution of the orbital period. 
Furthermore, these  extra modes and the modification to the propagation speed of gravitons affect the 
 quadrupolar  emission, which is important for systems where dipolar radiation is suppressed. 
Since binary pulsar observations of the orbital period's decay rate 
agree with GR's predictions within the observational uncertainties, 
they allow for stringent constraints to be placed on Lorentz-violating gravity. 

The modified orbital decay rate for NS binaries in Lorentz-violating theories, and more in general the motion of these systems,
 strongly depend on the {\textit{sensitivities}} of the stars~\cite{Foster:2007gr}\footnote{Foster's constraint on $c_{\pm}$ is only valid in the small coupling regime, precisely because the sensitivities had not been calculated until now.}. 
These quantities measure how much the binding energy of an isolated star changes with its motion relative
to the preferred frame (i.e.~relative to the \AE ther).
In order to calculate the sensitivities, we therefore need to find solutions describing NSs moving with respect to the \AE ther, 
as suggested in~\cite{Eling:2007xh}. 
More specifically, we will show that in order to extract the sensitivities, such solutions are only needed at linear order 
in a perturbative expansion in the velocity $v$, i.e.~without loss of generality,
we can restrict attention to solutions of stationary, non-spinning NS moving slowly with respect to the \AE ther.
We show that two seemingly different definitions of the sensitivities, one applicable in the
weak-field regime and another in the strong-field regime, lead to the same result.

To find these slowly-moving solutions, we first write down the most generic ansatz for the metric and \AE ther field at ${\cal O}(v)$, ensuring compatibility with the symmetries of
the problem (stationarity and rotational invariance around the motion's direction). With this ansatz, we write down the field equations at zeroth- and first-order in $v$. The resulting system of partial differential
equations is then expanded in tensor spherical harmonics, leading to an ordinary differential system that we solve numerically.  

The equations must be solved twice (in the interior and in the
exterior of the star) and then matched. The matching ensures that the 
potentials (defined in terms of the \AE ther and metric), which enter the equations 
through derivatives, are continuous everywhere in the spacetime, regular at the center 
of the star, and  produce an asymptotically flat geometry at spatial infinity. The interior
solution depends on the equation of state (EoS) for the NS matter. We here investigate
a variety of EoSs that are thought to represent realistic NS
configurations. We restrict attention to non-rotating, cold (and thus old) NSs, 
as these are appropriate simplifications for binary pulsar studies.

With this at hand, we use the numerical solutions that we obtained to compute the 
sensitivities for isolated NSs for the first time in the theories under consideration. 
We then construct fitting functions that allow us to
analytically model the sensitivities as a function of the compactness 
of the star and the coupling constants of the theory. 

Once the sensitivities have been found, we calculate the energy carried away by all propagating degrees of freedom 
in Einstein-\AE ther and khronometric theory. We show that dipolar radiation is produced and  generically constitutes the dominant
modification to the GR predictions~\cite{will-living}. Given the energy flux carried away by the propagating modes, we use conservation of energy and a balance law to compute the rate of change of a binary's binding energy, and from
this, the orbital decay rate. 
\begin{figure*}[htb]
\begin{center}
\begin{tabular}{lr}
\includegraphics[width=8.0cm,clip=true]{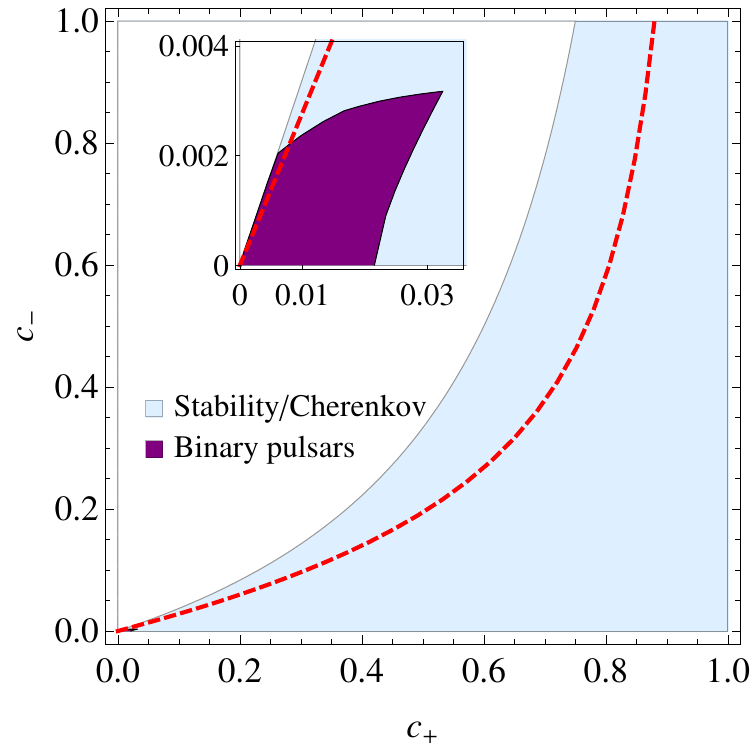}
& \includegraphics[width=8.0cm,clip=true]{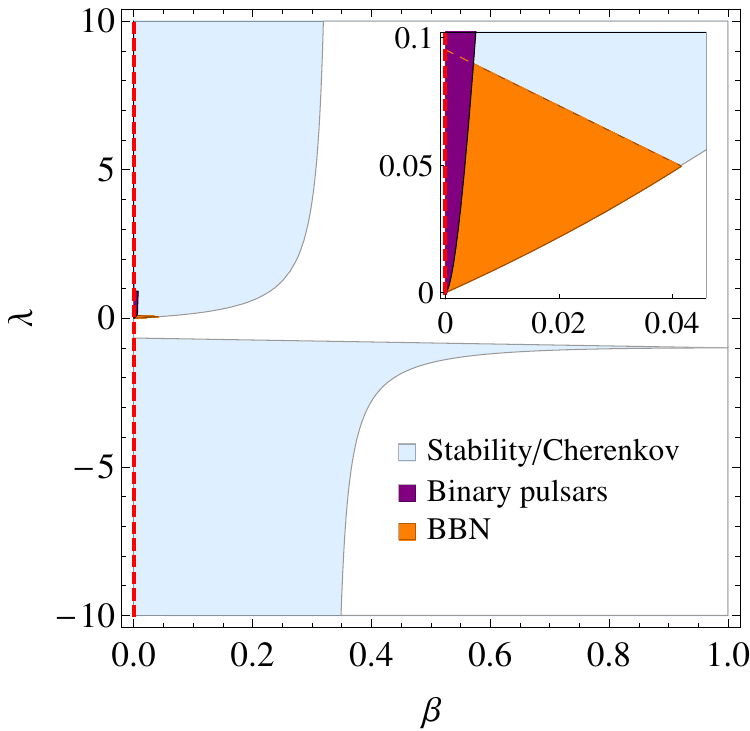}  
\end{tabular}
\caption{\label{fig:Pdot-constraint-summary} (Color online) 
Constraints on the $(c_{+},c_{-})$ plane in \AE ther theory (left) and $(\lambda,\beta)$ plane in khronometric theory (right) obtained by combining constraints derived from observations of  PSR J1141-6545~\cite{bhat}, PSR J0348+0432~\cite{2.01NS}, PSR J0737-3039~\cite{kramer-double-pulsar} and PSR J1738+0333~\cite{freire}. The areas outside the (allowed) shaded regions are ruled out by stability/Cherenkov considerations (light blue), BBN (dark orange) and the combined binary pulsar constraints (dark purple). The red dotted line corresponds to the values of the coupling constants required for the orbital decay rate to agree with the GR prediction in the zero-sensitivity/weak-field limit. Observe that the new constraints are much more stringent than all others.}
\end{center}
\end{figure*}

We compare our predictions for the orbital decay rate to observations
of binary pulsars PSR J1141-6545~\cite{bhat}, PSR J0348+0432~\cite{2.01NS}, and PSR J0737-3039~\cite{kramer-double-pulsar}. The first two are pulsars on a $0.17$-eccentricity, $4.74$-hour orbit and on a ${\cal O}(10^{-6})$-eccentricity, $2.46$-hour orbit respectively, around a white dwarf companion. The third is the relativistic double pulsar binary, on a $0.088$-eccentricity and $2.45$-hour orbit. These comparisons allow us to place constraints on the coupling constants of the theory. 

Another way to place constraints on Lorentz-violating theories is to consider modifications to the {\textit{conservative sector}}, controlled by the Hamiltonian, which for example affects the orbital shape and precession rate.    
Lorentz-violating corrections to the Hamiltonian induce precession of the spin and orbital angular momentum vectors. 
Since such non-GR precession is not found in pulsar observations, one can then place constraints on Lorentz-violation. 
The constraints are cast in a model-independent language by considering strong-field generalizations of the parametrized post-Newtonian (PPN) Hamiltonian. For example, observations
of PSR J1738+0333~\cite{freire} can be used to constrain the strong-field PPN parameters associated with preferred-frame effects. We here calculate these parameters for Einstein-\AE ther and khronometric theory, and then use PSR J1738+0333~\cite{freire} to place constraints on the couplings. 

Combining all of these constraints, we obtain the allowed coupling parameter space shown in 
Fig.~\ref{fig:Pdot-constraint-summary} (Einstein-\AE ther theory in the left panel 
and khronometric theory in the right panel).  The colored regions are those allowed after requiring stability and absence of
gravitational Cherenkov radiation~\cite{Blas:2010hb,Barausse:2011pu,Elliott:2005va} (light blue), 
BBN constraints~\cite{Audren:2013dwa,Carroll:2004ai,Zuntz:2008zz,Jacobson:2008aj} (dark orange) and binary pulsar constraints (dark purple). 
The red dashed line corresponds to the values of the coupling constants for which the orbital decay rate equals the GR prediction, 
assuming the sensitivities vanish and working at leading-order in a weak-field expansion~\cite{Foster:2005dk}. Observe 
that the new constraints obtained here are much stronger than all other constraints. The constraint found in this paper is consistent with the order of magnitude estimate of Foster's~\cite{Foster:2007gr} in the small coupling region, but  the constraint on $c_{-}$ found here is slightly stronger. 

Binary pulsar constraints lead to {\textit{regions}} of viable coupling parameter space. This is because in deriving these constraints one has to allow for different values of the coupling constants (within the Solar System constraints) and the sensitivities (because of the different possible EoSs), and account for the observational error in the orbital decay rate and the orbital period, as well as the error in the inferred masses of the binary. The particular shape of these viable regions is a result of the combination of the constraints associated with different binary pulsars. For example, dipole radiation is suppressed for double pulsar binaries relative to pulsar-white dwarf systems, because dipole radiation is proportional to the difference of the sensitivities and NSs have similar sensitivities. Therefore, for double pulsar systems quadrupolar radiation becomes comparable to dipole radiation,
which leads to a different shape for the allowed region of coupling constants.

\subsection{Layout and Conventions}

The remainder of this paper presents the details of the calculations
described above and is organized as follows:
Section~\ref{sec:MGT} presents the basics of Einstein-\AE ther and khronometric theory, 
including the action, field equations
and current constraints from low-energy phenomenology.
Section~\ref{sec:violations-SEP} describes how Lorentz-symmetry 
breaking leads to violations of the strong equivalence principle, how this is encoded
in the sensitivities, and how it affects the motion of compact objects.
Section~\ref{sec:NS-Sols}  and Section~\ref{sec:NS-SolsHL} are devoted
to the construction of slowly-moving NS solutions to first order in
velocity, in both Einstein-\AE ther and khronometric theory. 
Section~\ref{sec:numerical-results} presents results derived by
numerically solving the modified field equations, focusing on the
sensitivities. 
Section~\ref{sec:Constraints} presents constraints on Einstein-\AE
ther and khronometric theory based on binary pulsar
observations. 
Section~\ref{sec:Conclusions} concludes and points to future research.    
Appendices~\ref{app:separability},~\ref{app:mapping} and~\ref{app:fit}
present further mathematical details.  

Our conventions are as follows. We use natural units where
$c=1=\hbar$, where $\hbar$ is the (reduced) Planck constant and $c$ 
is the speed of light. We restore powers of $c$ when presenting
PN expressions. We use Greek letters in index lists to
denote spacetime components of tensors, while Latin letters in the
middle of the alphabet $(i,j,k,\ldots)$ denote purely spatial
components. We also employ the metric signature $(+,-,-,-)$. 
Indices in spatial vectors are raised and lowered with the Kronecker delta (except in Sec.~\ref{Sec:KH},  where the 3-metric $\gamma_{ij}$ is used to raise and lower indices). 

We use several gravitational constants, different masses, and different velocities throughout the paper, which we list here for convenience. Regarding gravitational constants:
\begin{itemize}
\item $G_{\AE}$ is the bare gravitational constant appearing in the action [Eq.~\eqref{ae-action}];
\item $G_N$ is the ``Newtonian'' gravitational constant measured locally by Cavendish-type experiments [Eq.~\eqref{eq:GN}]; 
\item $G_C$ is the ``cosmological'' gravitational constant that appears in the Friedmann equations [Eq.~\eqref{eq:GC-khronometric}]; 
\item $\mathcal{G}$ is the ``effective'' gravitational constant in a binary system [Eq.~\eqref{calG}]. 
\end{itemize}
Regarding the masses:
\begin{itemize}
\item $\tilde{m}_A$ is the gravitational mass of the $A$-th body in a point-particle approximation [Eq.~\eqref{taylor-pp}];
\item $m_A$ is the ``active'' gravitational mass of the $A$-th body in a point-particle approximation [Eq.~\eqref{active-mass}]; 
\item $M_\tot$ is the total gravitational mass of a star, which includes the gravitational, \AE ther and baryonic contributions [Eq.~\eqref{ae:mass}];
this mass generalizes $\tilde{m}_A$ to regimes where the point-particle approximation does not hold;
\item $M_{\obs}$ is the mass measured by Keplerian experiments, which turns out to coincide with $M_{\tot}$;
\item $M_* \equiv G_N M_\tot = G_N M_\obs$ is the length scale associated with the total mass $M_\tot=M_\obs$; 
\item  $M(r)$ is a function with dimension of  length, defined by Eq.~\eqref{M-def} and approaching $M_*$ as $r\to+\infty$;
\item $m \equiv m_1 + m_2$ is the total active mass of a binary system in the point-particle approximation;
\item $\mu \equiv m_1 m_2/m$ is the active reduced mass of a binary system in the point-particle approximation.
\end{itemize}
Regarding the velocities, we use
\begin{itemize}
\item $v^i$ or $v_A^i$ are both the $3$-velocity of an object relative to the \AE ther field;
\item $v_{12}^i = v_1^i - v_2^i$ is the relative velocity of the two bodies in a binary;
\item $V_{CM}^i$ is the center-of-mass velocity of the binary relative to the \AE ther.
\end{itemize}

\section{Modified Gravity Theories}
\label{sec:MGT}

In this section, we define the theories we focus on. We begin with a
description of Einstein-\AE ther theory and follow with khronometric
theory (the low-energy limit of Ho\v rava gravity).  In both cases, we
first introduce the action that defines the theory and then describe
its current experimental constraints. 

\subsection{Einstein-\AE ther Theory}

Einstein-\AE ther theory describes
gravity by means of a metric $g_{\alpha \beta}$  \textit{and} a
unit-norm timelike dynamical vector field $U^{\alpha}$ (the ``\AE ther field''). The latter  locally defines a preferred time direction,
which breaks  boost- and therefore Lorentz-invariance.
Up to total divergences, the
most generic covariant action that depends only on the fields and their
first derivatives, and is quadratic in the latter, is~\cite{Jacobson:2000xp,Eling:2004dk,Jacobson:2008aj}
\be 
\label{ae-action} S_{\AE} = \frac{1}{16\pi G_{\AE}}\int  ~d^{4}x \sqrt{-g}~ (-R
-M^{\a\b}{}_{\m\n} \nabla_\a U^\m \nabla_\b U^\n)\,,
\ee 
where $g$ and $R$ are the metric determinant and the Ricci scalar,
$U^{\alpha}$ satisfies the unit constraint $U^{\alpha} U_{\alpha}=1$, 
\be
M^{\a\b}{}_{\m\n} \equiv c_1 g^{\a\b}g_{\m\n}+c_2\d^{\a}_{\m}\d^{\b}_{\n}
+c_3 \d^{\a}_{\n}\d^{\b}_{\m}+c_4 U^\a U^\b g_{\m\n}\, ,
\ee 
($c_1$,
$c_2$, $c_3$ and $c_4$ being dimensionless coupling constants), and the
``bare'' gravitational constant $G_{\AE}$ is related to the
``Newtonian'' constant measured with Cavendish-type experiments
through~\cite{Carroll:2004ai} 
\be 
\label{eq:GN}
G_N=\frac{2 G_{\AE}}{2-(c_1+c_4)}\,.
\ee  

In order to enforce the weak equivalence principle\footnote{This principle states
that test particles, i.e.~bodies that are \textit{not} strongly
gravitating, follow trajectories independent of their mass if given
the same initial conditions. The same statement, but applied to strongly gravitating bodies (i.e. ones
with non-negligible gravitational binding energy), constitutes the \textit{strong} equivalence principle.} and the
absence of a fifth force in matter interactions,
one has to (minimally) couple the
matter fields to the metric but \textit{not} to the \AE ther.
Collectively denoting the matter fields by $\psi$, one can then write
the total action as 
\be 
\label{eq:ae_action}
S=S_{\AE}+S_{\MAT} (\psi,g_{\m\n})\,.
\ee
The matter action is  diffeomorphism invariant,
which implies the covariant conservation of the matter stress-energy
tensor $T^{\mu\nu}_\MAT$, i.e.~$\nabla_\mu
T^{\mu\nu}_\MAT=0$~\cite{Wald:1984cw}, and as a consequence test
particles follow geodesics~\cite{Poisson:2011nh}, thus satisfying the
weak equivalence principle.  However, as we will show in this paper,
in Einstein-\AE ther (and khronometric) theory, strongly gravitating bodies follow trajectories
that depend on the ratio between the body's binding energy and its
total mass.  This effect (known as the N\"ordvedt 
effect~\cite{nordvedt,nordvedt_dickes_interpretation}) generically
appears  when one introduces extra gravitational degrees of freedom
coupled non-minimally  to the metric (e.g.~it is present also in scalar
tensor theories~\cite{will-living}), and amounts to a violation of the
\textit{strong} equivalence principle. Its physical origin lies in the
fact that the matter fields couple to the extra gravitational fields
through the metric, and this coupling becomes important in
strong-gravity regimes.

The absence of a direct coupling of the matter fields to $U^{\alpha}$ in
Eq.~(\ref{eq:ae_action}) is also enforced 
by tests of Lorentz invariance in particle physics. Indeed, 
if the matter fields were directly coupled to $U^{\alpha}$, this would generically produce
Lorentz-violating effects in the standard model of particle physics. The
bounds on these effects are very tight \cite{Mattingly:2005re,Kostelecky:2008ts}, 
which means that for gravitational experiments it is reasonable to put all the
couplings between \AE ther and matter to zero\footnote{These couplings, however, may be significant for the
dark-matter sector \cite{Blas:2012vn}.}.
Naturalness arguments seem to be at odds with this choice, since they suggest that Lorentz violation may percolate from the gravitational sector into 
the standard model, thus producing
big (and experimentally forbidden) effects. 
However, this is not necessarily the case. As already mentioned, different
 mechanisms have been put forth leading to Lorentz-violating
effects in gravity that can be much larger than those in
particle physics, see e.g.~Ref.~\cite{Liberati:2013xla} for a review,  Ref.~\cite{Bednik:2013nxa} for recent developments, 
and Sec.~\ref{sec:LVG-intro} for further details. 

Variation of the action given in Eq.~\eqref{eq:ae_action} with respect to the metric and \AE ther 
field [imposing the unit constraint $U^{\alpha} U_{\alpha} = 1$
directly or by means of a Lagrange multiplier $\ell (U^\mu
U_\mu-1)$ in the action] provides a set of modified Einstein equations 
\be\label{E_def}
E_{\a\b}\equiv G_{\a\b} - T^{\AE}_{\a\b}-8\pi G_{\AE} T^\MAT_{\a\b}=0\,, 
\ee 
and the \AE ther equations
\be\label{AE_def} \AE_\mu\equiv\left(\nabla_\a
J^{\a\n}-c_4\dot{U}_\a\nabla^\n U^\a\right) \left(g_{\mu\nu}-U_\m
U_\n\right)=0\,.  
\ee 
Here, $G_{\a\b}=R_{\a\b}-R\, g_{\a\b}/2$ is the
Einstein tensor,
\ba\label{Tae}
T^{\AE}_{\a\b}&=&\nabla_\m\left(J^{\phantom{(\a}\m}_{(\a}U_{\b)}-J^\m_{\phantom{\m}(\a}U_{\b)}-J_{(\a\b)}U^\m\right)\nonumber\\ &&+c_1\,\left[
  (\nabla_\m U_\a)(\nabla^\m U_\b)-(\nabla_\a U_\m)(\nabla_\b U^\m)
  \right]\nonumber\\ &&+\left[ U_\n(\nabla_\m J^{\m\n})-c_4 \dot{U}^2
  \right] U_\a U_\b +c_4 \dot{U}_\a \dot{U}_\b\nonumber\\&&+\frac{1}{2} M^{\s\r}{}_{\m\n} \nabla_\s U^\m \nabla_\r U^\n
g_{\a\b}\,,\ea 
is the \AE ther stress-energy tensor,  where
\be
J^\a_{\phantom{a}\m}\equiv
M^{\a\b}_{\phantom{ab}\m\n} \nabla_\b U^\n\,,
\qquad\dot{U}_\n\equiv U^\m\nabla_\m U_\n\,.  
\ee
Finally, the matter
stress-energy tensor is defined as usual by 
\be
\label{eq:emtmatt}
T^{\a\b}_\MAT \equiv-\frac{2}{\sqrt{-g}} \frac{\delta S_{\MAT}}{\delta
  g_{\a\b}}\,.
\ee 

Experimental constraints and stability requirements greatly reduce the
viable parameter space for the four couplings $c_1$, $c_2$, $c_3$ and
$c_4$.  At 1PN order\footnote{A term proportional to $(v/c)^{2N}$ is said to be of Nth PN order.}, 
the dynamics of Einstein-\AE ther theory matches
that of GR, with the exception of two
``preferred-frame'' parameters $\alpha_1$ and $\alpha_2$, which are
exactly zero in GR but not in Einstein-\AE ther
theory~\cite{Foster:2005dk}. In the weak-field, Solar System observations constrain these
parameters to very small values $|\alpha_1|\lesssim 10^{-4}$ and
$|\alpha_2| \lesssim 10^{-7}$~\cite{will-living}, using lunar laser ranging
and solar alignment with the ecliptic. The strong-field counterparts of these parameters
(denoted here with an overhead hat) are constrained through binary and isolated pulsar 
observations also to very small values,
$|\hat{\alpha}_1| \lesssim 10^{-5}$ and
$|\hat{\alpha}_2| \lesssim 10^{-9}$~\cite{Shao:2013wga,Shao:2012eg}. 
We will discuss the Einstein-\AE ther form of $(\alpha_{1},\alpha_{2})$ in Sec.~\ref{sec:PNsols} and the relation to their strong-field
generalizations in Sec.~\ref{alpha-hat-sec}.

Since current constraint force $(\alpha_{1},\alpha_{2})$ to be small dimensionless numbers, we can 
expand the theory in these quantities and shrink the parameter space to just two parameters. In Einstein-\AE ther theory, 
this implies~\cite{Foster:2005dk,Jacobson:2008aj}
\begin{align}
\label{eq:Aecons}
c_2&=\frac{-2 c_1^2-c_1 c_3+c_3^2}{3 c_1}\,,
\\
\label{eq:Aecons2}
c_4&=-\frac{c_3^2}{c_1}\,,
\end{align}
to leading order in $\alpha_{1}$ and $\alpha_{2}$. The
only free parameters in Einstein-\AE ther theory are then $c_\pm\equiv c_1\pm c_3$.

Further constraints on $c_{\pm}$ come from requiring
that perturbations about a Minkowski background are stable and have
positive energy~\cite{Jacobson:2004ts}, and that matter does not emit
gravitational Cherenkov radiation~\cite{Elliott:2005va}. Such radiation would be emitted if
the speed of the \AE ther's and metric's propagating modes were smaller than the speed of light.  
These requirements result in the bounds
\ba
\label{cp_constraint} &&0\leq c_+\leq 1\,,\\ &&0\leq c_-\leq
\frac{c_+}{3(1-c_+)}\,,\label{cm_constraint} 
\ea 
to leading order in $\alpha_{1}$ and $\alpha_{2}$. In addition to these, 
Foster~\cite{Foster:2007gr} estimated that roughly $c_{\pm} < \mathcal{O}(10^{-2})$, assuming a small-coupling approximation ($c_{i} \ll 1$)
and using order-of-magnitude information about the orbital decay of binary pulsars. 

For arbitrary $c_{i}$, the dependence of the gravitational constant $G_{C}$ appearing in the Friedmann equation 
on the coupling constants is different than that of the gravitational constant $G_{N}$ measurable via Cavendish-type experiments.
For generic values of the couplings, therefore, $G_N$ and $G_C$ differ.
This difference leads to deviations from the metal abundances predicted by Big Bang Nucleosynthesis (BBN)~\cite{Carroll:2004ai} and to a modified growth of cosmological perturbations that affects the Cosmic Microwave Background (CMB) and the distribution of matter at large scales \cite{Audren:2013dwa}. However, 
once Eqs.~\eqref{eq:Aecons} and~\eqref{eq:Aecons2} are satisfied, $G_N \approx G_C$ up to terms of ${\cal{O}}(10^{-4})$, and thus, 
cosmological observations do not significantly reduce the viable
region defined by Eqs.~\eqref{cp_constraint} and~\eqref{cm_constraint}~\cite{Zuntz:2008zz,Jacobson:2008aj}.  
Similarly, although black hole solutions in Einstein-\AE ther theory differ from GR~\cite{Eling:2006ec,Barausse:2011pu,Barausse:2013nwa},
current electromagnetic observations of
black hole candidates are still not accurate enough to provide
constraints competitive with current bounds (see e.g.~Ref.~\cite{Bambi:2011jq}). 

\subsection{Khronometric Theory and Ho\v rava Gravity}
\label{Sec:KH}

As mentioned in the previous section, Einstein-\AE ther theory breaks Lorentz invariance by locally specifying a preferred time
direction through an \AE ther vector field $U^{\alpha}$. This vector field may appear from the 
existence of a  \textit{global} preferred time. In this case, it will be orthogonal to the hypersurfaces of constant preferred time, i.e. proportional to the gradient of a foliation-defining scalar field $T$:
\begin{equation}\label{U_dT}
U_\mu=\frac{\partial_\mu T}{\sqrt{g^{\mu\nu}\partial_\mu T \partial_\nu T}},
\end{equation}
which satisfies normalization condition $U^{\alpha} U_{\alpha} = 1$. The \AE ther vector field is timelike as a consequence
 of  the scalar defining a preferred time coordinate.  The scalar $T$ is often referred to as the ``khronon''\footnote{From the Greek 
$\chi\rho\acute o\nu o \varsigma$ (khronos) -- time. The romanization was chosen to avoid confusion with previous 
uses of the prefix `chrono', e.g. \cite{Levi,Segal}, which are not related to the theories under study here.}, and a metric theory in
 which Lorentz invariance is broken globally by such a scalar is called ``khronometric theory.''

The most generic action (quadratic in derivatives)
for khronometric theory
is given by Eq.~\eqref{ae-action} with the definition in Eq.~\eqref{U_dT}~\cite{Jacobson:2010mx,Blas:2010hb}.
The condition in Eq.~\eqref{U_dT} allows one to express one of the \AE ther terms in the action 
in terms of the other ones, and thus there are only three free independent \AE ther  terms. 
In particular, we can absorb the $c_1$ term into the other three terms, by multiplying the second, third and forth term respectively by the new couplings
\be\label{eq:EAtoKH}
\lambda\equiv c_2, \quad \beta\equiv c_3+c_1, \quad \alpha\equiv c_4+c_1.
\ee
Another form for the action can be derived by choosing the time coordinate to 
coincide with  $T=$ constant hypersurfaces. In this gauge, 
Eq.~\eqref{U_dT} becomes 
\be\label{U_lapse}
U_\alpha= (g^{TT})^{-1/2}\delta_{\alpha}^{T}=N\delta_{\alpha}^{T}  \,,
\ee
where $N=(g^{TT})^{-1/2}$ is the lapse. The action in Eq.~\eqref{ae-action} with the definition
in Eq.~\eqref{U_dT} then becomes 
\cite{Blas:2009ck}
\begin{multline}\label{action-K}
S_{K}=\frac{1-\beta}{16\pi G_{\AE}}\!\int dT d^3x \, N\sqrt{h} \, (K_{ij}K^{ij} - \frac{1+\lambda}{1-\beta} K^2
\\+  \frac{1}{1-\beta}{}^{(3)}\!R +  \frac{\alpha}{1-\beta}\, a_ia^i)+S_{\MAT}(\psi,g_{\mu\nu})\,,
\end{multline}
where $K^{ij}$, ${}^{(3)}\!R$ and $h^{ij}$ are respectively the extrinsic curvature, the $3$-Ricci curvature
and the 3-metric of the $T=$ constant hypersurfaces, and where we have defined the ``acceleration'' of the \AE ther flow\footnote{This vector is related
to the acceleration of \AE ther congruence  by $U^\m \nabla_\m U^\n=a^i \delta_i^\n$}, $a_i\equiv\partial_i \ln{N}$. Latin indices are manipulated
with the 3-metric of the $T=$ constant hypersurfaces.

An additional motivation for khronometric theory is that it coincides with the low-energy limit of a theory of
gravity that has remarkable properties at high energies. This theory is known as  Ho\v rava gravity~\cite{Horava:2009uw}, and achieves 
power-counting renormalizability by adding higher-order derivative terms to the khronometric action~\cite{Horava:2009uw,Blas:2009qj}:
\be
\label{action-H-full}
S_{H}= \frac{1-\beta}{16\pi G_{\AE}}\int dT d^3x \, N\sqrt{h}\left(L_2+\frac{1}{M_\star^2}L_4+\frac{1}{M_\star^4}L_6\right)\,,
\ee
where the $L_2$ term corresponds to the gravitational Lagrangian of 
 khronometric theory [cf.~Eq.~\eqref{action-K}], while 
$L_4$ and $L_6$ are terms that are suppressed by a suitable mass scale $M_\star$. These terms are respectively of fourth and sixth order in the spatial derivatives, but contain no derivatives with respect to $T$. 

Let us stress that the experimentally viable range for the mass scale $M_\star$ is rather broad. On the one hand, $M_\star$ is bound from above ($M_{\star}\lesssim 10^{16}$ GeV) to allow the theory to remain perturbative at all scales~\cite{Papazoglou:2009fj,Kimpton:2010xi,Blas:2009ck}, so that the power-counting renormalizability arguments of Ref.~\cite{Horava:2009uw} can be applied.
On the other hand, $M_\star$ is also bound from below from tests of Lorentz invariance. That lower bound, however, depends on the details
of the percolation of Lorentz violations from the gravity to the matter sector, which is not yet completely understood~\cite{Burgess:2002tb,Iengo:2009ix,Pospelov:2010mp,Blas:2010hb,Liberati:2012jf,Liberati:2013xla}. 
Taking instead into account only
the constraints from Lorentz violation in the gravity sector, one gets the relatively weak bound $M_\star\gtrsim 10^{-2}$~eV~\cite{will-living,Blas:2010hb}.

In this paper, we will only consider khronometric theory, i.e.~we will focus on
the low-energy limit of Ho\v rava gravity and neglect the higher-order terms $L_4$
and $L_6$. Neglecting
these terms is, however, an excellent approximation as far as astrophysical studies are concerned. In fact,
on purely dimensional grounds, the error introduced
by the $L_4$ and $L_6$ terms on a NS solution of mass $M_\tot$ in khronometric theory is of
${\cal{O}}(G_{N}^{-2} M_\tot^{-2} M_\star^{-2}) = {\cal{O}}(M_{\rm Planck}^4/(M_\tot M_{\star})^2)$. Taking the lowest conceivable value for $M_\star$, i.e $M_\star\sim 10^{-2}$ eV, one gets an error of roughly $10^{-16} (M_{\odot}/M_\tot)^2$ or smaller, when neglecting $L_{4}$ and $L_{6}$ in binary pulsar systems, which is clearly negligible. 

The field equations for khronometric theory can be obtained by varying the action in Eq.~\eqref{ae-action}, with the definition
in Eq.~\eqref{U_dT} replaced in it \textit{before} the variation. Varying with respect to
 $g_{\alpha\beta}$ and $T$ one finds 
\ba\label{hl1}
&&E_{\alpha\beta}-2\AE_{(\alpha}U_{\beta)}=0\,,\\
 \label{hleq}
&&\nabla_\mu \left(\frac{\AE^\mu}{\sqrt{\nabla^\alpha T \nabla_\alpha T}}  \right)=0\,,
\ea
where $E_{\alpha\beta}$ and $\AE_\alpha$ are defined as in Einstein-\AE ther theory [cf.~Eqs.~\eqref{E_def}--\eqref{AE_def}]. 

As can be seen, any hypersurface-orthogonal solution to Einstein-\AE ther theory
is also a solution to khronometric theory, but the converse is not necessarily true. It can be shown that spherically symmetric, static, asymptotically flat solutions are 
indeed the same in the two theories~\cite{Jacobson:2010mx,Blas:2011ni,Blas:2010hb,Barausse:2012ny}. However, the same is not true in more general cases; for example,
the sets of slowly rotating black hole solutions of the two theories do not overlap~\cite{Barausse:2012ny,Barausse:2012qh,Barausse:2013nwa}. Note that 
because of the Bianchi identity, Eq.~\eqref{hleq} is actually implied by Eq.~\eqref{hl1} and by the equations
of motion of matter (which imply $\nabla_\nu T^{\mu\nu}_{\MAT}=0)$, 
i.e.~the only independent equations of khronometric theory are actually the modified Einstein equations
and the equations of motion of matter~\cite{Jacobson:2010mx}.

Finally, let us discuss the current experimental bounds on the coupling constants of khronometric theory. 
Like in the case of Einstein-\AE ther theory, Solar System 
tests require $\alpha_1\lesssim 10^{-4}$ and $\alpha_2\lesssim 10^{-7}$, and similar stringent bounds
are imposed on $(\hat{\alpha}_1,\hat{\alpha}_2)$ by pulsar observations (see Sec.~\ref{sec:Constraints} for a detailed discussion). The fact that 
\be
\alpha_2=\frac{\alpha_1}{8+\alpha_1}\left[1+\frac{\alpha_1(1+\beta+2\lambda)}{4(\beta+\lambda)}\right],
\ee 
for khronometric theory \cite{Blas:2011zd}
 yields two possible ways 
to enforce the previous bounds. One way is to choose $\alpha_1$ to be small enough so that both bounds are simultaneously satisfied. This can be enforced by only one condition of the form
\be\label{option1}
|\alpha_{1}| = 4 \left|\frac{\alpha-2\beta}{1-\beta}\right|\lesssim 10^{-6} \left[1+\frac{\alpha_1(1+\beta+2\lambda)}{4(\beta+\lambda)}\right]^{-1}\,,
\ee 
which leaves the coupling parameters $\lambda$ and $\beta$ unconstrained. Another way is to saturate both bounds for $\alpha_1$ and $\alpha_2$, which leads to a smaller parameter space, i.e. one can require 
\begin{align}\label{option2a}
& 4 \left|\frac{\alpha-2\beta}{1-\beta}\right| \lesssim 10^{-4}\,,
\\
\label{option2b}
&\frac{\alpha - 2 \beta}{\alpha -2 }\left[1+\frac{(\alpha-2 \beta) (1+\beta+2\lambda)}{(\beta - 1) (\beta+\lambda)}\right] \lesssim 10^{-7}\,.
\end{align}
For our purposes, Eqs.~\eqref{option1} and~\eqref{option2a} are equivalent, since our bounds will certainly not
be sensitive to differences below ${\cal O}(10^{-4})$. Thus, the choice given by Eqs.~\eqref{option2a} and~\eqref{option2b}
is more restrictive, and without loss of generality we can neglect it [as it merely selects 
a very small subset of the parameter space allowed by Eq.~\eqref{option1} at leading order in $\alpha_1$ and $\alpha_2$].
We stress, however, that the formalism presented in this paper is perfectly suited to the constraints of Eqs.~\eqref{option2a} and~\eqref{option2b} as well, and we do investigate this choice in more detail in a later section (see e.g. Fig~\ref{fig:sens-C-HL}).

Requiring that the propagating modes are stable and have positive energy
in flat space, and not allowing gravitational Cherenkov emission by matter, one finds the conditions~\cite{Blas:2010hb,Barausse:2011pu,Elliott:2005va}
\ba
\label{con1}
&&0<\beta<1/3, \quad \lambda > \frac{\beta (\beta+1)}{1-3\beta}\, ,\\\label{con2}
&& 0<\beta<1/3, \quad \lambda<- \frac{2+\beta}{3}\, ,\\\label{con3}
&&1/3<\beta<1,\quad \frac{\beta (\beta+1)} {1-3\beta}<\lambda<- \frac{2+\beta}{3}\,.
\ea
Unlike in Einstein-\AE ther theory, cosmological constraints are rather strong in khronometric theory. In fact, one has
\be
\label{eq:GC-khronometric}
\frac{G_N}{G_C}=\frac{2+\beta+3\lambda}{2(1-\beta)}\, ,
\ee
to leading order in $\alpha_{1}$.  
The agreement between observations and the metal abundances predicted by BBN requires $|G_C/G_N-1|\lesssim 1/8$. Once this bound is combined with the stability/Cherenkov constraints of Eqs.~\eqref{con1}-\eqref{con3}, it selects a rather limited region of the $(\lambda,\beta)$ plane (cf.~the orange region in Fig.~\ref{fig:Pdot-constraint-summary}). Finally, when Eqs.~\eqref{option2a} and~\eqref{option2b} are imposed, the ratio $G_C/G_N$ is almost one, and BBN observations do not impose a new constraint, as in the case in Einstein-\AE ther theory.

Further constraints may come from observations of the CMB and large scale structure of the
Universe. As for BBN, those constraints are expected to be efficient except for the case $\beta=-\lambda$.
This has been demonstrated explicitly in Ref.~\cite{Audren:2013dwa}, which extended the action in Eq.~(\ref{ae-action}) by including a 
dynamical dark energy component. However,
because dynamical  dark-energy models affect the evolution of cosmological perturbations,
we will  not consider the bounds of Ref.~\cite{Audren:2013dwa} here.
Also, like in the Einstein-\AE ther case, black hole solutions in khronometric theory differ from the GR ones~\cite{Eling:2006ec,Blas:2011ni,Barausse:2011pu,Barausse:2013nwa,Barausse:2012ny,Barausse:2012qh,Barausse:2013nwa},
and may in principle allow to test the theory in the future.

Finally, Ref.~\cite{Jacobson:2013xta} recently found a relation between khronometric theory and Einstein-\AE ther theory at the level of the solutions
to the field equations. More precisely, Ref.~\cite{Jacobson:2013xta} showed that the solutions of khronometric theory can be obtained
from the Einstein-\AE ther theory solutions in the limit $c_1-c_3\to \infty$ (with $c_1+c_3$, $c_2$ and $c_1+c_4$ held fixed).
While we have \textit{not} used this correspondence in this paper to derive the solutions to the khronometric theory field equations,
we have used it to test the correctness of our results.

\section{Violations of the strong equivalence principle and the sensitivity parameters}
\label{sec:violations-SEP}

The prediction of observables in physical theories requires knowledge of the solution to the field equations for the system under consideration. 
However, when studying compact binary systems, e.g.~of  pulsars, exact solutions are not available in modified gravity theories, or for that matter, in GR. One is then forced to rely on approximations, such as the PN scheme, where the system's dynamics is expanded 
in the ratio between the characteristic velocity of the system
and the speed of light. In this scheme, compact objects are modeled \textit{effectively} with point particles.
Clearly, this approximation breaks down in the strong-field regions near the compact objects, but this will be
irrelevant in this paper; the PN scheme is ideal to study the orbital decay rate of binary pulsars.

As already mentioned in Sec.~\ref{sec:MGT}, the strong equivalence principle is violated in Einstein-\AE ther and khronometric theory
because of the N\"ordvedt
effect~\cite{nordvedt,nordvedt_dickes_interpretation}, i.e.~because an effective coupling between matter and the \AE ther appears in the strong-gravity regime.
To model this effect, the point-particle action used to describe compact objects must depend
on ``\AE ther charges'' or ``sensitivity parameters''~\cite{eardley,damour_esposito_farese} coupling the particles to the \AE ther.
In this section, we will explain how this comes about, what these parameters mean physically, and how they can be computed from the PN metric tensor. We then proceed with a strong-field definition of the sensitivities and conclude with a description of how the sensitivities percolate into observables, with a particular focus
on GW fluxes.

\subsection{A Post-Newtonian Route to the Sensitivities} 
\label{sec:PN-sensitivities}

\subsubsection{Point-Particle Action for Compact Objects}
\label{subsec:EOMs}

In PN theory, one models the motion of compact objects sufficiently far from each other
 {\textit{effectively}} through point particles.  
The way these particles couple to  the different fields of the theory encapsulates finite-size or strong-field effects.
For weakly gravitating objects (i.e. ones with negligible binding energy compared to
their total gravitational mass),
the point particles effectively describing them in the PN scheme can only couple to the metric [c.f. 
Eq.~(\ref{eq:ae_action}) and discussion in  Sec.~\ref{sec:MGT}].
However, even though the matter fields do not couple directly to the \AE ther, the metric
does so non-minimally. Therefore, because matter couples to the metric (although weakly) through gravity, 
an effective coupling appears between the \AE ther and the matter fields when gravity is strong (i.e.~when the metric perturbations produced
on the background by the presence of the object are large, as in the case of NSs). The existence of this effective coupling has long been known 
in the context of scalar-tensor theories (where it is called the N\"ordvedt
effect~\cite{nordvedt,nordvedt_dickes_interpretation}), and introduces deviations
away from geodesic motion for strongly gravitating objects, thus violating 
the strong equivalence principle. The sensitivities will parametrize this effective coupling.

In order to clarify how violations of the strong equivalence principle come about, 
let us briefly review the physical meaning of the sensitivities in scalar-tensor
theories~\cite{eardley}.
In those theories, the gravitational interaction is mediated by the metric and by
a gravitational scalar field $\phi$, coupled minimally to the matter but non-minimally to the metric.
The gravitational constant, and therefore the binding energy of a compact object, depend on the local value 
of the scalar field $\phi$. Because the binding energy contributes to
the gravitational mass,  one has to include an explicit dependence of the gravitational mass $\tilde{m}$ 
on the scalar field $\phi$ when using a point-particle model to describe compact 
objects, i.e.~$S_{\rm pp \; A} = - \int d\tau_A \; \tilde{m}_{A}(\phi)$, with $A$ the object's label.
From this, it is clear that the strong equivalence principle will be violated, because 
the scalar field depends on position, the mass becomes position-dependent, and the 
variation of $S_{\rm pp \; A}$ does not yield the geodesic equation.

We can parametrize deviations from geodesic motion by exploiting the fact that 
in the PN regime all motion is slow compared to the speed of light, and one expects that changes 
in scalar field are also generally small and slow. At leading PN order, 
it is sufficient to consider a leading-order Taylor expansion of the mass $\tilde{m}_A(\phi)$,
and parametrize deviations from geodesic motion through the sensitivity~\cite{eardley} 
\be
s^\phi_{A} \equiv \left.\frac{\partial \ln \tilde{m}_{A}}{\partial \ln \phi} \right|_{\phi=\phi_{0}}\,,
\label{s-def-ST}
\ee
where $\phi_{0}$ is the constant value of the scalar field far from the object.
The partial derivative in Eq.~\eqref{s-def-ST} is to be taken along
a reversible transformation of the object~\cite{eardley}. This is because
under a change of the local scalar field (and therefore of the local value of the gravitational constant), the binding energy also changes, resulting in the body expanding or shrinking. The extra
kinetic energy gained by the volume elements of the body is then
assumed to be transformed into potential energy 
without production of heat (i.e.~without production of entropy). In other words, the
body is thought to gradually adapt its structure to the change in the
scalar field. This is a good approximation if the scalar field
changes slowly enough, which is the case in the PN regime. 

In Einstein-\AE ther and khronometric theory the situation is similar. In general, because of the
effective coupling between the \AE ther vector field and matter in the strong field  regime, the compact object's 
structure, its binding energy and its gravitational mass will be a function of the motion relative to the 
\AE ther. This is exactly why Lorentz symmetry is violated in these theories, i.e. if Lorentz symmetry is broken, then 
the motion of any compact body with respect to the \AE ther should be experimentally detectable. 
Drawing inspiration from scalar-tensor theories, let us model systems of strongly gravitating 
objects with a point-particle action of the form~\cite{Foster:2007gr} 
\be 
\label{eq:actionpp}
S_{\rm pp \; A} = - \int d\tau_A \; \tilde{m}_{A}(\gamma_{A})\,, 
\ee
where $A$ labels the object, $\gamma_{A} \equiv U_{\m} u_{A}^{\m}$ (with $u_{A}^{\m}$ denoting the particle's four-velocity and $U^{\m}$, as usual, the \AE ther vector field) is the Lorentz factor of each 
particle relative to the \AE ther, and $d\tau_A$ is the proper length along the particle's trajectory.
Because PN theory is a perturbative expansion in the system's characteristic velocity,
and because $\gamma_{A} \sim 1$ corresponds to an object moving slowly relative to the \AE ther, 
we Taylor expand the action of Eq.~(\ref{eq:actionpp}) as
\begin{align}
\label{taylor-pp}
S_{\rm pp \; A} &=- \tilde{m}_{A}  \int d\tau_A \; 
\left\{1 + \sigma_{A} \left(1- \gamma_{A}\right) + \frac{1}{2} \sigma_{A}' \left(1 - \gamma_{A} \right)^{2}
\right.
\nn \\
&\left.
+ {\cal{O}}\left[\left(1-\gamma_{A}\right)^{3}\right] \right\}\,, 
\end{align}
where $\tilde{m}_{A}\equiv \tilde m_A(1)$ is a constant and we have defined the sensitivity parameters $\sigma_{A}$ and $\sigma'_{A}$ via
\begin{align}
\label{sigma-def}
\sigma_{A} &\equiv - \left.\frac{d \ln \tilde{m}_{A}(\gamma_A)}{d \ln \gamma_{A}}
\right|_{\gamma_{A} = 1}\,,\,
\\
\sigma_{A}' &\equiv \sigma_{A} + \sigma_{A}^{2} + \left.\frac{d^{2} \ln \tilde{m}_{A}(\gamma_A)}{d (\ln \gamma_{A})^{2}}\right|_{\gamma_{A} = 1}\,,
\end{align}
in analogy with scalar tensor theories. 
In what follows,
it will sometimes be convenient to use the   
rescaled sensitivity parameter
\be
\label{s-sigma-rel}
s_{A} \equiv \frac{\sigma_{A}}{1 + \sigma_{A}}\,.
\ee
Clearly, for weakly gravitating objects, $\sigma_{A}\approx s_A\approx 0$ and $\sigma_{A}'\approx 0$. 

Let us stress that the assumption that $v_A/c\ll 1$,  where $v_{A}\sim [2(\gamma_{A}-1)]^{1/2}$ is the velocity of the $A$-th compact object
relative to the \AE ther, is implicit in Eq.~\eqref{taylor-pp}.
The PN expansion, on the other hand, is in terms of the ratio $v_{12}/c$, 
where $v_{12}$ is the
the binary's relative velocity
(i.e. the binary system's characteristic speed).
In our analysis we will consider both $v_{12}/c\ll 1$ and $v_A/c\ll1$. 
This follows from the combination of these two facts: \textit{(i)} cosmologically, the \AE ther must be almost aligned
with the CMB frame (otherwise there would be strong and non-viable effects on the cosmological evolution; see also Ref.~\cite{Carruthers:2010ii} for a dynamical study of the alignment);
\textit{(ii)} the peculiar velocity of our Galaxy relative to the CMB is ${}\sim 10^{-3} c$, thus
the center of mass of binary systems moves slowly relative to the \AE ther. 

For an action of the form of Eq.~\eqref{eq:actionpp}, there is a subtlety in 
the derivation of the field equations.
In fact, the field equations presented in Sec.~\ref{sec:MGT} assume that the \AE ther does not couple directly to matter
[c.f. Eq.~\eqref{eq:ae_action}], but as mentioned above, this is not the case in a point-particle model aiming at
 effectively describing systems of compact objects.
Because of the presence of this effective coupling, the field equations get modified by terms
proportional to the sensitivities. In particular, for Einstein-\AE ther theory, 
the equations for the \AE ther [Eq.~\eqref{AE_def}]
are modified to include a source term
\begin{multline}
\label{eq:AEsens}
\tilde{\AE}^\mu\equiv \AE^\mu+\frac{8\pi G_{\AE}\tilde m_A}{u^0 \sqrt{-g}}\times\\\delta^{(3)}(x^i-x^i_A)(\sigma_A+\sigma_A'(1-\gamma_A))
 (u_{A}^\m-\gamma_AU^\m )=0\,,
\end{multline}
where $x_{A}^{i}$ is the worldline of the $A$-th point particle.
Also, the Einstein
equations [Eq.~\eqref{E_def}] remain valid, but the matter stress-energy tensor picks up additional terms with respect
to the definition Eq.~(\ref{eq:emtmatt}). Mathematically, these extra terms appear because the modified \AE ther equation [Eq.~\eqref{eq:AEsens}]
changes the value of the Lagrange multiplier
enforcing the unit constraint. Doing the calculation explicitly, it turns out that the matter
stress energy tensor becomes
\begin{align}
T^{\mu \nu}_{{\rm pp}{\AE}\,A} &= \frac{\tilde{m}_{A}}{u_{A}^{0} \sqrt{-g}} \delta^{(3)}\left(x^{i} - x^{i}_{A}\right) 
\left\{ \left[ 1 + \sigma_{A} + \frac{\sigma_{A}'}{2} \left(1 - \gamma^2_{A}\right) \right] 
\right.
\nn \\
&\left.
\times u^{\mu}_{A} u^{\nu}_{A} -  \left[\sigma_{A} + \sigma_{A}' \left(1 - \gamma_{A}\right) \right]( 2U^{(\mu} u_{A}^{\nu)}
-\gamma_AU^\m
U^\n) \right\}\,,\label{Tpp}
\end{align}
with the stress-tensor for the \AE ther field is still given by Eq.~\eqref{Tae}. 

Similarly, in  khronometric theory,
Eqs.~\eqref{hl1} and~\eqref{hleq} are still valid if $\AE^\mu$ is replaced by $\tilde{\AE}^\mu$ as defined by Eq.~\eqref{eq:AEsens}, and if
the point-particle stress-energy tensor is given by Eq.~\eqref{Tpp}. Alternatively (and equivalently), Eq.~\eqref{hl1} can be used
\textit{without} replacing  $\AE^\mu$ by $\tilde{\AE}^\mu$, \textit{but} with the modified point particle stress-energy tensor
\begin{align}
T^{\mu \nu}_{{\rm ppkh}\,A} &= \frac{\tilde{m}_{A}}{u_{A}^{0} \sqrt{-g}} \delta^{(3)}\left(x^{i} - x^{i}_{A}\right) 
\left\{ \left[ 1 + \sigma_{A} + \frac{\sigma_{A}'}{2} \left(1 - \gamma^2_{A}\right) \right] 
\right.
\nn \\
&\left.
\times u^{\mu}_{A} u^{\nu}_{A} -  \left[\sigma_{A} + \sigma_{A}' \left(1 - \gamma_{A}\right) \right]
\gamma_AU^\m
U^\n \right\}\,.
\end{align}
The resulting field equations imply, in particular, that hypersurface-orthogonal solutions of Einstein-\AE ther theory are also solutions of khronometric theory, although the 
converse is not always true.

\subsubsection{PN Solution to the Modified Field Equations}
\label{sec:PNsols}

In this section, we review the solution to the field equations 
for a binary system of compact objects. In particular, we will
solve the equations within a PN approximation ($v_{12}/c\ll 1)$, for binaries
moving slowly relative to the \AE ther ($v_A/c\ll 1$) and described by the effective action of Eqs.~\eqref{eq:actionpp} and~\eqref{taylor-pp}.
Note that the solution will intrinsically depend on the parameter $\tilde m_A$ and 
the sensitivities $\sigma_{A}$ and $\sigma_{A}'$, but 
$\sigma'_{A}$ will enter at higher order than $\sigma_{A}$ in the perturbative expansion [c.f. Eq.~\eqref{taylor-pp}].
The solution presented here for the Einstein-\AE ther theory was derived in Ref.~\cite{Foster:2007gr}, while
the solution for khronometric theory is novel (but see Ref.~\cite{Blas:2011zd} for
a restricted solution with vanishing couplings between the point particles and the \AE ther). 
The derivations are standard and we will simply present the results here for completeness. 

Let us work in the standard PN gauge and PN coordinates $(t',x',y',z')$ \cite{TEGP}. In both Einstein-\AE ther and khronometric theory, the 1PN metric takes the form 
\allowdisplaybreaks
\bw 
\begin{align}
\label{g00PN}
g_{0'0'} &= 1 - \frac{1}{c^2} \frac{2 G_{N} \tilde{m}_{1}}{r_{1}'} + \frac{1}{c^{4}} \left[ \frac{2 G_{N}^{2} \tilde{m}_{1}^{2}}{r_{1}'^{2}} + \frac{2 G_{N}^{2} \tilde{m}_{1} \tilde{m}_{2}}{r_{1}' r_{2}'} + \frac{2 G_{N}^{2} \tilde{m}_{1} \tilde{m}_{2}}{r_{1}' r_{12}'} - \frac{3 G_{N} \tilde{m}_{1}}{r_{1}'} v_{1}'^{2} \left(1 + \sigma_{1}\right) \right]+ 1 \leftrightarrow 2 + {\cal{O}}(1/c^{6})\,,
\\
\label{g0iPN}
g_{0'i'} &= -\frac{1}{c^{3}} \left[B_{1}^{-} \frac{G_{N} \tilde{m}_{1}}{r_{1}'} v_{1}'^{i} + B_{1}^{+} \frac{G_{N} \tilde{m}_{1}}{r_{1}'} v_{1}'^{j}\hat n_{1}'^{j} \hat n_{1}'^{i} \right]+ 1 \leftrightarrow 2 + {\cal{O}}(1/c^{4})\,,
\\
\label{gijPN}
g_{i'j'} &= -\left(1 + \frac{1}{c^{2}} \frac{2 G_{N} \tilde{m}_{1}}{r_{1}'}\right) \delta_{ij} + 1 \leftrightarrow 2 + {\cal{O}}(1/c^{4})\,,
\end{align}
\ew
where $\tilde{m}_{A}$ is the mass of the $A$-th point particle [cf. Eq.~\eqref{taylor-pp}], $v_A'^i$ its velocity, $r_{12}'$ the binary's separation, $r_{A}'$ the distance from the $A$-th particle to the field point, $\hat n_{A}'^{i}$ the unit vector associated with $r_{A}'$ and the symbol $1 \leftrightarrow 2$ means that one is to add all terms on the right-hand side of the equality with the exchange $1 \leftrightarrow 2$. We have restored here factors of $1/c$ to make the PN order counting clearer. A term proportional to $(\tilde{m}_{A}/r_{A}')^{N}$ and $(|v_{A}'{}^{i}|/c)^{2N}$ is said to be of $N$-th PN order or of ${\cal{O}}(1/c^{2N})$. 

In Einstein-\AE ther theory, the values of the constants $B_{A}^{\pm}$ are
\allowdisplaybreaks
\begin{align}
\label{B-AE}
B^{\pm}_{A} &\equiv  \pm \frac{3}{2} -2 \pm \frac{1}{4} (\alpha^{\AE}_1 -
2\alpha^{\AE}_2) \left( 1 + \frac{2-c_{14}}{2 c_{+}-c_{14}}\sigma_{A}^{\AE} \right)
\nn \\ 
& - \frac{c_{-}}{c_1} \sigma_{A}^{\AE} - \frac{1}{4} \alpha^{\AE}_1 \left( 1+ \frac{c_{-}}{2 c_1} \sigma_{A}^{\AE} \right)\,, 
\end{align}
where the weak-field PPN parameters are 
\ba
\label{alpha1}
\hspace{-.3cm}\alpha^{\AE}_1 &=& -\frac{8(c_3^2+c_1c_4)}{2c_1-c_{+} c_{-}}\,, \\
\label{alpha2}
\hspace{-.3cm}\alpha^{\AE}_2 &=& \frac{\alpha^{\AE}_1}{2} - \frac{(c_1+2c_3-c_4)
  (2c_1+3c_2+c_3+c_4)}{(2-c_{14})c_{123}}\,,
  \ea
and we have defined
\ba
\label{cpm}
c_{\pm} &\equiv& c_{1} \pm c_{3}\,,
\\
c_{14} &\equiv& c_{1} + c_{4}\,,\label{c14}
\\
\label{c123}
c_{123} &\equiv& c_{1} + c_{2} + c_{3}\,.
\ea
In khronometric theory, the values of the constants $B^{\pm}_{A}$ are
\ba
\label{B-HL}
B^{\pm}_{A} & \equiv & \pm \frac{3}{2} -2 \pm \frac{1}{4} (\alpha^\kh_1 -
2\alpha^\kh_2) \left( 1 + \frac{2-\alpha}{2 \beta-\alpha} \sigma_{A}^\kh \right)
\nn \\ & & 
- 2 \sigma_{A}^\kh -\frac{1}{4} \alpha^\kh_1 (1 + \sigma_{A}^\kh)\,, 
\ea 
where now the weak-field PPN parameters are
\begin{align}
\label{alpha1-HL}
\alpha^\kh_1 &=  \frac{4(\alpha-2 \beta)}{\beta-1}\,,
\\ 
\label{alpha2-HL}
\alpha^\kh_2 &=
\frac{(\alpha-2\beta)[-\beta^2+\beta(\alpha-3)+\alpha+\lambda(-1-3\beta+2\alpha)]}{(\beta-1)(\lambda+\beta)(\alpha-2)}\,.
\end{align}

The \AE ther field takes the form
\begin{align}
\label{ut_AE}
U^{0} &= 1 + \frac{1}{c^2}\frac{G_{N} \tilde{m}_{1}}{r_{1}'} + 1 \leftrightarrow 2 + {\cal{O}}(1/c^{4})\,,
 \\\label{ui_AE}
U^{i} &= \frac{1}{c^3} \frac{G_{N} \tilde{m}_{1}}{r_{1}'} \left(C_{1}^{-}  v_{1}'^{i} + C_{1}^{+} v_{1}'^j \hat n_{1}'^{j} \hat n_{1}'^{i} \right)+ 1 \leftrightarrow 2 + {\cal{O}}(1/c^{5})\,,
\end{align}
where  for Einstein-\AE ther 
\ba
C^{\pm}_{A} & \equiv & \frac{8+\alpha^{\AE}_1}{8c_1} [ c_{-} - (1-c_{-})\sigma_{A}^{\AE} ]
\nn \\ & & \pm \frac{2-c_{14}}{2} \left( \frac{2\alpha^{\AE}_2-\alpha^{\AE}_1}{2
  (c_1+2c_3-c_4)} + \frac{\sigma_{A}^{\AE}}{c_{123}}  \right)\,, 
\ea
while for khronometric theory
\ba
C^{\pm}_{A} & \equiv & \frac{8+\alpha^{\kh}_1}{4} (1+\sigma_{A}^\kh)
\nn \\ & & \pm \frac{2-\alpha}{4} \left( \frac{2\alpha^{\kh}_2-\alpha^{\kh}_1}{
(2\beta-\alpha) } + \frac{2\sigma_{A}^\kh}{\beta+\lambda}  \right)\,. 
\ea
%

\subsubsection{The Sensitivities from the PN Metric}

The PN solution presented in the previous section provides a way to calculate the sensitivities in practice. 
Consider in particular the solution for the metric tensor, but specialize it to a single 
object (rather than a binary system) moving relative to the \AE ther. Mathematically, this is achieved by setting one 
of the masses (e.g.~$\tilde{m}_2$) to zero (or the separation $r'_{12}$ to infinity), but physically it corresponds to 
placing oneself at a distance $r_1'$ from object 1, such that $R_*\ll r_1'\ll r'_{12}$ (with $R_*$ being the radius of the object). The existence of this ``buffer''
region is possible because the PN scheme requires $r'_{12}\gg R_*$, in order to model the stars as point particles.

The advantage of considering an (effectively) single-object system is 
that exact strong-field solutions that do not rely on the point-particle approximation might exist  for such a system.
Recall that this is not the case for binary systems, for which only approximate PN solutions exist, 
even in GR. If such a solution for a single object 
moving relative to the \AE ther can be found, the metric near spatial infinity will be given by
the PN solution of Eqs.~\eqref{g00PN}--\eqref{gijPN}, with one of the masses set to zero. Because this solution depends on $\sigma_{A}$,
one can in principle read off this quantity from the behavior of the exact solution near spatial infinity.

What components of the metric near spatial infinity should we use to extract $\sigma_{A}$? 
One cannot use the spatial part of the metric $g_{i'j'}$, because this does not depend on $\sigma_{A}$. 
We could use the $(t',t')$ piece of the metric, but $\sigma_{A}$ enters multiplied by a term of ${\cal{O}}(1/c^{4})$. One can find $\sigma_{A}$ more easily by using the gravitomagnetic sector of the metric 
[Eqs.~\eqref{g0iPN},~\eqref{B-AE} and~\eqref{B-HL}], and in particular the ${\cal O}(v)$ terms. 
By using Eq.~\eqref{B-AE} and Eq.~\eqref{B-HL}, one easily finds
\begin{align}
\label{sigmaA-def-EA}
\sigma_{A}^{\AE} &= -\frac{2 c_{1} \left[2\left(B^{+}_{A} + B^{-}_{A}\right) + 8 + \alpha_{1}^{\AE}\right]}{\left(c_{1} - c_{3}\right) \left(8 + \alpha_{1}^{\AE}\right)}\,,
\end{align}
in Einstein-\AE ther theory and
\begin{align}
\label{sigmaA-def-HL}
\sigma_{A}^{\kh} &= -\frac{\left[2\left(B^{+}_{A} + B^{-}_{A}\right) + 8 + \alpha_{1}^{\kh}\right]}{\left(8 + \alpha_{1}^{\kh}\right)}\,,
\end{align}
in khronometric theory. 

The coefficients $B_{A}^{\pm}$ that are needed to compute the sensitivities are to be extracted 
from the exact solution for an isolated NS moving relative to the \AE ther, as measured by an observer
near spatial infinity. The exterior solution will depend on the interior solution, since both must be properly
matched, as we will see in Sec.~\ref{sec:NS-Sols}. 
Therefore, the coefficients $B_{A}^{\pm}$ will 
depend on the strong-field behavior of the solution in the stellar interior. Note also that these coefficients
appear in the metric [Eqs.~\eqref{g0iPN},~\eqref{B-AE} and~\eqref{B-HL}] at ${\cal{O}}(v)$, and thus, 
to extract the sensitivities we only need the strong-field solution at linear order in velocities. 

Clearly, this procedure can in principle be pushed to next order in $v$, i.e.
$\sigma_A'$ can be extracted from the ${\cal O}(v^2)$ terms of the strong-field solution
near spatial infinity. However, as already mentioned, $\sigma_A'$ enters the dynamics
of compact binaries at higher order than $\sigma_A$ in a small-velocity expansion. Because
the orbital velocities of observed binary pulsars are at most $v_{12}/c \sim 10^{-3}$ 
(a value reached by the double binary pulsar PSR J0737-3039A~\cite{manchester,kramer-double-pulsar,burgay,kramer-wex}),
and because the center-of-mass velocities relative to the \AE ther are on the order of the peculiar velocity
of our Galaxy (i.e. $V_{\rm CM}/c \sim 10^{-3}$), effects proportional to $\sigma_A'$ will be negligible 
relative to effects proportional to $\sigma_{A}$. 

\subsection{A Strong-Field Route}
\label{sec:strongf}

As shown in the previous section, the sensitivity parameters can be calculated from the metric
describing a body moving with velocity $v$ relative to the \AE ther.
One unsatisfactory aspect of that derivation, however, is that it uses a ``weak-field''
PN approach that models bodies with point-particles. Here, we will 
relax the weak-field and point-particles assumptions, and show that Eqs.~\eqref{sigmaA-def-EA}
and \eqref{sigmaA-def-HL} also follow from a strong-field definition of the sensitivities 
(i.e.~one that relates the sensitivities to the interior structure of the star,
where the PN/point-particle treatment of the previous section breaks down).
In particular, we will confirm that the calculation of the sensitivities only requires 
knowledge of the metric near spatial infinity, and at linear order in the velocity. 

While the former is not that surprising (e.g.~in GR, genuinely strong-field quantities, such as
the ADM mass, can be defined through the asymptotic behavior of the metric or in terms of 
integrals over the extent of the body), the latter is. In fact, 
one can generalize the weak-field definition of the sensitivities 
[Eqs.~\eqref{taylor-pp} and ~\eqref{sigma-def}]
to the strong field by replacing the gravitational mass of the point-particle, $\tilde{m}(\gamma)$, with its strong-field counterpart, the 
total mass-energy of the body, $M_{\tot}$, and thus define
\begin{align}\label{sigma-def-2}
\sigma &=- \left.\frac{\partial \ln M_{\tot}}{\partial \ln
  \gamma}\right|_{v=0}\nn \\&=- 2 \left.\frac{\partial \ln M_{\tot}}{\partial
  (v^2)}\right|_{v=0}= - \left.\frac{\partial^2 \ln M_{\tot}}{\partial
  {v}^2}\right|_{v=0}\,.
\end{align}
It can be shown~\cite{Foster:2005fr,Eling:2005zq} that $M_{\rm tot}$ is indeed the strong-field generalization of the  point-particle mass $\tilde{m}(\gamma)$,
because far away from the star $g_{tt}= 1-2 G_N M_{\rm tot}/r+...$, which agrees with the point-particle solution of
Eq.~\eqref{g00PN} for $\tilde{m}(\gamma)=M_{\rm tot}$. As for $v$ (and $\gamma$), as mentioned above, they are the velocity (and Lorentz factor) of the body relative to the \AE ther, so in a coordinate system
comoving with the star (that is, one where the star is at rest), they are defined in terms of the behavior of
the \AE ther far from the star. 
In other words, in a system of asymptotically Cartesian coordinates where the motion is
along the $z$-axis, one has $U^{\mu}\partial_\mu=
(\partial_t-v \partial_z)/\sqrt{1-v^2} +{\cal O}(1/r)$ and
$g_{\mu\nu}=\eta_{\mu\nu}+{\cal O}(1/r)$.

What is confusing about Eq.~\eqref{sigma-def-2}, however, is that it would seem
to imply that the mass $M_{\tot}$ (and therefore the metric) need to be calculated at 
order ${\cal O}(v^2)$ to extract the sensitivities, in contrast with the results of the previous section.
In what follows, we will show that Eq.~\eqref{sigma-def-2} \textit{does} however lead to the same expressions [Eqs.~\eqref{sigmaA-def-EA} and \eqref{sigmaA-def-HL}]
for the sensitivities as the PN treatment of the previous section, 
essentially thanks to Gauss' theorem. This will confirm that one can extract the sensitivity from the ${\cal O}(v)$ pieces of the metric alone.
The procedure we outline below is similar to that used in scalar-tensor theories to extract the sensitivities of NSs
from the asymptotic behavior of the scalar field at spatial infinity~\cite{damour_esposito_farese}
(see also Ref.~\cite{Gralla:2013rwa} for an approach similar to ours.)

\subsubsection{Einstein-\AE ther Theory}

As mentioned above, we consider a star at rest, and the \AE ther moving relative to it, so that 
$U^{\mu}\partial_\mu=
(\partial_t-v \partial_z)/\sqrt{1-v^2} +{\cal O}(1/r)$ and
$g_{\mu\nu}=\eta_{\mu\nu}+{\cal O}(1/r)$ in asymptotically Cartesian coordinates. We also assume that
the system is in a stationary regime (i.e.~that the metric and \AE ther do not depend on the time coordinate), which
is expected to be the case after an initial transient. For such a system, we can write the total mass
as~\cite{damour_esposito_farese}
\begin{equation}\label{ae:mass}
M_{\tot}=-\int_\Sigma d^3 x ({\cal L}_{g}+{\cal L}_{\AE}+{\cal L}_{\MAT})\,,
\end{equation}
where ${\cal L}_{g}$ is the Einstein Lagrangian density\footnote{The Einstein Lagrangian
  density, unlike the Einstein-Hilbert one (from which it differs
  by a total divergence), depends only on first derivatives of the
  metric. Explicitly, ${\cal L}_{g}=-\frac{1}{16 \pi G_{\AE}}\sqrt{-g} g^{\mu\nu}
  (\Gamma^\alpha_{\mu\lambda}\Gamma^{\lambda}_{\nu\alpha}-\Gamma^\lambda_{\mu\nu}\Gamma^\alpha_{\lambda\alpha})$
  (cf.~e.g. Ref.~\cite{landau}).},  
${\cal L}_{\AE}$ is the \AE ther Lagrangian density
  \begin{equation}
{\cal L}_{\AE}= -\frac{1}{16 \pi G_{\AE}}\sqrt{-g}
M^{\alpha\beta}_{\phantom{ab}{\mu\nu}} \nabla_\alpha U^\mu\nabla_\beta
U^\nu\,,
\end{equation}
[the Lagrange multiplier term $\ell (U_\mu U^\nu-1)$ is not included
here because Eq.~\eqref{ae:mass} is to be evaluated on shell],
  ${\cal L}_{\MAT}$
is the matter's Lagrangian density,
and $\Sigma$ is a hypersurface of constant time.
We stress that one can write the total mass in terms of the Lagrangian densities alone because
the fields have no explicit time dependence (i.e.~in the absence of time derivatives, the Hamiltonian is simply $H=\dot{q} p -L =-L$),
and because the Einstein Lagrangian density ${\cal L}_{g}$
only depends on the metric and (quadratically) on its first derivatives.
We also note that the definition given by Eq.~\eqref{ae:mass}
was also used in Ref. \cite{Foster:2005fr} for the study of black hole mechanics, and 
later in Ref. \cite{Foster:2006az} in the context of GW emission. 
For khronometric theory, it
was used in Ref. \cite{Blas:2011zd}, where a brief summary with references to related work can be found.

The sensitivities can then be obtained by varying the mass of
Eq.~\eqref{ae:mass}.
When doing so, the bulk
terms evaluate to zero because of the Euler-Lagrange equations and one
is left with surface terms alone~\cite{damour_esposito_farese,Foster:2005fr,Blas:2011zd}. 
For a star, the hypersurface $\Sigma$ can be extended to
its center, and we thus have only surface terms at the
outer boundary (spatial infinity). 

More in detail, because we are taking
the difference between two neighboring solutions of the modified field
equations, the \AE ther and metric variations $\delta U^\mu$ and
$\delta g_{\mu\nu}$ preserve the unit constraint $U^\mu
U_\mu=1$. Thus, the bulk terms lead to the modified field
equations  and thus vanish, even though Eq.~\eqref{ae:mass} does not contain the
Lagrange multiplier term. Similarly, the surface terms coming from the
variation of ${\cal L}_{g_{{}}}$ vanish, because we are working in a gauge where
$g_{\mu\nu}=\eta_{\mu\nu}+{\cal O}(1/r)$~\cite{damour_esposito_farese}. Also, the surface terms
coming from the variation of the matter Lagrangian 
vanish because we assume the matter fields to be confined within the star, while the boundary of the
hypersurface $\Sigma$ can be pushed to spatial infinity.

As a result, the only surface terms that contribute in Einstein-\AE ther theory are those coming
from the \AE ther Lagrangian,
\be
\label{AE_variation}
\delta M_{\tot}=-\int_{\partial\Sigma} d^2S_i \delta U^\mu
\left(\frac{\partial {\cal L}_{U}}{\partial (\partial_i U^\mu)}
\right)\,.  \ee

Let us then evaluate Eq.~\eqref{AE_variation} for a moving stationary
configuration. Clearly, $\delta U ^{\mu}$ is the difference between
the \AE ther 4-velocities of two neighboring moving solutions with
velocities $v$ and $v+\delta v$, while the derivative $\partial {\cal
  L}_{U}/\partial (\partial_i U^\mu)$ must be calculated for a moving
solution with velocity $v$. It is more convenient to use
asymptotically spherical coordinates, in which
Eq.~\eqref{AE_variation} becomes 
\be
\label{AE_variation_explicit}
\delta M_{\tot}= \frac {1}{4 G_{\AE}}\lim_{r\to\infty} \int_{0}^\pi d\theta\, r^2
\sin\theta\, \delta U^\mu J^{r}_{\phantom{a}\mu}  \,.
\ee 

At lowest order and in Cartesian coordinates, the 4-velocity variation
is $\delta U^{\mu} = \delta v (v \delta^\mu_t-
\delta^\mu_z)+{\cal{O}}(1/r)+{\cal{O}}(\delta v^2)$. In spherical
coordinates, this becomes
\begin{align}
\delta U^{\mu} &= \delta v\left (v \delta^\mu_t- \cos\theta
\delta^\mu_r + \frac{\sin\theta}{r} \delta^\mu_\theta\right)
\left[1+{\cal O}\left(\frac1r,\delta v\right)\right]\nn\,.\\
\end{align}

To calculate $J^{r}_{\phantom{a}\mu}$, we 
need to solve the modified field equations 
for a system comprised of a star at rest 
and an \AE ther moving slowly relative to it.
In particular, expanding the equations at linear order in
the \AE ther's velocity and near spatial infinity, one obtains the solution  \ba
\label{metric-Foster-sph0text}
&& ds^2 = dt^2-dr^2 +\Bigg[-\frac{2M_*}{r}  (dt^2 + dr^2) \nn\\ &&
- r^2 d\theta^2 - r^2 \sin^2 \theta
d\varphi^2 \nn \\ & &   -2 v (B^{-} + B^{+}+4) \frac{M_*}{r} \cos \theta
dt dr\nn\\&& + v (7+2B^{-}) M_*  \sin\theta dt d\theta\Bigg]\times\left[1+{\cal O}\left(v,\frac1r\right)\right]\,,\nn\\ \\
\label{AE-Foster-sph0text}
&& U_\mu dx^\mu = dt+v  \cos\theta dr-vr \sin \theta d\theta\nn\\&&+\Bigg[-\frac{M_*}{r} dt- v
  (1+B^{-}+B^{+}+C^{-}+C^{+}) \frac{M_*}{r} \cos\theta dr\nn\\&&+
  vr (3+2 B^{-}+2C^{-}) \frac{M_*}{2r}  \sin \theta d\theta\nn\Bigg]\\&&\times\left[1+{\cal O}\left(v,\frac1r\right)\right]\,,  
 \ea
  with 
    \begin{align}
\label{B-AE_strongfield}
B^{\pm} &\equiv  \pm \frac{3}{2} -2 \pm \frac{1}{4} (\alpha^{\AE}_1 -
2\alpha^{\AE}_2) \left( 1 + \frac{2-c_{14}}{2 c_{+}-c_{14}}\bar{\sigma}_{\AE} \right)
\nn \\ 
& - \frac{c_{-}}{c_1} \bar{\sigma}_{\AE} - \frac{1}{4} \alpha^{\AE}_1 \left( 1+ \frac{c_{-}}{2 c_1} \bar{\sigma}_{\AE} \right)\,, \\
\label{C-strongfield}
C^{\pm} & \equiv  \frac{8+\alpha^{\AE}_1}{8c_1} [ c_{-} - (1-c_{-})\bar{\sigma}_{\AE}]
\nn \\ &  \pm \frac{2-c_{14}}{2} \left( \frac{2\alpha^{\AE}_2-\alpha^{\AE}_1}{2
  (c_1+2c_3-c_4)} + \frac{1}{c_{123}} \bar{\sigma}_{\AE} \right)\,, 
\end{align}
where the length scale $M_*$ can be shown to be related to
the total mass of the star, $M_{\tot}$, by $M_*=G_N M_{\tot}$~\cite{Foster:2005fr,Eling:2005zq}, while
  $\bar{\sigma}_{\AE}$ is a  free parameter characterizing the asymptotic behavior of the
  solution, and which a priori might have no relation to the sensitivity of Eq.~\eqref{sigma-def-2}.
  To compute this parameter, one   needs to solve the modified field equations in the stellar interior and match
to an exterior solution, as we   do in the next section. 
Not surprisingly, Eqs.~\eqref{metric-Foster-sph0text} and \eqref{AE-Foster-sph0text}
can also  be obtained directly 
by transforming the PN solution in Eqs.~\eqref{g00PN}, \eqref{g0iPN}, \eqref{gijPN}, \eqref{ut_AE}, \eqref{ui_AE} to the appropriate gauge
and identifying $\sigma\mapsto \bar \sigma$.

Using these expressions, one gets
\begin{align}
J^{r}_{\phantom{a}\mu} &=\Bigg\{\delta_\mu^t c_3 \frac{M_*}{r^2} -v
\delta_\mu^r \frac{M_* \cos \theta}{r^{2}} \left[c_2 \left(C_- - C_+ - 3\right)
\right.  
\nn \\
&+ \left. 
c_+ \left(C_-+C_+-2\right)-c_4\right] - v \delta_\mu^\theta \frac{M_* \sin \theta}{2r} 
\nn \\
&\times 
\left[2 \left(-c_1 C_-+c_3 C_++c_3\right)-\left(B_++B_-\right) c_-\right] \Bigg\}
\nn \\
&\times
\left[1+{\cal{O}}\left(\frac1r,v\right)\right]\,,
\end{align}
with which Eq.~\eqref{AE_variation_explicit} can be evaluated explicitly to find
\be \delta M_{\tot} = -\bar{\sigma}_{\AE} M_{\tot} v \; \delta v  \left[1 +{\cal O}(v,\delta
  v)\right] \,, \ee
where we have used the relation $M_*=G_N M_{\tot}$ mentioned above.
We therefore have
\be \frac{\partial \ln M_{\tot}}{\partial {v}} ({v})= -\bar{\sigma}_{\AE} \; v \,, \ee
and the sensitivity is
\begin{align}
\label{sigmaA-def-EA2}
\sigma_{\AE} &=-  \left.\frac{\partial^2 \ln M_{\tot}}{\partial  {v}^2}\right|_{v=0} \nn \\
 &= \bar{\sigma}_{\AE} 
 = -\frac{2 c_{1} \left[2\left(B^{+} + B^{-}\right) + 8 + \alpha^{\AE}_{1} \right]}{\left(c_{1} - c_{3}\right) \left(8 + \alpha^{\AE}_{1}\right)}\,, 
\end{align}
which reduces exactly to Eq.~\eqref{sigmaA-def-EA}.  

\subsubsection{Khronometric Theory}

As discussed in Sec.~\ref{sec:MGT}, khronometric theory can be thought of as
Einstein-\AE ther theory with hypersurface orthogonality [Eq.~\eqref{U_dT}]
being imposed in the action before variation.
Equivalently, khronometric theory can be derived from the same action as
Einstein-\AE ther theory, plus four Lagrange multipliers $\ell_\alpha
\omega^\alpha$ that enforce that the vorticity vector $\omega^\alpha=
\epsilon^{\alpha\beta\gamma\delta} U_\beta \partial_\gamma U_\delta$
vanishes. Here, $\epsilon^{\mu\nu\alpha\beta}=\tilde{\epsilon}^{\mu\nu\alpha\beta}/\sqrt{-g}$, where 
$\tilde{\epsilon}^{\mu\nu\alpha\beta}$ is the Levi-Civita symbol.

Such a formulation of the action has the advantage of containing only first
(and not second) derivatives of the fields (and in particular of the \AE ther field $U_\alpha$). This allows one to
define Lagrangian densities in the usual way and proceed as in the previous section for Einstein-\AE ther theory 
to define the sensitivities in the strong field regime. Because
the Lagrange multiplier terms vanish on shell, the mass of a
time-independent configuration is given again by Eq.~\eqref{ae:mass},
exactly as in the Einstein-\AE ther case. When taking the difference
between the mass of two neighboring solutions, the bulk terms vanish
as a result of the modified field equations, and one is left with
surface terms alone. Exactly as in the Einstein-\AE ther case, the
surface terms coming from ${\cal L}_{g_{{}}}$ vanish in a gauge
where $g_{\mu\nu}=\eta_{\mu\nu}+{\cal O}(1/r)$, and so do the terms
coming from the variation of the matter Lagrangian, because the matter 
field are confined within the star.

As for the surface terms coming from the Einstein-\AE ther Lagrangian,
using Eq.~\eqref{U_dT} we can express the \AE ther variation $\delta U_\mu$ in
terms of the variation $\delta T$ of the foliation as
\be \delta U_\mu= -\frac{1}{2} U_\mu U_\nu U_\lambda \delta
g^{\nu\lambda} +\frac{\delta^\nu_\mu -U^\nu U_\mu}{\sqrt{g^{\alpha\beta}\partial_\alpha T
    \partial_\beta T}} \partial_\nu \delta T\,.  \ee
A simple calculation then yields
\be \delta M_{\tot} = - \int d^3 x\, \partial_i \left[ \frac{\partial {\cal
      L}_{U}}{\partial (\partial_i U_\mu)} \delta U_\mu+
  \frac{(\delta^i_\mu -U^i U_\mu)  \bar{\AE}^\mu \delta
    T}{\sqrt{g^{\alpha\beta}\partial_\alpha T \partial_\beta T}}
  \right] \ee
with 
\begin{align}
\bar{\AE}^\mu&=\frac{\partial {\cal L}_{U}}{\partial
  U_\mu}-\partial_i\left[\frac{\partial {\cal L}_{U}}{\partial
    (\partial_i U_\mu)}\right]\,\nn \\&=-\frac{1}{8 \pi G_{\AE}} \sqrt{-g}
\left(\alpha \dot{U}^\nu \nabla^\mu U_\nu-\nabla_\alpha J^{\alpha \mu}\right).
\end{align}
From Gauss' theorem, we then have 
\begin{equation}
\delta M_{\tot} = - \int d^2S_i \left[ \frac{\partial {\cal L}_{U}}{\partial
    (\partial_i U_\mu)} \delta U_\mu+ \frac{(\delta^i_\mu -U^i U_\mu)
    \bar{\AE}^\mu \delta T}{\sqrt{g^{\alpha\beta}\partial_\alpha T
      \partial_\beta T}}  \right]\,. \label{surfaceIntHL}
\end{equation}
This expression must be evaluated with boundary conditions
$g_{\mu\nu}=\eta_{\mu\nu}+{\cal O}(1/r)$
and $U^{\mu}\partial_\mu=
(\partial_t-v \partial_z)/\sqrt{1-v^2} +{\cal O}(1/r)$.

In a suitable gauge, the asymptotic solution for a slowly moving star is given by 
\ba
\label{metric-HL-asymp_2}
&& ds^2 = dt^2-dr^2+
\Bigg\{-\frac{2M_*}{r} (dt^2+dr^2)  \nn \\ & & - r^2 \left( d\theta^2 +
\sin^2 \theta d\varphi^2 \right) \nn \\ & &   -2 v \left[ (B^{-} +
B^{+}+4) \frac{M_*}{r} \right] \cos \theta dt dr  \nn \\ & & + 2v
r \left[  (3+B^{-}-J) \frac{M_*}{r} \right] \sin\theta dt
d\theta\Bigg\}\times \left[1+{\cal O}\left(v,\frac1r\right)\right]\,,\nn\\ \\
&&{U}_{\m} dx^\m =(dt+v \cos\theta
dr  -vr\sin\theta d\theta)\nn \\ &&
\times\left[1-\frac{M_*}{r}+{\cal O}\left(\frac{1}{r^2}\right)\right]+{\cal O}(v^2)\,,
\label{HL-Foster-sph}
\ea
with 
\ba
\label{B-HL-strongfield}
B^{\pm} & \equiv & \pm \frac{3}{2} -2 \pm \frac{1}{4} (\alpha^{kh}_1 -
2\alpha^{kh}_2) \left( 1 + \frac{2-\alpha}{2 \beta-\alpha} \bar{\sigma}_{\kh} \right)
\nn \\ & &
- 2 \bar{\sigma}_{\kh} -\frac{1}{4} \alpha^{kh}_1 (1 + \bar{\sigma}_{\kh})\,,\\
\bar{J}& \equiv&
\frac{(2+3\lambda+\beta)[2\beta+\alpha(-1- \bar{\sigma}_{\kh}) + 2 \bar{\sigma}_{\kh}]}{2(\lambda+\beta)(\alpha-2)}\,,
\ea
where $M_*=G_N M_{\tot}$~\cite{Foster:2005fr,Eling:2005zq} and $\bar{\sigma}_{\kh}$ is again a constant (a priori unrelated to the sensitivity) 
that characterizes the solutions and which
needs to be calculated from the full slowly moving solution (i.e.~from the solution describing both the interior and exterior of the star).
As in the Einstein-\AE ther case, Eqs.~\eqref{metric-HL-asymp_2} and \eqref{HL-Foster-sph} can be obtained by solving 
the modified field equations directly at ${\cal O}(v)$ near spatial infinity, or also by transforming
Eqs.~\eqref{g00PN}, \eqref{g0iPN}, \eqref{gijPN}, \eqref{ut_AE}, \eqref{ui_AE} after identifying $\sigma\mapsto \bar \sigma$.

From Eqs.~\eqref{metric-HL-asymp_2} and \eqref{HL-Foster-sph}, one obtains
  $\bar{\AE}^\mu={\cal O}(v)$ and $T=t+v r \cos\theta$, so one can rewrite Eq.~\eqref{surfaceIntHL} as 
\begin{align}
\label{HL_variation_explicit}
\delta M_{\tot} &= \frac {1}{4 G_{\AE}}\lim_{r\to\infty} \int_{0}^\pi
d\theta\, r^2 \sin\theta  \nn \\ &\times\left[ \delta U_\mu
  J^{r\mu} +\left(\alpha \dot{U}^\nu \nabla^r U_\nu-\nabla_\alpha J^{\alpha
    r}\right) {\delta T}\right]+{\cal O}(v^2) \,.
\end{align}
Finally, we can write 
\begin{align}
\delta U_{\mu} &= \delta v\left (v \delta_\mu^t+ \cos\theta
\delta_\mu^r - r {\sin\theta} \delta_\mu^\theta\right)\left[1+{\cal O}\left(\frac1r,\delta v\right)\right]\,,\nn\\
\delta T& = \delta v \,r \cos\theta\,,
\end{align}
and use Eqs.~\eqref{metric-HL-asymp_2} and \eqref{HL-Foster-sph} to explicitly compute
\begin{align}
J^{r \mu} &=  \Bigg\{\delta^\mu_t \beta \frac{M_*}{r^2} -v \delta^\mu_r \frac{M_* \cos \theta}{r^{2}} 
\left[ (2 + B^- + B^+) \beta + \alpha 
\right. 
\nn \\ 
&+ \left.
\lambda (3 + B^- - B^+ - 2 \bar{J}) \right] + v \delta^\mu_\theta \frac{\beta M_* \sin \theta}{2 r^{3}} 
\nn \\
&\times \left(2 + B^- - B^+ - 2 \bar{J}\right) \Bigg\}
\left[1+{\cal{O}}\left(\frac1r,v\right)\right]\,,  
\\ 
\alpha \dot{U}^\nu & \nabla^r U_\nu-\nabla_\alpha J^{\alpha r} =\nn\\ & v \frac{  M \cos \theta}{r^{3}} 
\left[2 \lambda \left(-B_- +B_+ +2 \bar{J}-3\right)
\right. 
\nn \\
&+\left. 
\beta \left(-B_-+3 B_+ +4 \bar{J}-2\right)-2 \alpha\right]
\left[1+{\cal{O}}\left(\frac1r,v\right)\right]\,.
\end{align}

Inserting these expressions in Eq.~\eqref{HL_variation_explicit}, one
easily obtains
\be
\label{dMKH}
\delta M_{\tot} = -\bar{\sigma}_{\kh} M_{\tot}  v \; \delta v \left[1 +{\cal O}(v,\delta v)\right]
\,, \ee
where we have used again $M_*=G_N M_{\tot}$. We then have
\be 
\frac{\partial \ln M_{\tot} }{\partial {v}} = -\bar{\sigma}_{\kh} {v} \,, 
\ee
and the sensitivity 
\begin{equation}\label{sigmaA-def-HL2}
\sigma_{\kh} = - \left. \frac{\partial^2 \ln M_{\tot}}{\partial
  {v}^2}\right|_{v=0}=\bar{\sigma}_{\kh}=-1 - \frac{2 \left(B^{+} + B^{-}\right)}{8 + \alpha^{kh}_{1}}\,,
  \end{equation}
which reduces exactly to Eq.~\eqref{sigmaA-def-HL}. 

\subsection{Effect of the Sensitivities on the Motion of Binary Systems}
\label{sec:cons-sect}

The strong equivalence principle is defined as the universality of free fall for strongly gravitating bodies. GR satisfies this principle, but this is clearly not the case for theories in which the sensitivities are not zero. This is because
the sensitivities, as described previously, characterize how the structure of
a compact object (i.e.~a NS) changes with the motion relative to the ambient field in which the object
is immersed (i.e.~the \AE ther field in Einstein-\AE ther or khronometric theory). As we will show later, the sensitivities depend on the particular object under consideration, 
and in particular, on its compactness. Therefore, unless the sensitivities are exactly zero as in GR, different bodies
respond differently to  motion relative to the ambient field, and thus move along different trajectories, 
violating the strong-equivalence principle. 

In the case of Einstein-\AE ther and khronometric theory, the sensitivities affect
both the conservative and dissipative sectors. 
As for the former, the sensitivities modify Newton's universal gravitation law, i.e.~at Newtonian
order the motion of a binary is described by~\cite{Foster:2007gr}
\begin{equation}
\label{Newtonian-a}
\dot{v}_A^i= -\frac{G_N \tilde{m}_B \hat n_{AB}^i}{(1+\sigma_A) r_{AB}^2}\,,
\end{equation}
where $r_{AB}=|\boldsymbol{x}_A-\boldsymbol{x}_B|$ and $\hat n_{AB}^i=(x^i_A-x^i_B)/r_{AB}$.
Let us rewrite this expression as~\cite{Foster:2007gr}
\begin{equation}
\label{eq:New_active}
\dot{v}_A^i= -\frac{{\cal G} {m}_B \hat n_{AB}^i}{r_{AB}^2}\,,
\end{equation}
where we define the {\textit{active}} gravitational masses as
\be
\label{active-mass}
m_B\equiv\tilde{m}_B (1+\sigma_B)
\ee
and the 2-body coupling constant
\be
\label{calG}
{\cal G}\equiv\frac{G_N}{(1+\sigma_A) (1+\sigma_B)}\,.
\ee
Similarly, the sensitivities enter the equations of motion also at higher PN order in the conservative sector~\cite{Foster:2007gr}.

When it comes to the dissipative sector, and in particular to binary pulsars, 
the most well-known observable is the rate of change
of the orbital period. In fact, it was the monitoring of this quantity that led to the 
first indirect detection of GWs by Hulse and 
Taylor~\cite{Hulse:1974eb,Taylor:1982zz,Taylor:1989sw}. For orbits satisfying Eq.~\eqref{eq:New_active}, this observable can be written as
\be
\label{Pdot-eq}
\frac{\dot{P_{b}}}{P_{b}} = - \frac{3}{2} \frac{\dot{E}_{b}}{E_{b}}\,,
\ee
where $P_{b}$ is the orbital period and $E_{b}$ is the binary's binding energy. 
In deriving this relation, we have used the fact that the conservative
sector is corrected only as in Eq.~\eqref{Newtonian-a} to leading PN order.
Thus, the binding energy is given by $E_{b} = - {\cal{G}} \mu m/(2 a)$ and the orbital period
by $P_{b} = 2 \pi a^{3/2}/({\cal{G}} m)^{1/2}$, as in Newtonian orbital mechanics, but in terms
of the 2-body coupling constant ${\cal{G}}$ and the active gravitational masses $m_{A}$. Here, $a$ 
is the semi-major axis and
\be
\label{eq:defM}
\mu \equiv m_{1} m_{2}/m, \quad m \equiv m_{1} + m_{2},
\ee 
are the reduced (active) mass 
and the total (active) mass respectively.

Equation~\eqref{Pdot-eq} can be further manipulated 
by relating the rate of change of the binding energy to the total flux of energy ${\cal{F}}$ carried
away from the system, 
\be
\dot{E}_{b} = - {\cal{F}}\,.
\label{balance-law}
\ee 
In GR, $\cal{F}$ is only
due to the propagation of tensor modes (i.e.~GWs), but in modified theories one normally 
finds radiation from scalar and vector modes. The balance law in Eq.~\eqref{balance-law} 
should hold because the total energy of the system is a conserved quantity \cite{Blas:2011zd,Foster:2005fr,Eling:2005zq}.

The energy flux carried by all propagating degrees of freedom can be
computed by investigating the Noether charges and currents in the
theory under consideration. These charges and currents depend on the solution to the
evolution equations for the metric perturbations on a  Minkowski
background in the \textit{far-zone}, i.e. at distances much larger than the 
GW wavelength as measured in the system's center
of mass frame. In what follows, we will summarize how to compute this
flux in Einstein-\AE ther and khronometric theory, which then
leads to a prediction for the orbital decay rate via Eq.~\eqref{Pdot-eq}.

\subsubsection{Dissipative PN Dynamics in Einstein-\AE ther Theory}

The energy flux in Einstein-\AE ther theory and for a matter action of the form of Eq.~\eqref{taylor-pp} was first calculated by Foster~\cite{Foster:2007gr}. After re-deriving these results we have found important discrepancies with Foster's work, which justifies showing the calculation in some detail here. 
In the following, we present the correct formulas that correct the miscalculations in Refs.~\cite{Foster:2007gr}.

The energy flux is carried by degrees of freedom that propagate away from the source. In this region, it is convenient to treat the metric
and \AE ther fields as perturbations around a Minkowski background $\eta_{\m\nu}$ with the \AE ther pointing in the time direction and decompose
their spatial components as 
\be
\label{eq:metric_dec}
\begin{split}
&h_{0i}=\gamma^h_i+\gamma^h_{,i}, \quad U^i=\nu^i+\nu_{,i},\\
&h_{ij}=\phi_{ij}+\frac{1}{2\Delta}\left(\delta_{ij}\Delta F^h-F^h_{,ij}\right)+2\phi_{(i,j)}+\phi^h_{,ij},
\end{split}
\ee
where $\Delta$ is the Laplacian differential operator and
\be
\gamma^h_{i,i}=\nu^i_{,i}=\phi_{ii}=\phi_{i,ij}=\phi_{i,i}=0.
\ee
Once the previous decomposition is used in the field equations, Eqs.~\eqref{E_def} and \eqref{eq:AEsens}, one can follow the calculation in 
Ref.~\cite{Foster:2007gr} to find 
the waveform at a distance $r$ from the source (note the opposite sign due to different conventions)
\be
\phi_{ij}=-\frac{2 G_{\AE}}{r} \ddot Q_{ij}^{TT}(t-r/w_2), 
\ee
where 
\be
\label{w2-def}
w_2^2 \equiv \frac{1}{1-c_{+}}, 
\ee
is the speed of propagation of the $\phi_{ij}$ modes and
\be
\label{Q-mom}
Q_{ij} =I_{ij}-\frac{1}{3} \delta_{ij} I_{kk}\,, 
\ee
with $I_{ij}$ given by 
\begin{align}
\label{I-mom}
I_{ij} &= \sum_{A}  m_A x^i_A x^j_A  \left[1+{\cal{O}}\left({1}/{c^{2}}\right)\right]\,,
\end{align}
for a system of $A$ bodies. Recall that here $x_{A}^{i}(t)$ are the trajectories of the $A$-th point
particle, while $v_{A}^{i}(t) = \dot{x}_{A}^{i}$ are their  3-velocities.

The transverse-traceless projector is built with the unit-norm vector $\hat n^i\equiv r^i/r$, where $r^i$ are the coordinates of the point
where the fields are being computed \cite{Blas:2011zd}.
To derive the previous expressions we used the conservation properties of an improved  energy-momentum tensor 
as defined in Ref.~\cite{Foster:2007gr}.
  For the vector modes, we choose the gauge $\phi_i=0$. The field $\gamma_i^h$ can be solved as
\be
\gamma_i^h=c_+ \nu^i,
\ee
up to terms with vanishing time derivative.
 Finally, 
the equation for $\nu^i$ yields
\begin{multline}
\label{eq:wave_v}
\nu^i=\frac{-2 G_{\AE}}{(2c_1-c_+ c_-)\,r}\times\\\left(\frac{\hat n^j}{w_1}\left[\frac{c_+}{1-c_+}\ddot Q_{ij}+\ddot{\cal Q}_{ij}+{\cal V}_{ij}\right]-2\Sigma^i\right)^T.
\end{multline}
Here, the different tensors on the right-hand side are to be evaluated at a retarded time $t-r/w_1$,
where $w_1$ is the speed of propagation of the vector modes,
\be
\label{w1-def}
w_1^2 \equiv \frac{2 c_1 - c_{+} c_{-}}{2(1-c_{+}) c_{14}}\,.
\ee 
Also, ${\cal Q}_{ij}$ is the trace-free part of the rescaled mass  quadrupole moment
\begin{align}
{\cal I}_{ij} &=\sum_{A} \sigma_A \tilde m_A x_A^i x_A^j (1+{\cal{O}}(1/c^{2})) \,, 
\end{align}
and we have also introduced the tensors\footnote{The tensor ${\cal V}_{ij}$ is absent in the calculations of
Ref.~\cite{Foster:2007gr}, which is a mistake. For the bounds derived later, this
term plays a subleading role.}
\begin{align}
{\cal V}_{ij}&=2\sum_A \sigma_A \tilde m_A \dot v^{[i}_A x_A^{j]} \left[1+{\cal{O}}\left({1}/{c^{2}}\right)\right]\,,\\
\label{Sigma-mom}
\Sigma^i &=-\sum_A\sigma_A \tilde m_A v^i_A \left[1+{\cal{O}}\left({1}/{c^{2}}\right)\right]\,.
\end{align}

Finally, for the scalar sector, once we impose the gauge $\nu=\gamma^h=0$, the equations for the
fields $\phi_h$ and $h_{00}$ in the far-zone read\footnote{At the PN order we work to, we do not need the ${\cal O}(1/c^4)$  in Eq.~\eqref{eq:scalarswaveEA} since it corresponds to a conserved  quantity, see Ref.~\cite{Foster:2007gr}.} 
\begin{align}
\label{eq:scalarswaveEA}
&\Delta h_{00} =\frac{\Delta F}{c_{14}}-\frac{16 G_{\AE}}{c_{14}} \sum_A \tilde m_A  
\nonumber \\ 
&\qquad \; \;
\times \delta^{(3)}(x^i-x^i_A(t)) \left[1+{\cal{O}}\left({1}/{c^{2}}\right)\right]\,,
\\
&c_{123}\Delta \dot \phi^h+(1+c_2)\dot F=-16 \pi G_{\AE} \hat n^i\sum_A m_A  v^i_A
\nonumber \\
&\quad \quad \times \delta^{(3)} (x^i-x^i_A(t)) \left[ 1 + {\cal{O}} \left({1}/{c^{2}} \right)\right]\,,  
\end{align}
 while
\begin{align}
\label{F-mode}
F&=\frac{4 G_{\AE} c_{14}}{(c_{14}-2)r}\left[\frac{3}{2}(Z-1) \hat n^i 
\hat n^j \ddot Q_{ij}+\frac{1}{2} Z \ddot I_{kk} \right.  \nn \\ &\left.
-\frac{\hat n^i\hat n^j}{c_{14} w_0^2}(\ddot{\cal Q}_{ij}+\frac{1}{3}
\delta_{ij}\ddot{\cal I}_{kk})+\frac{2}{c_{14} w_0} \hat n^i
\Sigma_i\right]+{\cal O}(1/c^{5})\,,
\end{align}
where we have defined
\be 
\label{Z-def}
Z \equiv \frac{(\alpha^{\AE}_1-2\alpha^{\AE}_2)(1-c_+)}{3(2c_+-c_{14})}\,,
\ee
and the speed of the $F$ mode is
\be
\label{w0-def}
w_0^2 \equiv \frac{(2-c_{14}) c_{123}}{(2+3c_2+c_{+}) (1-c_{+}) c_{14}}\,.
\ee
These constants depend on $(c_{\pm},c_{13},c_{14},c_{123})$ [Eqs.~\eqref{cpm}-\eqref{c123}]
and $(\alpha_{1}^{\AE},\alpha_{2}^{\AE})$ [Eqs.~\eqref{alpha1}-\eqref{alpha2}]. 
Note that although $w_{0}$ is the same as that defined in Ref.~\cite{Foster:2005dk,Foster:2006az,Foster:2007gr},
$Z$ and the structure of Eq.~\eqref{F-mode} differ from Ref.~\cite{Foster:2006az,Foster:2007gr}. We have checked\footnote{Another way to check the 
consistency of the approach is to use the limit of Ref.~\cite{Jacobson:2013xta} to compare with the khronometric results to be shortly derived. Also 
in this case, our results are consistent.} that our 
expressions are consistent when taking the weak-field limit~\cite{Foster:2006az}, which is not the case for Ref.~\cite{Foster:2007gr}.

Using  the previous equations and the expressions in Ref.~\cite{Foster:2006az,Foster:2007gr}  we can compute the total flux ${\cal{F}}$, and via Eq.~\eqref{Pdot-eq} the rate of change of the orbital period:
\begin{align}
\label{pdot-AE}
\frac{\dot{P}_{b}}{P_{b}}&=-\frac{3 a G_{\AE}}{{\cal{G}} \mu m}
\Big\langle \frac{{\cal A}_1}{5}\dddot Q_{ij}\dddot Q_{ij}+\frac{{\cal A}_2}{5}\dddot
     {\cal Q}_{ij}\dddot Q_{ij}  +\frac{{\cal A}_3}{5}\dddot{\cal Q}_{ij}\dddot{\cal Q}_{ij}
\nn \\ 
& 
+{\cal B}_1\dddot I \dddot
     I +{\cal B}_2\dddot {\cal I} \dddot I +{\cal B}_3\dddot {\cal I}
     \dddot {\cal I} +{\cal C}\dot \Sigma_i \dot\Sigma_i+{\cal D}\dot {\cal V}_{ij} \dot{\cal V}_{ij}\Big\rangle\,.
\end{align}
In these equations, the angled-brackets stand for an average over several
wavelengths and we have defined the shorthands
\begin{align}
\label{A-funcs-1}
{\cal A}_1 &\equiv \frac{1}{w_2}+\frac{2 c_{14} c_+^2}{(2c_1-c_- c_+)^2 w_1}+\frac{3 c_{14}(Z-1)^2}{2 w_0(2-c_{14})}\,,  
\\ 
{\cal A}_2 &\equiv \frac{2(Z-1)}{(c_{14}-2)w_0^3}+\frac{2 c_+}{(2c_1-c_+c_-)w_1^3}\,,  
\\ 
\label{A-funcs-3}
{\cal A}_3 &\equiv\frac{1}{2 w_1^5 c_{14}}+ \frac{2}{3c_{14}(2-c_{14})w_0^5}, \quad {\cal B}_1\equiv
\frac{c_{14} Z^2}{4 w_0(2-c_{14})}\,,  
\\ 
{\cal B}_2 &\equiv \frac{Z}{3 w_0^3(c_{14}-2)},\quad  {\cal B}_3 \equiv \frac{1}{9c_{14}
  w_0^5(2-c_{14})}\,, 
\\ 
\label{C-func}
{\cal C} &\equiv \frac{4}{3 w_0^3
  c_{14}(2-c_{14})}+\frac{4}{3 c_{14} w_1^3},\quad {\cal D}\equiv \frac{1}{6 w^5_1 c_{14}}\,.
\end{align}

Let us now evaluate the different terms in Eq.~\eqref{pdot-AE} for a compact binary system in a generic orbit in the center of mass frame. Using basic results from Newtonian and PN orbital mechanics, we can evaluate the moments in Eqs.~\eqref{Q-mom}-\eqref{I-mom} (cf. Ref.~\cite{Foster:2007gr}). The only expression not found in Ref.~\cite{Foster:2007gr} is
\be
\dot {\cal V}_{ij}=\frac{2{\cal G} \m m}{r_{12}^3} (s_2-s_1)\left[\dot{\hat n}_{12}^{[i}X_{CM}^{j]}-2\hat n_{12}^{[i}X_{CM}^{j]}\dot r_{12}+r_{12}^{[i}\dot X_{CM}^{j]}\right],
\ee
where $X^i_{CM}\equiv (m_1 x_1^i+m_2 x_2^i)/m$ are the center-of-mass coordinates [see also Eq.~\eqref{eq:defM} and the definitions around Eq.~\eqref{Newtonian-a}].
These moments are to be used in Eq.~\eqref{pdot-AE}. As explained in Ref.~\cite{Foster:2007gr}, the secular terms that depend on $X_{CM}^i$ can be neglected, 
as they are canceled by an opposite contribution in the action. The final result is 
\newpage
\bw
\begin{align}
\label{Pdot-AE}
\frac{\dot{P}_{b}}{P_{b}} &= - 3  \Big\langle  \left(\frac{ {\cal G} G_{\AE}a \m \,m}{r_{12}^4}\right) 
  \Bigg\{ \frac{8}{15}({\cal A}_1+{\cal
  S}{\cal A}_2+ {\cal S}^2{\cal A}_3)(12v_{12}^2-11 \dot
r_{12}^2)+4({\cal B}_1+{\cal S}{\cal B}_2+ {\cal S}^2{\cal B}_3)\dot
r_{12}^2 \nn
\\ &+\left(s_1-s_2\right)^2\left({\cal
  C}+\left(\frac{18}{5}{\cal A}_3+2{\cal D} \right) V_{CM}^{j} V_{CM}^{j}+\left(\frac{6}{5}{\cal A}_3+36 {\cal B}_3-2{\cal D}\right)
(V_{CM}^{i} \hat n_{12}^i)^2\right) \nn
\\ &+\left(s_1-s_2\right)\left[12({\cal
    B}_2+2{\cal S}{\cal B}_3)V_{CM}^i \hat n_{12}^i v^j_{12} \hat n_{12}^j+\frac{8}{5}({\cal A}_2+2{\cal S}{\cal
    A}_3)V_{CM}^i(3v_{12}^i-2\hat n_{12}^i v_{12}^j \hat n_{12}^j)
  \right]\Bigg\}\Big\rangle\,,
\end{align}
\ew 
where recall that $v_{12}^{i}\equiv \dot r_{12}^i$ is the relative 3-velocity,  $V_{CM}^{i}\equiv \dot X^i_{CM}$ is the center-of-mass velocity of the system relative to the \AE ther field and
${\cal{S}} \equiv s_1 m_2/m +s_2 m_1/m.$

At this juncture, let us stop to discuss some qualitative features of the orbital decay rate predicted in Einstein-\AE ther theory [Eq.~\eqref{Pdot-AE}]. First, note that in the limit in which all the coupling parameters go to zero, i.e. $c_{i} \to 0$, Eq.~\eqref{Pdot-AE} reduces to the GR result. Second, the Einstein-\AE ther terms can in principle dominate over the GR
ones, under appropriate conditions (i.e. unequal and sufficiently large sensitivities, and sufficiently large couplings).
This is because the leading-order Einstein-\AE ther term scales with fewer powers of $\tilde{m}/r_{12}$ than the GR result. Indeed, the Einstein-\AE ther terms enter first at absolute order ${\cal{O}}(1/c^{8})$, whereas the GR prediction is of absolute order ${\cal{O}}(1/c^{10})$. Such scalings are typical of dipolar radiation, and those terms are proportional to the difference of the sensitivity  parameters  squared. The existence of a preferred frame is clear in the previous expression from the dependence on the velocity of the center of mass with respect to the \AE ther, $V_{CM}^i$. Finally, note that for circular orbits, $\dot{r}_{12} = 0$ and $\hat n_{12}^i v_{12}^i=0$, and the above expressions simplify greatly. 

\subsubsection{Dissipative PN Dynamics in Khronometric Theory}

The energy flux in khronometric theory was first calculated in Ref.~\cite{Blas:2011zd}, which focused on weak-field sources. We will here present the main features of their analysis, together with the generalization to the strong field case.

The total energy flux carried by propagating degrees of freedom can be computed in
khronometric theory, using the decomposition in Eq.~\eqref{eq:metric_dec}, to be~\cite{Blas:2011zd}
\be 
\label{angle-int-Edot}
\langle {\cal F}\rangle=-\frac{1}{32\pi G_{\AE}}\oint_S d \Omega \;
r^2 \left\langle \frac{1}{c_t}\dot \phi_{ij}\dot
\phi_{ij}-\frac{(\alpha-2)}{2\alpha c_s}\dot F \dot F \right\rangle
+ \langle\dot O\rangle\,, 
\ee 
where $c_{t}^{-2}= 1-\beta$ is the speed of propagation of the tensor modes
and  
\be
c_s^2=\frac{(\alpha-2)(\beta+\lambda)}{\alpha(\beta-1)(2+\beta+3\lambda)},
\ee
is the speed of propagation of the scalar modes. The term $\dot O$ 
corresponds to a boundary term given by a time derivative. This term has two components: one that cancels when averaged over the orbit and another one that cancels against secular terms (cf.~Ref.~\cite{Foster:2007gr}).

The far-zone fields appearing in Eq.~\eqref{angle-int-Edot}
can be obtained by solving the modified field equations  from Sec.~\ref{subsec:EOMs}, 
and using the method described in Ref.~\cite{Blas:2011zd} (see also Ref.~\cite{Foster:2007gr}). The 
results are
\begin{align}
\phi_{ij}&=-\frac{2 G_{\AE}}{r} \ddot Q_{ij}^{TT}(t-r/c_t), \nn \\
\label{psikh}
F&= \frac{4G_{\AE}\alpha}{ (\alpha-2) 
  r}\left[\frac{3}{2}({\cal Z}-1) \hat n^i \hat n^j \ddot
Q_{ij}+\frac{1}{2} {\cal Z} \ddot I_{kk} \right.  \nn \\ &\left.  -\frac{\hat n^i\hat n^j}{\alpha c_s^2}(\ddot{\cal Q}_{ij}+\frac{1}{3}
\delta_{ij}\ddot{\cal I}_{kk})+\frac{2}{\alpha c_s} \hat n^i
\Sigma^i\right]\,.
\end{align}
In the expression for $F$, the different moments should be evaluated
at the retarded time $t-r/c_s$, and we have defined
\be
{\cal Z}\equiv \frac{(\alpha_1^{kh}-2\alpha_2^{kh})(1-\beta)}{3(2\beta-\alpha)}.
\ee

After performing the angular integration in Eq.~\eqref{angle-int-Edot}, the orbital decay rate is still
given by Eq.~\eqref{pdot-AE},  but now with the coefficients 
\begin{align}
\label{A1-def-HL}
{\cal A}_1 &\equiv \frac{1}{c_t}+\frac{3 \alpha({\cal Z}-1)^2}{2 c_s(2-\alpha)},
\quad {\cal A}_2\equiv \frac{2({\cal Z}-1)}{(\alpha-2)c_s^3}, 
\\ 
\label{A3-def-HL}
{\cal A}_3 &\equiv \frac{2}{3\alpha(2-\alpha)c_s^5}, \quad {\cal B}_1\equiv
\frac{\alpha{\cal Z}^2}{4 c_s(2-\alpha)}, 
\\ 
{\cal B}_2 &\equiv \frac{{\cal Z}}{3 c_s^3(\alpha-2)},\quad  {\cal B}_3\equiv \frac{1}{9\alpha
  c_s^5(2-\alpha)},
\\ 
{\cal C} &\equiv \frac{4}{3 c_s^3\alpha(2-\alpha)}, \quad {\cal D}=0.
\end{align}
Clearly then, the same observations made after 
Eq.~\eqref{pdot-AE} are still valid here. In particular, the orbital decay rate is of dipolar structure and depends on the sensitivities. 

\section{Neutron Star Solutions in Einstein-\AE ther Theory}
\label{sec:NS-Sols}

In this section, we construct Einstein-\AE ther
theory solutions that  describe isolated non-spinning NSs moving slowly relative to the \AE ther.
We begin by presenting the metric and \AE ther ansatz, and expand the field
equations at zeroth- and first-order in the velocity $v$ relative to the \AE ther. We then
explain how to solve these equations numerically.

\subsection{Metric Ansatz, \AE ether Field Ansatz and Neutron Star Model}
\label{subsec:ansatz-AE}

Let us consider a generic ansatz for the metric and \AE ther fields, for a
stationary configuration describing a non-spinning NS
in slow motion with velocity $v^i$ relative to the \AE ther.

One can adopt a coordinate system comoving
with the fluid elements of the NS by aligning the time coordinate vector to the fluid's 4-velocity $u^\mu$.
More specifically, we can choose a spacelike hypersurface $\Sigma$ and assign spatial
coordinates on it. One can then define the spatial coordinates of an event $p$ as
those of the intersection between $\Sigma$ and the worldline of the fluid element passing through $p$.\footnote{Note that
this construction yields a \textit{bona fide} coordinate chart in the absence of caustics for the fluid flow (which is a reasonable assumption for the system that we are considering)
and in the absence of closed timelike curves (which would in any case violate causality and make the spacetime pathological).}
In these comoving coordinates,  the fluid elements are at rest, while the \AE ther is moving. In particular, the fluid 4-velocity field is
\be 
u^\mu = \frac{\delta^\mu_t}{\sqrt{g_{tt}}}\,.
\label{4-velocity0}
\ee

Because the system is invariant under spatial rotations around the direction of
the motion relative to the \AE ther, it is convenient to adopt 
asymptotically spherical coordinates $(t,r,\theta,\varphi)$ in which the \AE ther's velocity
direction corresponds to the polar axis. Clearly, in these coordinates
both the metric and \AE ther will not depend on the azimuthal coordinate $\varphi$. 
Noting then that the system's configuration must be invariant
under the simultaneous reflection of 
\ba
\label{reflection-v}
v^i &\mapsto & -v^i\,, \\
\label{reflection-t}
t &\mapsto & -t\,, \ea
it is clear that at first order in the velocity,
the only non-zero components of the metric are the cross-terms
$g_{t r}$ and $g_{t \theta}$, and the only non-vanishing components of the \AE ther field are
$U_r$ and $U_\theta$. (Note that a contribution to $U_t$ at ${\cal O} (v)$ is forbidden
by the normalization condition $U^\mu U_\mu=1$.)

More explicitly, the most generic ansatz at linear order in $v$ for the metric and \AE ther is given by 
\begin{align}
\label{metric-ansatz}
ds^2 &= e^{\nu (r)} dt^2 - e^{\mu (r)} d r^2 - r^2 \left( d\theta^2
+ \sin^2 \theta d \varphi^2 \right) 
\nn \\ 
& + 2 v V(r,\theta) dt dr +
2 v r S(r,\theta) dt d\theta + \mathcal{O}(v^2)\,,  
\\
\label{aether-ansatz} U_\mu &= e^{\nu(r)/2} \delta^t_\mu+ v W(r,\theta ) \delta_\mu^r+ v Q(r,\theta ) \delta_\mu^\theta +
\mathcal{O}(v^2)\,, 
\end{align}
and the fluid 4-velocity Eq.~\eqref{4-velocity0} becomes
\be u^\mu =  e^{-\nu/2} \delta^\mu_t\,.
\label{4-velocity}
\ee
Note that for a star, the \AE ther cannot have any $U_r$ component at $\mathcal{O}(v^0)$, as shown explicitly in Ref.~\cite{Eling:2007xh} using the field equations.

The ansatz of Eqs.~\eqref{metric-ansatz}--\eqref{4-velocity} therefore depends on two potentials $\mu(r)$ and $\nu(r)$
at order $\mathcal{O}(v^0)$, and on four potentials $V(r,\theta)$, $S(r,\theta)$, $W(r,\theta)$, $Q(r,\theta)$ at $\mathcal{O}(v)$.
This ansatz, however, has been derived by choosing spatial coordinates comoving with the fluid without specifying the time coordinate. 
In particular, one is free to perform a coordinate transformation of the form  
\be t'= t + v H(r,\theta)\,,
\label{t-transform}
\ee
which can be used to set any one of the potentials $V(r,\theta)$, $S(r,\theta)$, $W(r,\theta)$, $Q(r,\theta)$
to zero while keeping the ansatz of Eqs.~\eqref{metric-ansatz}--\eqref{4-velocity} valid at $\mathcal{O}(v)$
(modulo a redefinition of the remaining three potentials).
In what follows, we will set therefore $Q=0$ without loss of generality.

We use a perfect fluid stress-energy tensor to describe the NS matter:
\be T_{\mu\nu}^\MAT = \left[ \tilde{\rho} (r) + \tilde{p}(r) \right]
u_\mu u_\nu - \tilde{p}(r) g_{\mu\nu}\,+{\cal O}(v)^2,
\label{Tmunu-mat}
\ee
where $\tilde{\rho} (r)$ and $\tilde{p}(r)$ are
the fluid's energy density and pressure at $\mathcal{O}(v^0)$, respectively. 
The density and pressure do not have any
$\mathcal{O}(v)$ contributions because they must be invariant under
the transformations in Eqs.~\eqref{reflection-v}
and~\eqref{reflection-t}. Moreover, one can explicitly show that any
$\mathcal{O}(v)$ pieces in the density and pressure must vanish
from the $\mathcal{O}(v)$ pieces of the $(t,t)$ and $(r,r)$ components
of the modified field equations. Note also that, in spite of these facts,
the $(t,r)$ and $(t,\theta)$ components of $T_{\mu\nu}^\MAT$ have
$\mathcal{O}(v)$ contributions due to the ${\cal O}(v)$ terms in the metric.

The relation between internal pressure and energy density is
parametrized by the EoS. We here investigate four different,
realistic EoSs: APR~\cite{APR}, SLy~\cite{SLy},
Shen~\cite{Shen1,Shen2} and LS220~\cite{LS}. For the Shen and LS220 EoSs, we use
a temperature of $0.1$ MeV and consider neutrino-less,
$\beta$-equilibrium. These EoSs are not the only ones that are observationally viable, but they
represent a sufficiently large sample of the EoSs to allow us
to investigate how the different observables depend on this choice. 

\subsection{Field Equations}
\subsubsection{${\cal{O}}(v^0)$ Equations}
\label{eqs_Ov0}

Let us first look at the field equations at $\mathcal{O}(v^0)$. The
$(t,t)$, $(r,r)$, and $(\theta,\theta)$ components of the field
equations give three independent equations
\bw
\begin{align}
\label{Ein-tt-0}
&16 \frac{dM}{dr}  -4 c_{14} r (r-2M) \frac{d^2\nu}{dr^2}- c_{14} r
(r-2M) \left( \frac{d\nu}{dr} \right)^2  +4 c_{14} \left( r
\frac{dM}{dr}-2r +3M \right) \frac{d\nu}{dr} = 64 \pi \rho r^2\,,  \\
\label{Ein-rr-0}
&c_{14} r^2(r-2M) \left( \frac{d\nu}{dr} \right)^2 +8 r (r-2M)
\frac{d\nu}{dr}-16 M = 64 \pi r^3 p\,,  \\
\label{Ein-thth-0}
& 4 r^2 (r-2M) \frac{d^2\nu}{dr}-(c_{14}-2) r^2 (r-2M) \left(
\frac{d\nu}{dr} \right)^2  \-4 r \left( r \frac{dM}{dr} -r+M
\right) \frac{d\nu}{dr}-8r\frac{dM}{dr}+8M = 64 \pi r^3 p\,,
\end{align}
\ew
respectively, where 
$p(r)$ and $\rho(r)$ are the rescaled NS pressure and density 
\be p \equiv \frac{2-c_{14}}{2} G_N \tilde{p}\,, \quad \rho \equiv
\frac{2-c_{14}}{2} G_N \tilde{\rho}\,, 
\label{prho-rescale}
\ee
with the length scale $M(r)$ defined by
\be e^{-\mu(r)} \equiv 1-\frac{2 M(r)}{r}\,.  
\label{M-def}
\ee
With these definitions, $M(r \to \infty) = M_{*}  = G_{N} M_{\tot}$, 
where $M_{\tot}$ is given by Eq.~\eqref{ae:mass}~\cite{Foster:2005fr,Eling:2005zq}. Because $G_N$ is the value of 
the gravitational constant measured with a Cavendish-type experiment, thus entering Kepler's laws, 
$M_{\tot}$ also corresponds to the NS mass measured at spatial infinity by a Keplerian experiment. To make this last fact clear, 
we define $M_\obs \equiv M_{\tot}$.

By using Eqs.~\eqref{Ein-tt-0} and~\eqref{Ein-thth-0}, one can
eliminate $(d\nu/dr)^2$ from Eq.~\eqref{Ein-rr-0} to yield
\begin{align}
\label{dnudr}
\frac{d\nu}{dr} &= \frac{1}{(c_{14}-2) r (r-2M)} \left\{ 2 (c_{14}-2)
r \frac{dM}{dr} \right.  \nn \\ &+ \left.  2(c_{14} -2) M+16 \pi r^3
[(2 c_{14}-1) p+ \rho ] \right\}\,.
\end{align}
By using this equation, $\nu$ can be eliminated from
Eq.~\eqref{Ein-tt-0} to give
\begin{align}
\label{dMdr}
\frac{dM}{dr} &= \frac{1}{ c_{14} (c_{14}-2) r} \left\{- (8-6 c_{14}+c_{14}^2) M   \right.  \nn \\ &\hspace{-.7cm}\left.  + 2 (c_{14}-2)
(r-2 M)^{1/2}  (r -2 M+c_{14} M+4 \pi
c_{14} r^3 p)^{1/2}  \right.  \nn \\ &
\left.  - 2r [4 \pi c_{14} (2 c_{14}-1) r^2  p+4 \pi c_{14} r^2  \rho
  + c_{14}-2] \right\}\,,
\end{align}
and substituting this equation into Eq.~\eqref{dnudr}, one has
\be
\label{dnudr2}
\frac{d\nu}{dr} =- \frac{4}{c_{14} r}\left[1- \left(\frac{r-2 M+c_{14}
    M+4 \pi c_{14}  r^3 p}{r-2M}\right)^{1/2}\right]\,.  \ee
One can use Eqs.~\eqref{dMdr} and~\eqref{dnudr2} to eliminate $dM/dr$
and $\nu$ from Eq.~\eqref{Ein-thth-0} to yield
\be
\label{dpdr}
\frac{dp}{dr} = \frac{2}{c_{14}r} (\rho +p) \left[1 -
  \left(\frac{r-2M+c_{14} M+4 \pi c_{14} r^3 p}{r-2M}
  \right)^{1/2}\right]\,.  \ee
This equation corresponds to the modified Tolman-Oppenheimer-Volkoff
(TOV) equation~\cite{Oppenheimer:1939ne,Tolman:1939jz}. Solving the field equations to zeroth-order in
velocity then reduces to solving Eqs.~\eqref{dMdr}--\eqref{dpdr},
together with an EoS, for the 4 unknown functions $\nu(r)$, $M(r)$,
$\rho(r)$ and $p(r)$. 

One may find it instructive to expand these equations further in a
small coupling approximation, i.e.~$c_{14} \ll 1$. Doing so, the above
equations become
\begin{align}
\frac{dM}{dr} &= 4 \pi r^2 \rho - \frac{c_{14}}{4r (r-2M)} \left[M^2 +
  8 \pi (7 p+ 2\rho ) M r^3  \right.  \nn \\ &\left.  + 8 \pi (2\pi
  r^2 p^2- \rho-3 p) r^4\right] + \mathcal{O}(c_{14}^2)\,,
\\ \frac{d\nu}{dr} &= 2\frac{4 \pi r^3 p +M}{r (r-2M)}- c_{14}
\frac{(4 \pi r^3 p +M)^2}{2r (r-2M)^2 } + \mathcal{O}(c_{14}^2)\,,
\\ \frac{dp}{dr} &= -\frac{(4\pi r^3 p +M) (\rho +p)}{r (r-2M)} \nn
\\ &+ c_{14}\frac{(4 \pi r^3 p +M)^2 (\rho + p)}{4r(r-2M)^2} +
\mathcal{O}(c_{14}^2)\,.
\end{align}
The first terms in the above equations agree with the GR result for the TOV equation, see
e.g. Ref.~\cite{Oppenheimer:1939ne,Tolman:1939jz}.

\subsubsection{${\cal{O}}(v)$ Equations}

Let us now look at the field equations at $\mathcal{O}(v)$. The
modified field equations and the \AE ther field equations become a system
of partial differential equations for the unknown functions
$V(r,\theta)$, $S(r,\theta)$ and $W(r,\theta)$. One can separate the
variables $r$ and $\theta$ by using the following Legendre
decomposition:
\ba
\label{LD:1}
V(r,\theta) &=& \sum_n k_n (r) P_n (\cos \theta)\,, \\ S(r,\theta) &=&
\sum_n s_n (r) \frac{d P_n (\cos \theta)}{d\theta}\,, \\ W(r,\theta)
&=& \sum_n w_n (r) P_n (\cos \theta)\,, 
\label{LD:3}
\ea
where $P_n$ is the $n$-th Legendre polynomial. This structure can be
inferred from symmetry principles by looking at how each component of
the metric behaves under spatial rotations and reflections with respect 
to the equatorial plane, i.e.~this is nothing but a tensor spherical harmonic 
decomposition (see e.g.~Ref.~\cite{Thorne:1980rm}).

With this decomposition at hand, the $(t,r)$ component of the modified
field equations and the $r$ and $\theta$ components of the \AE ther field
equations become
\allowdisplaybreaks \bw \ba
\label{dsdr}
\frac{ds_n}{dr} &=& \frac{1}{4} \frac{1}{n (n+1) r^2} \left\{ -4 n
(n+1) c_{-} r e^{\nu /2} j_n  - c_{14} r^2 (r-2 M) \left(
\frac{d\nu}{dr} \right)^2 k_n  -4 [ 2( r-2 M) k_n+ n (n+1) (c_{14}-1)
  r s_n] r \frac{d\nu}{dr} \right. \nn \\ & & \left. + 4 \left[ 4 M+16
  \pi p r^3 +n(n+1) (1- c_{-}) r \right] k_n-4 n (n+1) r s_n
\right\}\,, \\
\label{dwdrr}
\frac{d^2j_n}{dr^2} &=& \frac{1}{2} \frac{1}{c_{123} r^2(r-2 M)}
\left\{ \left[ -(c_2+c_3-c_4) r^2  (r-2 M) \frac{d^2\nu}{dr^2}  j_n +2
  c_{123} r^2\frac{d^2M}{dr^2} j_n \right. \right. \nn \\ & &
  \left. \left. +r \left[ -c_{123} r (r-2 M) \frac{dj_n}{dr} + \left[
      (2c_{123}-c_{14}) r \frac{dM}{dr} + (2 c_{+} -3c_{14} -2 c_2) M-
      2 (c_3-c_4) r  \right] j_n \right] \frac{d\nu}{dr}
  \right. \right. \nn \\ & & \left. \left. + 2 c_{123} \left( 3r
  \frac{dM}{dr} - 2 r+ M \right)  r \frac{dj_n}{dr} +2 \left[ 2 c_2 r
    \frac{dM}{dr} -2 (2 c_{123}+c_2) M+ [n (n+1) c_1 +2 c_{123}] r
    \right] j_n \right] \right. \nn \\ & & \left. + n (n+1) r
e^{-\nu/2} \left[ (c_{14}+c_2) r   \frac{d\nu}{dr} s_n - (c_{123} +
  c_2) r \frac{ds_n}{dr} + (3 c_{+}+2 c_2) s_n + c_{-} k_n \right]
\right\}\,, \\
\label{dkdr}
\frac{d k_n}{dr} &=& \frac{1}{2} \frac{1}{c_{-} r (r-2 M)} \left\{
\left[ 2  (c_2+c_3+c_4) r (r-2 M) \frac{d\nu}{dr} j_n-4 (c_2 + c_3) r
  \frac{dM}{dr} j_n +4 (c_2+c_3) r (r-2 M) \frac{dj_n}{dr}
  \right. \right. \nn \\ & & \left. \left. + 4 [ - (3c_{123} + c_1) M+
    2 c_{123} r] j_n \right] e^{\nu/2} -2 c_{+} r^2 (r-2 M) \frac{d^2
  s_n}{dr^2} \right. \nn \\ & & \left. + \left[- (c_{-}+2 c_4) r
  \frac{ds_n}{dr} + (c_{-}+2 c_4) s_n +  (c_{+}+2 c_4) k_n \right] r
(r-2 M) \frac{d\nu}{dr} + 2  c_{+} r \left( 3 M-2 r+r \frac{dM}{dr}
\right) \frac{ds_n}{dr} \right. \nn \\ & & \left. - 2 (c_{+} s_n -
c_{-} k_n) r \frac{dM}{dr} -2 (3c_{+} s_n+ c_{-} k_n) M+ 4 n (n+1)
c_{123} r  s_n \right\}\,, \ea \ew
respectively, where we have defined [recall also the definitions in Eqs.~\eqref{cpm}-\eqref{c123}]
\ba j_n &\equiv& w_n - e^{-\nu/2} k_n\,. \ea
Note that the $n$-th mode is independent of all other modes,
i.e.~there is no mode coupling (see Appendix \ref{app:separability} for an explanation).
 Note also that, as expected, the new
unknown functions at ${\cal{O}}(v)$ depend on the solutions to
the ${\cal{O}}(v^0)$ equations of the previous subsection. 

\subsection{Asymptotic Solutions, Matching Conditions and Numerical Techniques}
\label{subsec:AS}

The above system of differential equations must be solved order by
order in velocity twice: once in the interior of the NS with a given
EoS, and once in the exterior where the density and pressure
vanish. We solve these equations as an initial value problem. The
solutions will generically depend on integration
 constants. These are determined by 
imposing that the solutions be  continuous and differentiable at some
matching radius, usually
the NS surface.

Let us first concentrate on the initial conditions for the
${\cal{O}}(v^{0})$ equations. By imposing regularity at the NS center,
the  behavior of the solution to the ${\cal{O}}(v^{0})$
equations near the center is 
\ba
\label{rho0}
\hspace{-.5cm}\rho(r) &=& \rho_c + \rho_2 r^2 + \mathcal{O}(r^3)\,, \\
\label{p0}
\hspace{-.5cm}p(r) &=& p_c + 4 \pi \frac{(\rho_c +3 p_c) (\rho_c + p_c) }{3
  (c_{14}-2)}r^2 + \mathcal{O}(r^3)\,, \\
\label{M0}
\hspace{-.5cm}M(r) &=& - 4 \left(\frac{2 \pi \rho_c + 3 \pi c_{14} p_c}{3(c_{14}-2)} \right)r^3 +
\mathcal{O}(r^5)\,, \\
\label{nu0}
\hspace{-.5cm}\nu(r) &=& \nu_c  -8 \pi \frac{\rho_c+3 p_c}{3 (c_{14}-2)} r^2 +
\mathcal{O}(r^3) \ (r\to 0^+)\,, \ea
where $\rho_c$, $p_c$ and $\nu_{c}$ are the values of the density,
pressure and $\nu$ at the NS center. The quantity $\rho_2$ is a
constant that can be expressed in terms of $\rho_c$ and $p_c$ via
Eqs.~\eqref{rho0},~\eqref{p0} and the EoS. Similarly, by requiring
asymptotic flatness, the asymptotic behavior of the solution near
spatial infinity is 
\ba
\label{M-inf}
M(r) &=& M_* + \frac{c_{14}}{4} \frac{M_*^2}{r} + \frac{c_{14}}{4}
\frac{M_*^3}{r^2} + \mathcal{O} \left( c_{14} \frac{M_*^4}{r^3}
\right)\,, \\
\label{nu-inf}
e^{\nu (r)} &=& 1 -\frac{2 M_*}{r}-\frac{c_{14}}{6} \frac{M_*^3}{r^3}
+ \mathcal{O} \left( c_{14} \frac{M_*^4}{r^4} \right) \ (r \to
\infty)\,, \nn \\ \ea
where $M_* \equiv M(\infty) = G_N M_\obs$, with $M_\obs$ the NS mass observed by a Keplerian experiment.

Given these asymptotic solutions, we numerically solve the
${\cal{O}}(v^{0})$ differential equations as follows. First, we choose
a specific value of the NS central density $\rho_c$, and with it, we
obtain $p_c$ from the EoS $p_c = p(\rho_c)$. We obtain the interior
solution by solving Eqs.~\eqref{dMdr} and~\eqref{dpdr}, together with
the EoS, from a core radius $r_\epsilon \ll R_*$ up to the NS surface,
defined as the radius $R_*$ at which the pressure vanishes. We then use
the values obtained from this interior solution evaluated at the NS
surface as initial conditions for a new evolution of the equations in
the exterior. The exterior solution for $M(r)$ is then evaluated at a
large boundary radius $r_b \gg R_*$ and set equal to the asymptotic
expression in Eq.~\eqref{M-inf}, evaluated at the same radius $r_{b}$, 
to solve for the observed NS mass $M_{\obs}$.

Next, we solve Eq.~\eqref{dnudr2} for $\nu(r)$ in the interior (from
$r=r_\epsilon$ to the NS surface) using a trial value for $\nu_c$. As
before, we then use the value of $\nu(r=R_*)$ as an initial condition to
solve for $\nu(r)$ in the exterior. Since Eq.~\eqref{dnudr2} is
shift-invariant, the trial solution plus a constant
$\nu^\mrm{tr}(r)+C_\nu$ is also a solution. We choose $C_\nu$ such
that $\nu(r)$ satisfies the asymptotic behavior of Eq.~\eqref{nu-inf}
at $r=r_b$~\cite{yunes-CSNS,Yagi:2013mbt}
\be e^{\nu^\mrm{tr}(r_b) + C_\nu} =  1 -\frac{2
  M_*}{r_b}-\frac{c_{14}}{6} \frac{M_*^3}{r_b^3}  + \mathcal{O} \left(
c_{14} \frac{M_*^4}{r_b^4} \right)\,.  \ee
Note that in practice $r_{b} < \infty$, but we choose it to be
sufficiently large such that the error incurred is smaller than other
numerical errors, e.g.~those due to the discretization of the differential
equations. 

Let us now discuss the initial conditions for the
${\cal{O}}(v)$ equations. Equations~\eqref{g0iPN} and \eqref{metric-Foster-sph0text}
imply that we are only interested in the $n=1$ component of the Legendre decomposition, since
these functions determine the sensitivities, and thus, the
orbital period decay rate. Imposing regularity at the NS center, the
 solution to Eqs.~\eqref{dsdr}-\eqref{dkdr} near the center
is
\bw \ba
\label{k1-0}
k_1 (r) &=& C + \frac{1}{30} \frac{1}{c_{123} (c_{14}-2)}   \left\{ 3
(c_{14}-2) [2(6- c_{-}) c_2+6c_{+} + 6 c_3+c_1^2-c_3^2] e^{\nu_c/2} D
\right. \nn \\ & &  \left. - 24 \pi \left[ \left( 3 + 2c_{+} - 4 c_2
  \right) c_4 + \left( 4- 4 c_1+ 9 c_{14} \right) c_2+2 c_1^2+ \left(
  1+2c_3+3c_{14} \right) c_1+ \left( 3 c_{14}-2 \right) c_3 \right]
p_c C \right.  \nn \\ & & \left. - 8\pi  \left[ \left( 3 + 2c_{+} -4
  c_2 \right)  c_4+ 2( 11-2 c_1) c_2+ 2c_1^2+\left( 2c_3+7 \right) c_1
  + 4 c_3 \right] \rho_c C \right\} r^2 + \mathcal{O}(r^3)\,, \\
\label{s1-0}
s_1 (r) &=& C - \frac{1}{30} \frac{1}{c_{123} (c_{14}-2)}  \left\{ 3
(c_{14}-2) \left[ 4( c_{-}-1 ) c_2- 2c_{+} - 2 c_3 + 3 c_1^2 - 3 c_3^2
  \right] e^{\nu_c/2} D \right. \nn \\ & & \left. - 24\pi \left[
  \left( 6c_{+} + 8 c_2 - 1 \right) c_4+ \left( 8 c_1 -3 c_{14} -8
  \right)  c_2+ 6c_1^2 + \left( 6c_3 - c_{14} - 7 \right) c_1-
  (6+c_{14}) c_3 \right] p_c C \right.  \nn \\ & & \left. -  8\pi
\left[ \left( 6 c_{+} + 8 c_2- 1 \right) c_4+ 2 \left( 4 c_1- 7
  \right)  c_2+ 6c_1^2 + 3 \left( 2c_3-3 \right) c_1- 8 c_3 \right]
\rho_c C \right\} r^2  + \mathcal{O}(r^3)\,, \\
\label{w1-0}
w_1(r) &=& D r^2 + \mathcal{O}(r^3)\,, \ea 
where $C$ and $D$ are constants of integration. Similarly, we can
impose asymptotic flatness at spatial infinity and
obtain the asymptotic solution 
\ba
\label{asymp-v}
k_1(r) = k_1^\infty (r) & \equiv & -1+A \frac{M_*}{r} + (k_{A_2} A +
k_{c_2}) \frac{M_*^2}{r^2} + \left[ B +  (k_{A_{31}} A + k_{c_{31}} )
  \ln \left( \frac{r}{M_*} \right) \right] \frac{M_*^3}{r^3}
+\mathcal{O} \left( \frac{M_*^4}{r^4} \right)\,, \\
\label{asymp-s}
s_1(r) = s_1^\infty (r) & \equiv & -1+ (s_{A_1} A + s_{c_1} )
\frac{M_*}{r} + (s_{A_2} A + s_{c_2} ) \frac{M_*^2}{r^2} \nn \\ & &+
\left[ s_{A_{30}} A - \frac{B}{2} + s_{c_{30}} + (s_{A_{31}} A +
  s_{c_{31}} ) \ln \left( \frac{r}{M} \right) \right]
\frac{M_*^3}{r^3}   +\mathcal{O} \left( \frac{M_*^4}{r^4}  \right)\,,
\\
\label{asymp-w}
w_1(r) = w_1^\infty (r) & \equiv & \frac{A+2(c_{14}-1)}{c_{-}}
\frac{M_*}{r} + (w_{A_2} A + w_{c_2} ) \frac{M_*^2}{r^2}  +
(w_{A_{30}} A +w_{c_{30}}) \frac{M_*^3}{r^3}  +\mathcal{O} \left(
\frac{M_*^4}{r^4}  \right)\,,  \ea \ew
where $A$ and $B$ are constants of integration, $s_{A_1}$ and
$s_{c_1}$ are given by
\ba s_{A_1} &\equiv& \frac{1}{4} \frac{ (3 c_1 +2 c_2 -4) c_1 - 2 c_2 - (2
  c_2 +3 c_3 +2) c_3 }{c_{-} c_{123}}\,, \nn \\ \\ s_{c_1} &\equiv& -
\frac{1}{2} \frac{1}{c_{-} c_{123}} \left[(c_1-c_2 +3 c_3 + 3 c_4 -4)
  c_1 \right. \nn \\ & & \left. + (3 c_2+2 c_3+3 c_4 -2) c_3+ 2 c_2
  (c_4-1) \right]\,,  \ea
and the other coefficients, $k_{A_2}$, $k_{c_2}$, $k_{A_{31}}$,
$k_{c_{31}}$, $s_{A_2}$, $s_{c_2}$, $s_{A_{30}}$, $s_{c_{30}}$,
$s_{A_{31}}$, $s_{c_{31}}$, $w_{A_2}$, $w_{c_2}$, $w_{A_{30}}$ and
$w_{c_{30}}$, similarly depend only on $c_i$ $(i=1,2,3,4)$. We do not
present explicit expressions for these quantities here because they
are lengthy and unilluminating. 

One might be worried that the first terms in Eqs.~\eqref{asymp-v}
and~\eqref{asymp-s} seem to be inconsistent with the asymptotic
flatness condition, because they lead to terms in the metric of the
form $g_{ti} \sim v dx^i dt$. However, such terms are coordinate
artefacts; for example, the Schwarzschild metric in
Eddington-Finkelstein coordinates has a component that goes as $g_{vr}
\sim dvdr$.  An inverse coordinate transformation
$t' \approx t - vz$ with $z= r \cos \theta$
 would then eliminate
such terms in Eqs.~\eqref{asymp-v} and~\eqref{asymp-s}, but we choose
not to do so here because the equations of structure are simpler in
these coordinates. The asymptotic behavior above is consistent with
the one found by Foster~\cite{Foster:2007gr}, as explained in Appendix~\ref{app:mapping}.

The numerical solution to the modified field equations at
${\cal{O}}(v)$, i.e.~Eqs.~\eqref{dsdr}--\eqref{dkdr}, can be obtained
by exploiting the linearity and homogeneity of the differential
system. First, let us arbitrarily choose two sets of values for the constants
$(C,D)$ in Eqs.~\eqref{k1-0}--\eqref{w1-0} and solve
Eqs.~\eqref{dsdr}--\eqref{dkdr} from $r=r_\epsilon$ to the NS
surface. Then, we evaluate these interior solutions at the NS surface
and use them as initial conditions to numerically find the exterior
solution. Doing so, we obtain two homogeneous solutions $A_i^{(1)}$
and $A_i^{(2)}$, where $A_i \equiv (k_1,s_1,w_1)$, everywhere in the
numerical domain. With this at hand, the general solutions $A_i$  can
be obtained by linear superposition
\be A_i = C' A_i^{(1)} + D' A_i^{(2)}\,.  
\label{general-sol}
\ee
The new constants $C'$ and $D'$ must be determined, together with $A$
and $B$ in Eqs.~\eqref{asymp-v}--\eqref{asymp-w}, by requiring the
following matching conditions at spatial infinity:
\ba k_1 (r_b) &=& k_1^\infty (r_b)\,, \\ s_1 (r_b) &=& s_1^\infty
(r_b)\,, \\ w_1 (r_b) &=& w_1^\infty (r_b)\,, \\ w_1' (r_b) &=&
w_1^\infty{}' (r_b)\,.  \ea As before, we evaluate the matching at a
fixed radius $r_{b} \gg R_*$ that is large enough for the errors to be
 smaller than those due to the discretization of the
equations. We will later check that our results are insensitive to the
choice of core and boundary radii. 

The numerical techniques used to solve the ${\cal{O}}(v^0)$ and
${\cal{O}}(v)$ differential system are the following. As already
mentioned, we treat the system as an initial value problem, with
initial conditions given by the asymptotic behavior of the solution
about the NS center. We then numerically
solve the equations with an adaptive step-size, 4th-order Runge-Kutta
method~\cite{gsl}. All numerical solutions presented in this paper
are obtained with this numerical algorithm. In all cases, we have
checked that increasing the numerical resolution does not affect the
results presented.  

Finally, to obtain the sensitivities we transform the asymptotic solution for the metric [Eqs.~\eqref{asymp-v} and \eqref{asymp-s}]
to the gauge of Eqs.~\eqref{g00PN}--\eqref{gijPN} (see Appendix \ref{app:mapping} for the explicit calculation). This relates the integration
constant $A$, which we can determine from our numerical solution, to the constants $B^{\pm}$ appearing in Eqs.~\eqref{sigmaA-def-EA} and \eqref{sigmaA-def-EA2} through
\be A = -(B^{-} + B^{+}+2)\,.
\label{Eq:A-text}
\ee
With this relation at hand, we can then write the sensitivities for an isolated star in Einstein-\AE ther theory in terms of the constant $A$ as 
\begin{align}
\sigma_{\AE} &= \frac{2 c_{1} \left(2 A - 4 - \alpha_{1}^{\AE}\right)}{c_{-} \left(8 + \alpha_{1}^{\AE}\right)}\,.
\end{align}
%
\section{Neutron Star Solutions in Khronometric Theory}
\label{sec:NS-SolsHL}

In this section, we construct solutions of khronometric theory
that describe non-spinning NSs moving slowly relative to the \AE ther.  
We follow the same organizational structure as in
Sec.~\ref{sec:NS-Sols}, and thus begin by introducing the metric and
khronon field ansatz for a generic non-spinning stationary and slowly moving system,
and then present the field equations and explain how to solve them numerically.

\subsection{Metric Ansatz, Khronon Field Ansatz and Neutron Star Model}

In khronometric theory, one can construct the \AE ther field from the
khronon field $T$ through the relation in Eq.~\eqref{U_dT}.
The most generic ansatz for the metric, \AE ther and for the 4-velocity of the NS
fluid is then still given by Eqs.~\eqref{metric-ansatz}--\eqref{4-velocity}. However, Eq.~\eqref{U_dT} (hypersurface orthogonality) implies the relation
\be \epsilon^{\alpha\beta\gamma\delta} U_{\beta} \nabla_\gamma
U_\delta =0\,,
\ee
between the different components of the  \AE ther field. 
Thus, in the generic 
 ansatz of Eq.~\eqref{aether-ansatz} one has  \be
 {\partial  (Q e^{-\nu/2})}/{\partial r} =e^{-\nu/2}
{\partial  W}/{\partial \theta}
\ee
in the khronometric case. 
This means that one can choose the
function $H(r,\theta)$ in Eq.~\eqref{t-transform} such that it
satisfies the conditions
\be \frac{\partial H}{\partial r} = We^{-\nu/2}, \quad  \frac{\partial
  H}{\partial \theta} = Qe^{-\nu/2}\,.  \ee
With this, one can eliminate \textit{both} $U_r$ and $U_\theta$ and thus set $W=Q=0$ by
performing the coordinate transformation in Eq.~\eqref{t-transform}
with a judicious choice of $H$. This transformation corresponds to using a foliation adapted to
the khronon $T=$ constant hypersurfaces [i.e.~$t=T+{\cal O}(v)^2$], and leaves the ansatz of 
Eqs.~\eqref{metric-ansatz}--\eqref{4-velocity} unchanged modulo a redefinition
of the potentials $S$ and $V$.

\subsection{Field Equations}

The modified field equations at $\mathcal{O}(v^0)$ are exactly the
same as those in \AE ther theory~\cite{Jacobson:2010mx,Blas:2011ni,Blas:2010hb,Barausse:2012ny}, with the substitutions of Eq.~\eqref{eq:EAtoKH}. At $\mathcal{O}(v)$, the only non-vanishing components of the
modified Einstein equations are the $(t,r)$ and $(t,\theta)$ ones,
which, after performing the Legendre decomposition of
Eqs.~\eqref{LD:1}-\eqref{LD:3}, lead  to 
\bw \ba
\label{dkdrr}
\frac{d^2 k_n }{dr^2} &=& \frac{1}{8} \frac{1}{(\beta + \lambda) r^2
  (r-2 M)} \left\{ 4  \alpha r^2 (r-2 M) \frac{d^2 \nu}{dr^2} k_n + 8
(\beta + \lambda) r^2  \frac{d^2 M}{dr^2}  k_n +  \alpha r^2  (r-2 M)
\left( \frac{d \nu}{dr}  \right)^2 k_n \right. \nn \\ & & \left. - 4
\left[ - (\beta + \lambda) r (r-2 M)\frac{d k_n}{dr} +   (\alpha +
  \beta + \lambda) r \frac{d M}{dr} k_n + [ [3(\alpha + \lambda) -
      \beta +4] M-2 (\alpha + \lambda +1) r] k_n \right. \right. \nn
  \\ & & \left. \left. +n (n+1) (\lambda+1)  r s_n \right] r  \frac{d
  \nu}{dr} +8 (\beta + \lambda) r \left( -2 r +3 r \frac{d M}{dr}  + M
\right) \frac{d k_n}{dr} + 16 \lambda  r \frac{d M}{dr}  k_n +4 n
(n+1) (\beta + 2 \lambda +1) r^2\frac{d s_n}{dr} \right. \nn \\ & &
\left. + [- 16 (2 \beta + 3 \lambda + 1) M + 4 [-16 \pi r^2 p + n
    (n+1) (\beta -1) + 4 (\beta + \lambda ) ] r] k_n -4 n (n+1)  (3
\beta + 2\lambda  -1) r s_n  \right\}\,, \\
\label{dsdrr}
\frac{d^2 s_n }{dr^2} &=& \frac{1}{4} \frac{1}{(\beta -1) r^2  (r-2
  M)} \left\{ 4 \alpha r^2 (r-2 M) \frac{d^2 \nu}{dr^2} s_n + \alpha
r^2  (r-2 M) \left( \frac{d \nu}{dr} \right)^2 s_n \right. \nn \\ & &
\left. + 2 r \left[ (\beta -1) r (r-2 M) \frac{d s_n}{dr} -2 \alpha r
  \frac{d M}{dr} s_n + [-2(3 \alpha - \beta +1) s_n - 2(\beta -1) k_n
  ] M \right. \right. \nn \\ & & \left. \left. +r [ (4 \alpha -\beta
    +1) s_n + (\beta -1) k_n ] \right] \frac{d \nu}{dr} +4  (\beta -1)
r \left( 3 M+ r \frac{d M}{dr} -2 r \right) \frac{d s_n}{dr}
\right. \nn \\ & & \left.  +4 r [-(\beta +3) s_n +  (\beta + 2\lambda
  +1) k_n] \frac{d M}{dr} -4 (\beta + 2 \lambda +1) r (r-2 M) \frac{d
  k_n}{dr} \right. \nn \\ & & \left. + [-12 (\beta -1) s_n + 4  (7
  \beta + 6 \lambda -1) k_n] M+8 [ [-8 \pi r^2 p+ n (n+1) (\beta +
    \lambda)] s_n - 2 (\beta + \lambda)  k_n] r \right\}\,. \ea 
%
\subsection{Asymptotic Solutions, Matching Conditions and Numerical Techniques}

The initial conditions for the ${\cal{O}}(v^{0})$ equations are
exactly the same as in Einstein-\AE ther theory, as described in
Sec.~\ref{subsec:AS}, since the differential equations are
identical. The initial conditions for the ${\cal{O}}(v)$ equations, on
the other hand, will be different. Requiring that the metric be
asymptotically flat, the asymptotic behavior of the solution to
Eqs.~\eqref{dkdrr} and~\eqref{dsdrr} for the $n=1$ modes is
\ba
\label{asymp-v-HL}
k_1^{\infty}(r) &=& -1 + A \frac{M_\NS}{r} + \frac{ ( 7 + 11 \beta
  +18 \lambda ) A - 6 (6+ \alpha ) \lambda - 2(2 \alpha + 11) \beta -2
  (\alpha +7)}{8 (\beta + \lambda ) }  \frac{M_\NS^2}{r^2} \nn \\ & &
+ \left[B  + \frac{2 (\beta +3 \lambda +2) [ (\alpha + 5)A - (3 \alpha
      + 10) ] }{15(\beta + \lambda )} \ln \left( \frac{r}{M_\NS}
  \right)  \right]  \frac{M_\NS^3}{r^3} + \mathcal{O} \left(
\frac{M_\NS^4}{r^4} \right)\,, \\
\label{asymp-s-HL}
s_1^{\infty}(r) &=& -1 + \frac{(-1+3 \beta +2 \lambda) A +2 (1+2
  \beta+ 3 \lambda )}{4(\beta + \lambda)} \frac{M_\NS}{r} + \frac{ -
  (3 + \beta + 4 \lambda) A + 2(1+ \lambda) \alpha +6(1- \beta )}{8
  (\beta + \lambda ) }  \frac{M_\NS^2}{r^2} \nn \\ & & + \left\{
\frac{ -8(1 - 4 \beta - 3 \lambda ) \alpha + 5(13 + 38 \beta + 51
  \lambda) }{120 (\beta + \lambda)}  A - \frac{1}{2} B -  \frac{ 3( 1
  + 15 \beta + 17 \lambda ) \alpha + 5 (13  +38 \beta  +51 \lambda)
}{60 (\beta + \lambda )} \right. \nn \\  & &  \left. - \frac{ (\beta
  +3 \lambda +2) [ (\alpha + 5)A - (3 \alpha + 10) ] }{15(\beta +
  \lambda )} \ln \left( \frac{r}{M_\NS} \right)  \right\}
\frac{M_\NS^3}{r^3} + \mathcal{O} \left( \frac{M_\NS^4}{r^4}
\right)\,.  \ea \ew
Imposing regularity at the NS center, the
solution near the center is
\begin{align}
\label{k1-0-HL}
k_1(r) &= C + D r^2 + \mathcal{O}(r^3)\,, \\
\label{s1-0-HL}
s_1 (r) &= C  +  \frac{r^2}{3 (\beta -2 \lambda  -3) (\alpha -2 )}
\nn \\ &\times  \left( -3 \alpha \left\{ 8 \pi \left[ \left(\beta + 3
  \lambda + 3\right) p_c   \right. \right.  \right.  \nn
  \\ &\left. \left. \left.  + \rho_c \right] C + \left( 3\beta +4
\lambda +1 \right) D\right\}   \right.  \nn \\ &\left.  -16 \pi \left[
  \left(\beta + 3 \lambda  -1\right) \rho_c -3 p_c \right] C  \right.
\nn \\ &\left.  +6 \left(3 \beta + 4 \lambda +1\right) D \right)  +
\mathcal{O}(r^3)\,.
\end{align}

The matching conditions and procedure are identical to
those described in Sec.~\ref{subsec:AS} for Einstein-\AE ther theory. Similarly, the numerical
techniques are the same, and the extraction of 
the sensitivity is done by transforming the asymptotic solution for the metric 
[Eqs.~\eqref{asymp-v-HL} and \eqref{asymp-s-HL}]
to the gauge of Eqs.~\eqref{g00PN}--\eqref{gijPN} 
(see Appendix \ref{app:mapping} for the explicit calculation), which again yields Eq.~\eqref{Eq:A-text}.
Thus, one can calculate the sensitivity with the numerically-derived value of $A$ and Eqs.~\eqref{sigmaA-def-HL} or \eqref{sigmaA-def-HL2} via
\begin{align}
\sigma^{\kh} &= \frac{2 A - 4 - \alpha_{1}^{\kh}}{8 + \alpha_{1}^{\kh}}\,.
\end{align}

\section{Numerical Neutron Star Solutions and Sensitivities}
\label{sec:numerical-results}

In this section we present the results obtained by numerically solving
the modified field equations. In particular, we concentrate on
deriving numerical results for the sensitivities and developing an
analytic fitting formula. We first tackle the Einstein-\AE ther 
case, and then move on to khronometric theory. 

\subsection{Einstein-\AE ther Theory}

Let us first focus on the numerical solutions at
${\cal{O}}(v^{0})$.  At this order, the main observable is
the relation between the NS mass and its radius, for a sequence of NSs
in a given EoS family.  As can be seen from the equations of Sec.~\ref{eqs_Ov0} (and as already noted in Ref.~\cite{Eling:2007xh}),
these solutions depend only on the $c_{14}$ combination of the coupling constants [see Eq.~\eqref{c14}].

\begin{figure}[h]
\begin{center}
\includegraphics[width=8.5cm,clip=true]{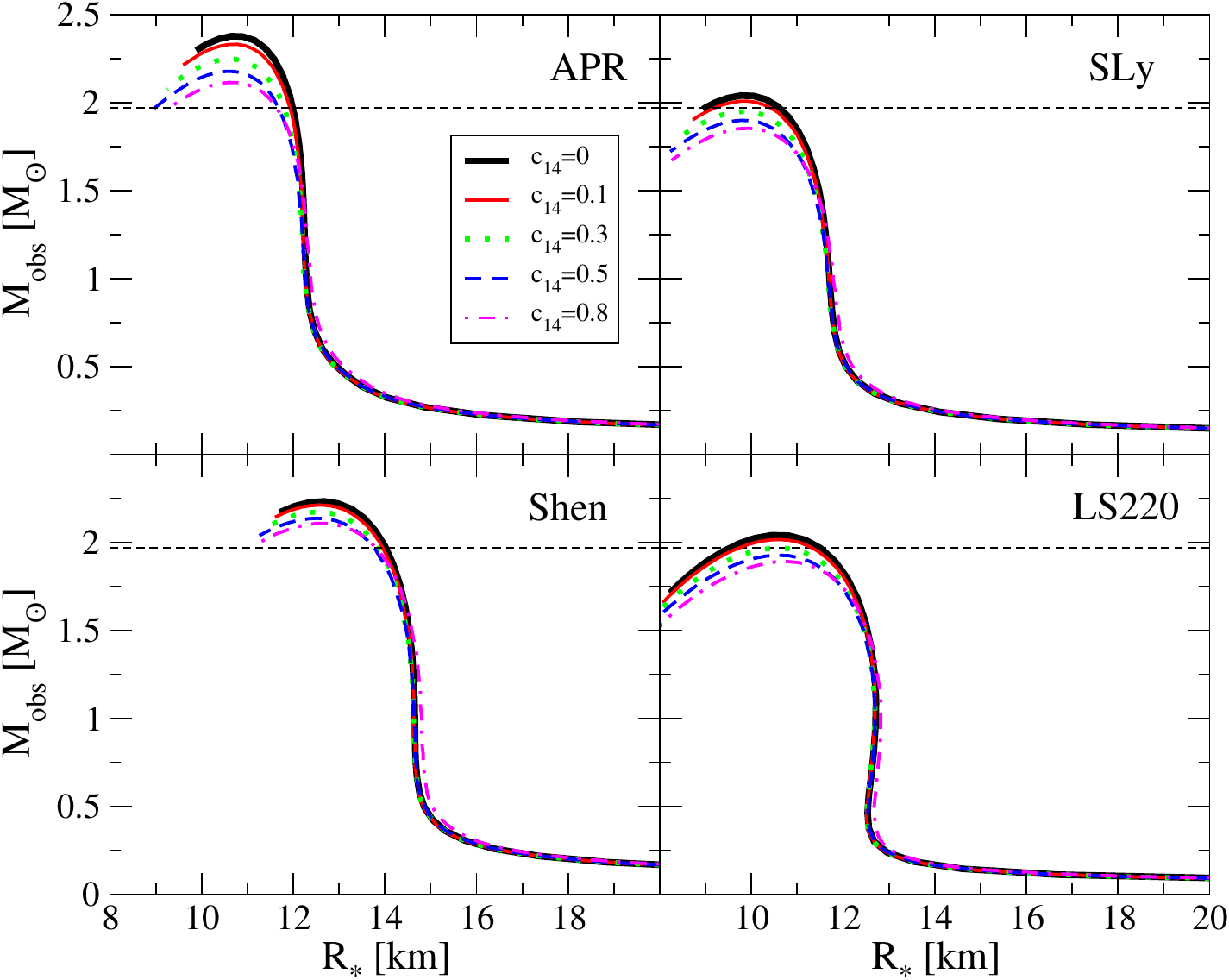}  
\caption{\label{fig:MR-AE} (Color online) Mass-radius relations in
  Einstein-\AE ther theory with different coupling strengths
  ($c_{14}=0.1,0.3,0.5,0.8$), where the thick black curve corresponds
  to the GR result. Each panel corresponds to a different EoS:  APR
  (top left), SLy (top right), Shen (bottom left) and LS220 (bottom
  right). The horizontal dashed line corresponds to the lower mass
  bound provided by observations of PSR
  J0348+0432~\cite{2.01NS}. Observe that as $c_{14}$ is increased, the
  NS mass decreases for a fixed radius.}
\end{center}
\end{figure}

Figure~\ref{fig:MR-AE} shows the mass-radius relation in
Einstein-\AE ther theory for different values of $c_{14}$. The horizontal line at
$M=1.97M_\odot$ is the lower mass bound derived from observations of
PSR J0348+0432~\cite{2.01NS}. Observe that, as one increases $c_{14}$,
the NS mass decreases for a fixed radius, which is consistent with the
conclusions in Ref.~\cite{Eling:2007xh}. 

This figure is a perfect example of the strong degeneracy
between the EoS and modified gravity effects, which in turn prevents us
from constraining modified theories with the mass-radius relation alone.
For example, if we knew that the LS220 EoS was the correct one, then
we could argue that the observation of PSR J0348+0432~\cite{2.01NS}
requires that $c_{14} < 0.1$. However, we do not know the correct EoS
and, for instance, the APR EoS could be the correct one. If that were the
case, we would not be able to place competitive constraints on $c_{14}$. We then conclude
that the observation of PSR J0348+0432~\cite{2.01NS} (or any other system
for that matter) is ineffective at constraining Einstein-\AE ther theory through
the mass-radius relation.

Let us now consider the ${\cal{O}}(v)$ solutions, and in particular, the
sensitivities in Einstein-\AE ther~theory. In the weak-field limit, i.e.~expanding
in the ratio of the binding energy $\Omega$ to the NS mass $M_{\obs}$, one can show
that the sensitivity scales as~\cite{Foster:2007gr}
\be
\label{eq:weak-field-AE-s}
s_{\AE}^{\wf} = \left( \alpha_1^{\AE} - \frac{2}{3} \alpha_2^{\AE} \right)
\frac{\Omega}{M_\obs} + \mathcal{O} \left( \frac{\Omega^2}{M_{\obs}^2}
\right)\,, \ee
with $\alpha_{1}^{\AE}$ and $\alpha_{2}^{\AE}$ given by Eqs.~\eqref{alpha1} and~\eqref{alpha2}, 
while~\cite{Foster:2005dk,Foster:2006az}
\be 
\Omega = - \frac{1}{2}  G_N \int d^3x \rho (r) \int d^3x' \frac{\rho(r')}{|\bm{x}-\bm{x'}|}\,, 
\ee
with $r=|\bm{x}|$ and $r'=|\bm{x'}|$. When plotting the weak-field sensitivity using Eq.~\eqref{eq:weak-field-AE-s},
we evaluate $\Omega$ by using the Legendre expansion of the Green's function of the Laplacian 
operator~\cite{willwiseman}. Of course, this integral depends on the EoS through $\rho(r)$.

\begin{figure}[h]
\begin{center}
\begin{tabular}{r l}
\includegraphics[width=8.5cm,clip=true]{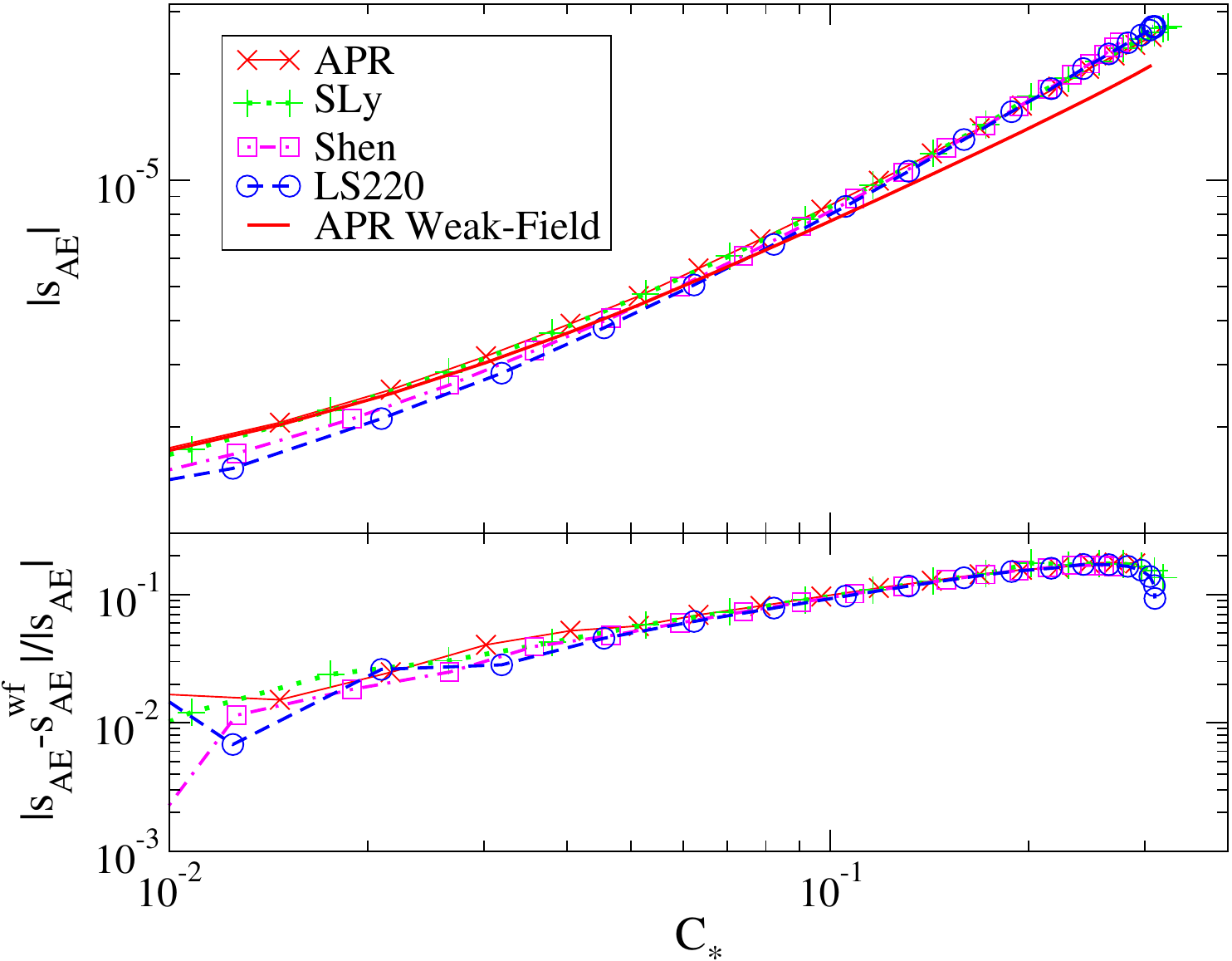}  
\end{tabular}
\caption{\label{fig:sens-alpha1} (Color online)    
The absolute magnitude of the sensitivity in Einstein-\AE ther~theory against the NS compactness for
various EoS and the weak-field expression in Eq.~\eqref{eq:weak-field-AE-s} with the APR EoS. 
  The bottom panel shows the fractional difference between the sensitivity and the weak-field
  expression in Eq.~\eqref{eq:weak-field-AE-s}. We saturate the PPN parameters $\alpha_1^{\AE}$ and $\alpha_2^{\AE}$ 
such that they satisfy the Solar System constraint, and $c_+=10^{-4}=c_-$. Observe that, as expected, the weak-field result is highly
  accurate for small compactness, but it becomes inaccurate for realistic 
  NS compactnesses.}
\end{center}
\end{figure}

Figure~\ref{fig:sens-alpha1} shows the absolute magnitude
of the sensitivity in Einstein-\AE ther theory, calculated from the numerical solution to the
${\cal{O}}(v)$ modified field equations, for different EoS as a function of NS compactness $C_{*} = M_{*}/R_* =G_N M_\obs/R_*$.
For comparison, we also plot the weak-field expression for the sensitivity [Eq.~\eqref{eq:weak-field-AE-s}] with the APR EoS.
The bottom panel shows the fractional difference between the actual sensitivity 
and its weak field approximation. Observe that as $C_{*}$ increases, 
Eq.~\eqref{eq:weak-field-AE-s} becomes highly inaccurate, with errors of roughly $15$--$30\%$
for realistic NS compactnesses, i.e.~for compactnesses $0.1 \lesssim C_{*} \lesssim 0.3$.  

\begin{figure}[h]
\begin{center}
\includegraphics[width=8.5cm,clip=true]{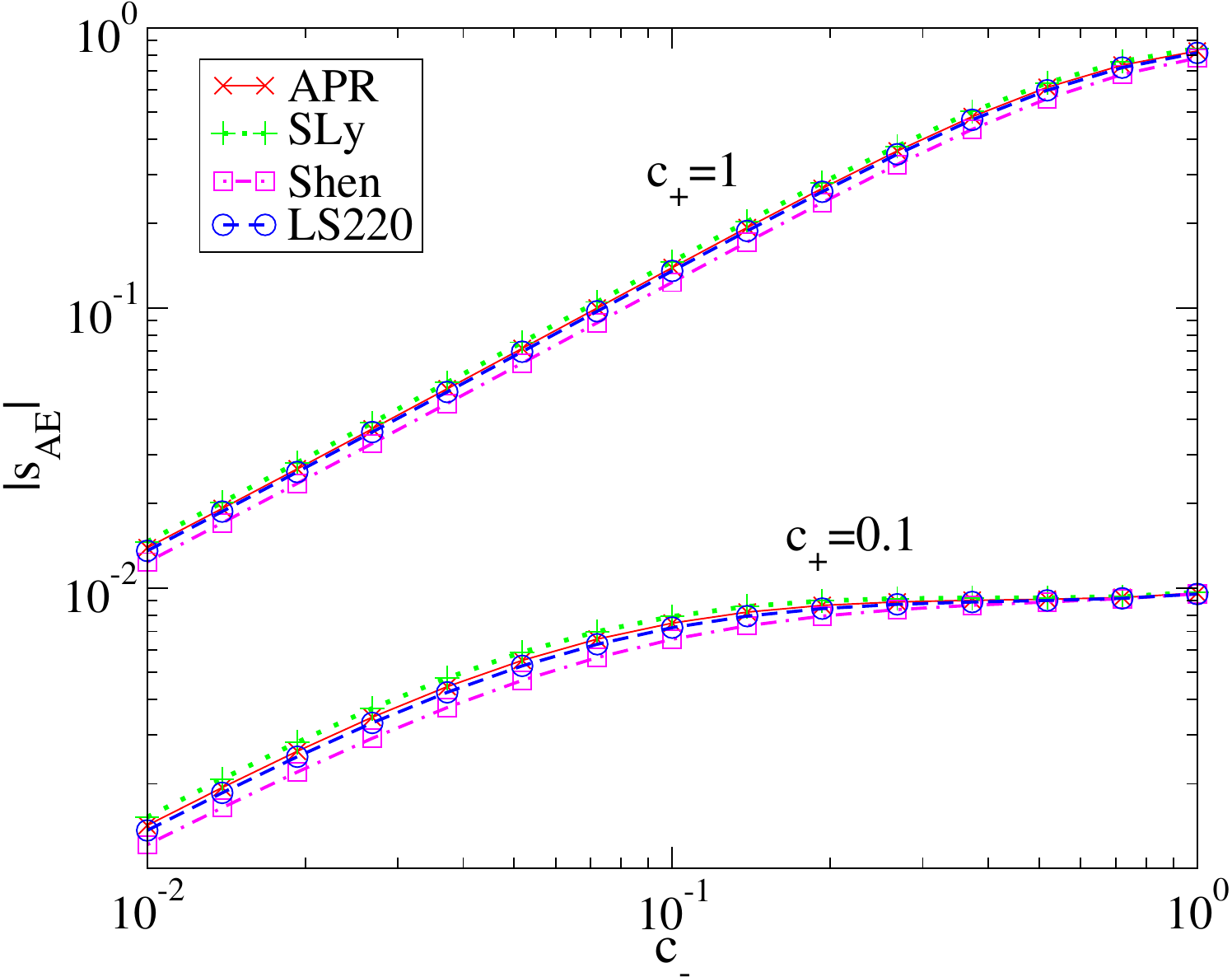}  
\caption{\label{fig:sens-c-} (Color online) 
  The absolute magnitude of the sensitivity in Einstein-\AE ther~theory against $c_{-}$ for
  $c_{+}=1$ and $c_{+}=0.1$. We saturate the PPN parameters $\alpha_1^{\AE}$ and $\alpha_2^{\AE}$ and set the NS
  mass to $M_\obs=1.4M_\odot$.}
\end{center}
\end{figure}

Figure~\ref{fig:sens-c-} shows the absolute magnitude of the sensitivity as a function 
of $c_{-}$ for $c_{+} = 1$ and $c_{+} = 0.1$, saturating the Solar System
constraint for $\alpha_{1}^{\AE}$ and $\alpha_{2}^{\AE}$. Observe that the sensitivity 
increases monotonically with $c_{-}$ and $c_{+}$. 

\begin{figure}[htb]
\begin{center}
\includegraphics[width=9cm,clip=true]{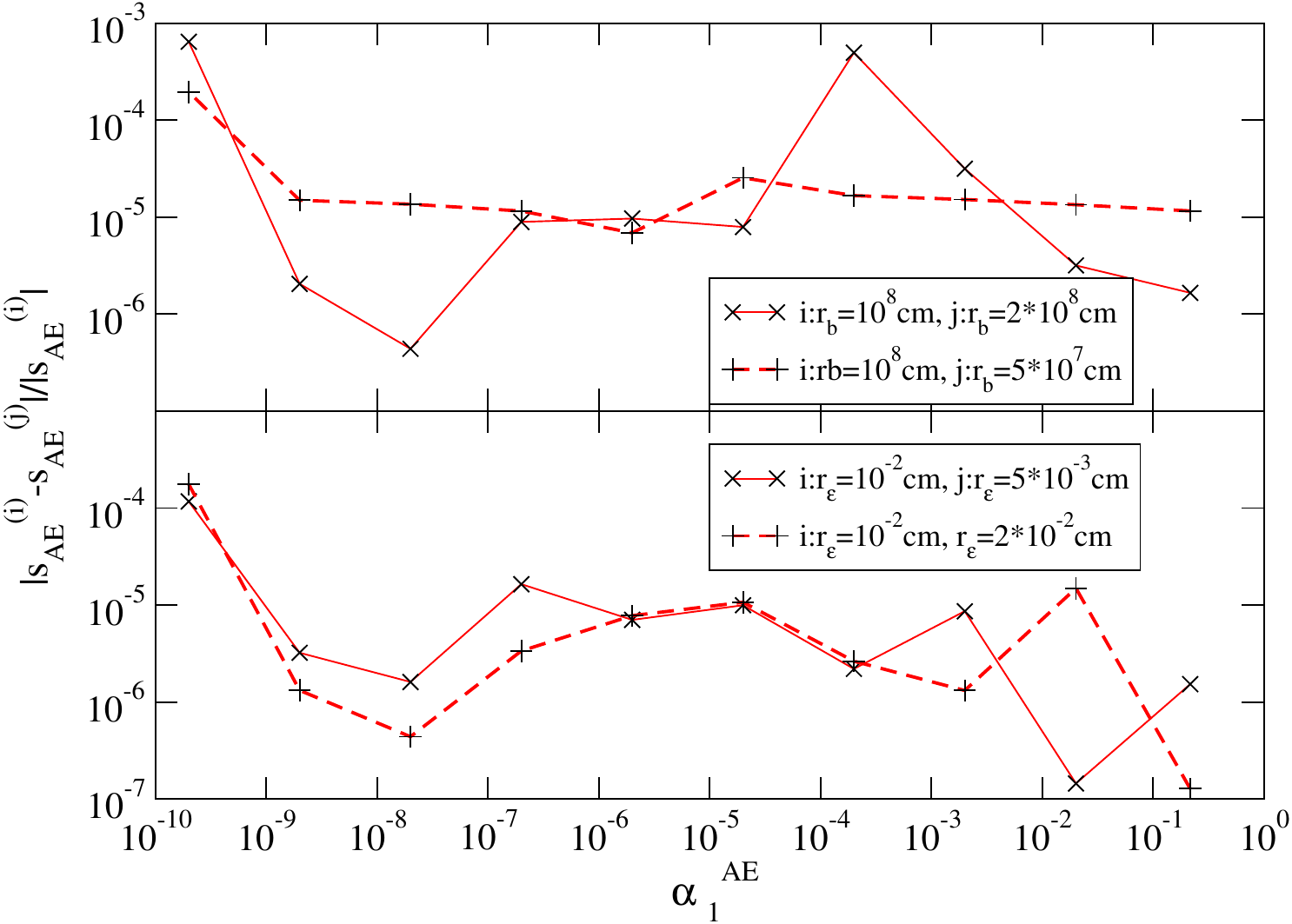}  
\caption{\label{fig:sens-core-boundary} (Color online) The fractional
  difference in sensitivity against the PPN parameter $\alpha_1^{\AE}$ in Einstein-\AE ther~theory for different boundary radius $r_b$ (top) and core radius
  $r_\epsilon$ (bottom) with the APR EoS. We set
  the coupling constants to $c_1=2c_2=2c_3=-2c_4$, with
  $M_\obs  = 1.4M_\odot$.}
\end{center}
\end{figure}

One may wonder whether the results presented here are
robust to the method used to obtain the numerical solutions of the
modified field equations. Two key features of this method are the choice
of core radius (where the interior integration is numerically
started) and the value of the boundary radius (where the matching is
carried out). Figure~\ref{fig:sens-core-boundary} shows the
sensitivity as a function of the matching radius (top panel) and the
core radius (bottom panel) for the APR EoS. Observe that the
fractional error on the sensitivities is always smaller than $0.1
\%$. 

Figures~\ref{fig:sens-alpha1} and~\ref{fig:sens-c-} show that the sensitivities
in Einstein-\AE ther theory do not depend strongly on the EoS. Because of this, 
it is natural to develop an EoS-independent fitting formula. First, we choose
the LS220 EoS as a representative EoS with which to compute sensitivities
and we set $\alpha_{1}^{\AE} = 10^{-4}$ and $\alpha_{2}^{\AE} = 4 \times 10^{-7}$, thus
saturating Solar System constraints. With this, we then numerically compute
the sensitivity in the range $c_{+} \in 3 \times (10^{-4},10^{-1})$, 
$c_{-} \in 3 \times  (10^{-5},10^{-3})$ and $C_{*} \in (0.11,0.28)$ for a total 
of $10^{3}$ points. We choose this region in the parameters $(c_{+},c_{-})$ as it will still be allowed after imposing
binary pulsar constraints (see Fig.~\ref{fig:Pdot-constraint-summary}). 

With the data in hand, we  choose a functional form for the fitting function. 
By plotting the sensitivity as a function of $c_{+}$, $c_{-}$ and $C_{*}$, we empirically
find that $s_{\AE}$ is well-described by a polynomial in these quantities 
(at least in the region of parameter space described above). We could use a fitting
function that corrects Eq.~\eqref{eq:weak-field-AE-s} with higher powers of $C_{*}$, 
but we find that this is not as good as  
\be 
\label{s-fit}
s_{\AE} = \sum_{\ell,m,n=0}^{2} c_{\ell, m, n} c_{+}^{\ell} c_{-}^{m} C_{*}^{n}\,.
\ee
We fit the data described above with Eq.~\eqref{s-fit}, using the fact that each numerical point 
is known to at least a fractional numerical accuracy of $1\%$. The fit returns an $r^{2}$ value of $0.959$, 
with fitted coefficients given in Table~\ref{fitting-table}. We also carried out a more accurate fit 
by keeping more terms in the polynomial expansion of Eq.~\eqref{s-fit}; the $r^{2}$ is then $0.9994$ 
and the coefficients are given in Appendix~\ref{app:fit}.  Note that $c_{0,0,0}  = {\cal{O}}(\alpha_{1,2}^{\AE}) \neq 0$, 
because we have chosen $\alpha_{1,2}^{\AE}$ such that Solar System constraints are saturated. Thus, the
fitting function does not go to zero as $c_{+} \to 0$ and $c_{-} \to 0$. 

{\renewcommand{\arraystretch}{1.2}
\begin{table*}
\begin{centering}
\begin{tabular}{cccccccccccccccccccccccccccc}
\hline
\hline
\noalign{\smallskip}
\AE & \multicolumn{1}{c}{$c_{0,0,0}$} 
& \multicolumn{1}{c}{$c_{0,0,1}$} 
& \multicolumn{1}{c}{$c_{0,0,2}$} 
& \multicolumn{1}{c}{$c_{0,1,0}$}
& \multicolumn{1}{c}{$c_{0,1,1}$} 
& \multicolumn{1}{c}{$c_{0,1,2}$} 
& \multicolumn{1}{c}{$c_{0,2,0}$} 
& \multicolumn{1}{c}{$c_{0,2,1}$} 
& \multicolumn{1}{c}{$c_{0,2,2}$} 
\\
\hline
\noalign{\smallskip}
& $1.95 \times 10^{-5}$  & 
$-3.15\times 10^{-4}$ & 
$4.60 \times 10^{-4}$ & 
$7.58 \times10^{-2}$ & 
$-1.07$ & 
$4.34$ &
$-3.19 \times 10^{1}$ & 
$4.37 \times 10^2$ & 
$-1.6 \times 10^3$ \\

\noalign{\smallskip}
\AE & \multicolumn{1}{c}{$c_{1,0,0}$}  
& \multicolumn{1}{c}{$c_{1,0,1}$} 
& \multicolumn{1}{c}{$c_{1,0,2}$} 
& \multicolumn{1}{c}{$c_{1,1,0}$}
& \multicolumn{1}{c}{$c_{1,1,1}$}
& \multicolumn{1}{c}{$c_{1,1,2}$} 
& \multicolumn{1}{c}{$c_{1,2,0}$} 
& \multicolumn{1}{c}{$c_{1,2,1}$} 
& \multicolumn{1}{c}{$c_{1,2,2}$} 
\\
\hline
& $-2.14 \times 10^{-2}$ & 
$2.90 \times 10^{-1}$ & 
$-9.86 \times 10^{-1}$ & 
$6.39 \times 10^{1}$ & 
$-8.34 \times 10^2$ & 
$2.68 \times 10^3$ & 
$-4.57 \times 10^{3}$ & 
$5.7 \times 10^{4}$ & 
$-1.51 \times 10^5$ \\

\noalign{\smallskip}
\AE & \multicolumn{1}{c}{$c_{2,0,0}$} 
& \multicolumn{1}{c}{$c_{2,0,1}$} 
& \multicolumn{1}{c}{$c_{2,0,2}$} 
& \multicolumn{1}{c}{$c_{2,1,0}$}
& \multicolumn{1}{c}{$c_{2,1,1}$} 
& \multicolumn{1}{c}{$c_{2,1,2}$} 
& \multicolumn{1}{c}{$c_{2,2,0}$} 
& \multicolumn{1}{c}{$c_{2,2,1}$} 
& \multicolumn{1}{c}{$c_{2,2,2}$}
\\
\hline
& $5.67 \times 10^{-1}$ & 
$-7.67$ & 
$2.65 \times 10^1$ & 
$-1.87 \times 10^3$ & 
$2.49 \times 10^4$ & 
$-8.04 \times 10^4$ & 
$2.32 \times 10^5$ & 
$-2.99 \times 10^6$ & 
$8.91 \times 10^6$ \\
\hline
\hline
\noalign{\smallskip}
$kh$ & \multicolumn{1}{c}{$c_{0,0,0}$} 
& \multicolumn{1}{c}{$c_{0,0,1}$} 
& \multicolumn{1}{c}{$c_{0,0,2}$} 
& \multicolumn{1}{c}{$c_{0,1,0}$}
& \multicolumn{1}{c}{$c_{0,1,1}$} 
& \multicolumn{1}{c}{$c_{0,1,2}$} 
& \multicolumn{1}{c}{$c_{0,2,0}$} 
& \multicolumn{1}{c}{$c_{0,2,1}$} 
& \multicolumn{1}{c}{$c_{0,2,2}$} \\
\hline
& 
$-2.87\times 10^{-8}$ &
$-1.82\times 10^{-6}$ & 
$6.34 \times 10^{-6}$ & 
$3.20 \times 10^{-6}$ & 
$2.09 \times 10^{-5}$ &
$-9.83 \times 10^{-5}$ & 
$-2.62 \times 10^{-5}$ & 
$-1.14\times 10^{-4}$ & 
$5.95 \times 10^{-4}$ \\
\noalign{\smallskip}
$kh$ & \multicolumn{1}{c}{$c_{1,0,0}$} 
& \multicolumn{1}{c}{$c_{1,0,1}$} 
& \multicolumn{1}{c}{$c_{1,0,2}$} 
& \multicolumn{1}{c}{$c_{1,1,0}$}
& \multicolumn{1}{c}{$c_{1,1,1}$} 
& \multicolumn{1}{c}{$c_{1,1,2}$} 
& \multicolumn{1}{c}{$c_{1,2,0}$} 
& \multicolumn{1}{c}{$c_{1,2,1}$} 
& \multicolumn{1}{c}{$c_{1,2,2}$} \\
\hline
& 
$ 0.313$ &
$-1.01$ & 
$ 1.25$ & 
$-0.506$ & 
$1.76$ &
$-0.983$ & 
$ 0.148$ & 
$ 0.491$ & 
$-2.86$ \\
\noalign{\smallskip}
$kh$ & \multicolumn{1}{c}{$c_{2,0,0}$} 
& \multicolumn{1}{c}{$c_{2,0,1}$} 
& \multicolumn{1}{c}{$c_{2,0,2}$} 
& \multicolumn{1}{c}{$c_{2,1,0}$}
& \multicolumn{1}{c}{$c_{2,1,1}$} 
& \multicolumn{1}{c}{$c_{2,1,2}$} 
& \multicolumn{1}{c}{$c_{2,2,0}$} 
& \multicolumn{1}{c}{$c_{2,2,1}$} 
& \multicolumn{1}{c}{$c_{2,2,2}$} \\
\hline
& 
$1.03$ & 
$-0.268$ & 
$-8.38$ & 
$-9.29$ & 
$-39.3$ &
$206$ & 
$51.6$ &  
$254$ & 
$-1250$ \\
\noalign{\smallskip}
\hline
\hline
\end{tabular}
\end{centering}
\caption{\label{fitting-table} Estimated numerical coefficients for the fitting formulas of the sensitivity in Einstein-\AE ther (\AE) theory and khronometric gravity (kh).}
\end{table*}
}

\subsection{Khronometric Theory}

Spherically symmetric solutions in khronometric theory are identical
to those of Einstein-\AE ther theory [after the substitution in Eq.~\eqref{eq:EAtoKH}]~\cite{Jacobson:2010mx,Blas:2011ni,Blas:2010hb,Barausse:2012ny} . Thus, the mass-radius relation for NSs
will also be identical to what we presented in Fig.~\ref{fig:MR-AE}. 
We recall, however, that stability/Cherenkov requirements, Solar System experiments and BBN bounds 
restrict the couplings $\lambda$, $\beta$ and $\alpha$ to very small values (cf. Fig.~\ref{fig:Pdot-constraint-summary}),
so the deviations away from the GR mass-radius relation will be even smaller than in Einstein-\AE ther~theory.
For this reason, and because of the degeneracies between the mass-radius relation and the EoS,
khronometric theory cannot
be constrained by measurements of NS pulsar masses alone.

Let us then focus on the numerical solutions to the modified field equations 
at ${\cal{O}}(v)$. In the weak-field limit, one can show
that the sensitivities in khronometric theory scale in the same exact way as
in Einstein-\AE ther theory [Eq.~\eqref{eq:weak-field-AE-s}], except that now $\alpha_{1}^{\kh}$ and
$\alpha_{2}^{\kh}$ are given by Eqs.~\eqref{alpha1-HL} and~\eqref{alpha2-HL}. 

\begin{figure*}[htb]
\begin{center}
\begin{tabular}{r l}
\includegraphics[width=8.5cm,clip=true]{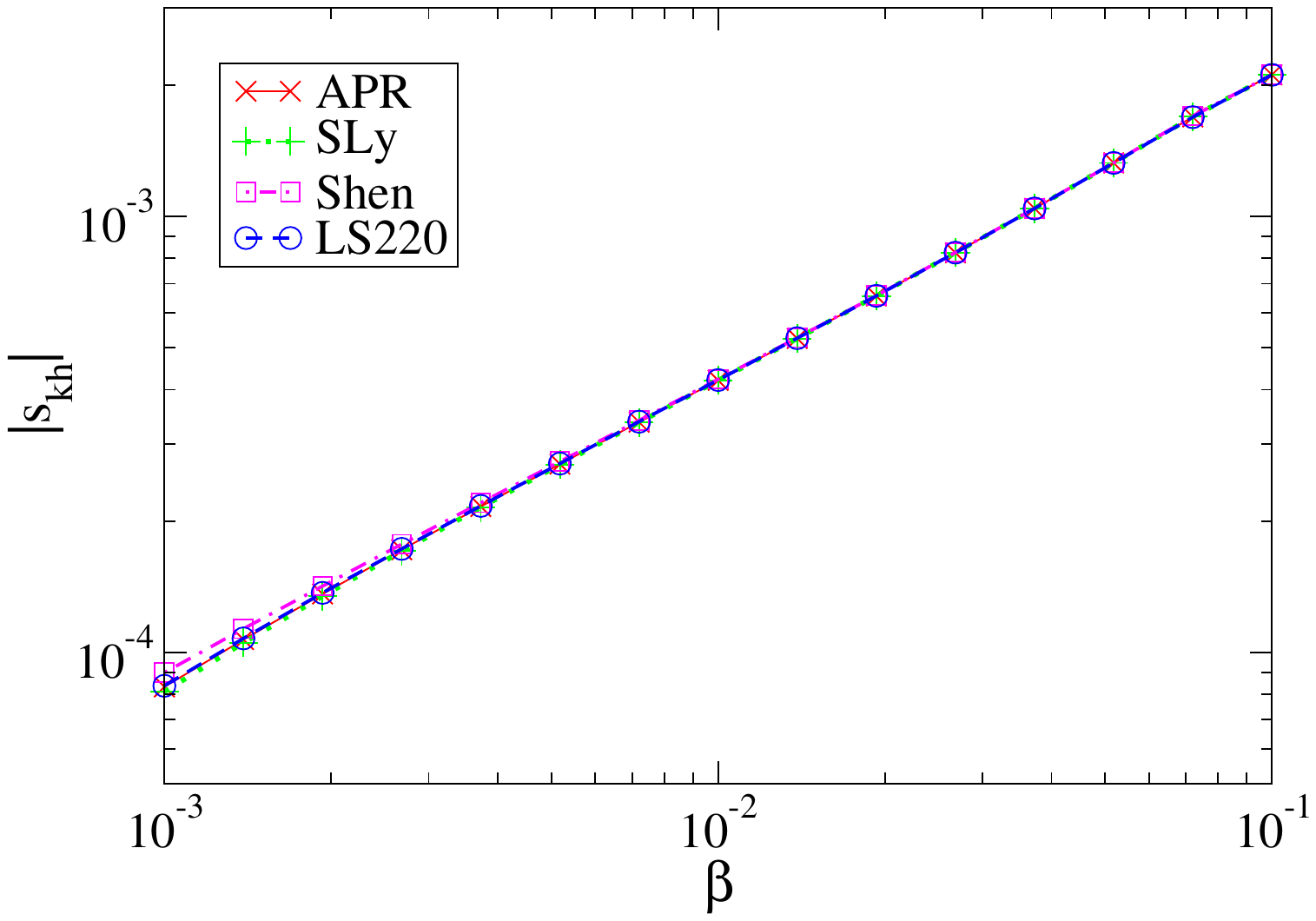}  &
\includegraphics[width=8.5cm,clip=true]{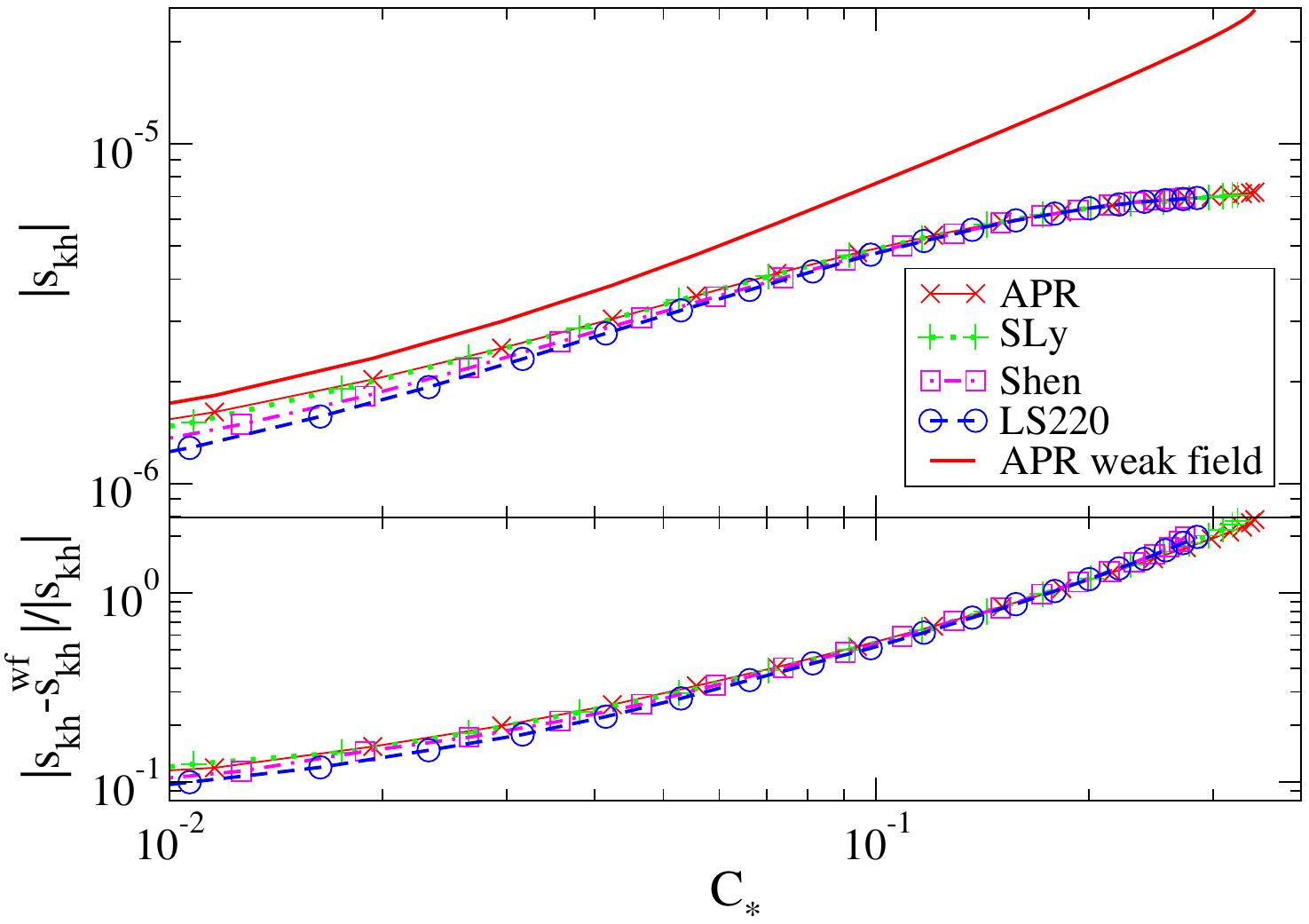}  
\end{tabular}
\caption{\label{fig:sens-C-HL} (Color online) 
  Absolute value of the sensitivity in khronometric theory as a function of the
  coupling constant $\beta$ for a star with $M_{\obs} = 1.4
  M_{\odot}$ (left panel) and as a
  function of compactness with $\beta = 10^{-4}$ (right panel). In
  both cases, we set the PPN parameters $\alpha_{1}^{\kh}$ and $\alpha_{2}^{\kh}$ by saturating the
  Solar System constraint [we saturate Eqs.~\eqref{option2a} and~\eqref{option2b}]. The bottom right panel shows the fractional
  difference between the numerical sensitivity and its weak-field
  value. Observe that for large compactnesses, the weak-field result
  can be very inaccurate. }
\end{center}
\end{figure*}

Figure~\ref{fig:sens-C-HL} shows the absolute magnitude
of the sensitivity in khronometric theory, calculated from the numerical solution to the
${\cal{O}}(v)$ modified field equations, for different EoSs as a function of $\beta$ (left panel) 
and NS compactness $C_{*}$ (right panel). Here we have chosen $\alpha$ and $\lambda$ by saturating 
Eqs.~\eqref{option2a} and ~\eqref{option2b}.
The bottom right panel shows the fractional difference 
between the actual sensitivity and its weak field approximation, where we recall that the latter is given by Eq.~\eqref{eq:weak-field-AE-s} with $\alpha_1^{\AE}$ and $\alpha_2^{\AE}$ replaced by $\alpha_1^{\kh}$ and $\alpha_2^{\kh}$. 
Observe first that the behavior of the sensitivity in khronometric theory as a function of $\beta$  is 
rather different from that of the sensitivity in Einstein-\AE ther~theory as a function of $c_{-}$. 
Observe also that as $C_{*}$ increases, the weak-field result becomes highly inaccurate, 
with errors of roughly $50$--$100\%$ for realistic NS compactnesses. 

As in the Einstein-\AE ther case, Fig.~\ref{fig:sens-C-HL} shows that the sensitivities
in khronometric theory are also rather insensitive to the EoS. Because of this, 
it is natural to develop an EoS-independent fitting formula. Again, we choose a fitting function of the form 
\be 
s_{\kh} = \sum_{\ell,m,n=0}^{2}  c_{\ell, m, n} \; \beta^{\ell} \lambda^{m} C_{*}^{n} \,, 
\ee
and fit this to sensitivities computed numerically using the LS220 EoS and choosing $\alpha = 2 \beta$ due to current Solar System constraints [see Eq.~\eqref{option1}]. We carry out this fit with data in the region $\beta \in (5 \times 10^{-5},5\times 10^{-3})$, $\lambda \in (3 \times 10^{-3},10^{-1})$ and $C_{*} \in (0.11,0.26)$ with a total of $10^{3}$ points. This range in $\beta$ and $\lambda$ for the fits is chosen in this way because these values remain viable after placing binary pulsar constraints. As in the Einstein-\AE ther case, when carrying out the fits, we use the fact that each numerical point is known to at least a fractional accuracy of $1\%$. The fit returns an $r^{2}$ value of $0.99996$, with fitted coefficients given in Table~\ref{fitting-table}. Notice that this fit is significantly better than the Einstein-\AE ther one, so we do not need to re-do the fit with a higher-order polynomial. 

We will not repeat Fig.~\ref{fig:sens-core-boundary} for khronometric theory, but we have checked
that indeed our results are accurate to better than $0.1\%$, and thus, unaffected
by the way the numerical solution is obtained. 

\section{Pulsar Constraints}
\label{sec:Constraints}

In this section, we discuss how to place constraints on Lorentz-violating theories with binary pulsar observations. We begin with an overview of how binary pulsars can be used to test GR. We then specialize our discussion to Lorentz-violating theories. We conclude with an implementation of these ideas to constrain Einstein-\AE ther and khronometric theory.    

\subsection{Binary Pulsars As Laboratories for Fundamental Physics}

Einstein-\AE ther and khronometric theory predict that the orbital period decay of binary systems should generically be faster than that predicted by GR because of the emission of dipolar radiation, i.e.~due to a correction to the dissipative sector of GR. All observations of binary pulsars to date, however, have verified the GR prediction. Therefore, binary pulsars can be used to place constraints on these theories, by essentially requiring that the non-GR effects be smaller than observational uncertainties~\cite{stairs}. 

Binary pulsar astrophysics is carried out by fitting a suitably-averaged set of observed pulses to a given pulsar model. The output of this procedure is then a set of best-fit model parameters that describe the observation (e.g.~the rate of change of pericenter, the orbital decay rate, the rate of change of the line of nodes, etc.) with a given observational error estimate for each of these parameters. This observational error, of course, decreases with observation time, as more data is accumulated.   

Some parameters of the binary system, however, are not directly measurable, but are rather inferred from other observables. Let us for example consider the individual masses of the binary. These are inferred by noting that other observables are functions of the individual masses, once one chooses a gravitational theory. Since these observables are measured up to some uncertainty, this error also propagates into the inferred masses.

More specifically, the inferred individual masses can be determined from two binary observables that depend only on the conservative sector of the theory, e.g.~the rate of change of the pericenter and the Shapiro time-delay. This inference can be carried out with the leading-order, Newtonian expressions for these post-Keplerian observables, since 1PN order corrections and higher will be greatly subdominant. In Einstein-\AE ther and in khronometric theory, the conservative sector is modified to leading PN order only through the substitutions $G_{N} \mapsto {\cal{G}}=G_{N}[1+{\cal{O}}(\sigma_{A},\sigma_{B})]$ and $\tilde{m}_{A} \mapsto m_{A}=\tilde{m}_{A} [1+{\cal{O}}(\sigma_{A})]$ [cf. Eq.~\eqref{Newtonian-a}]. This leads to corrections of ${\cal{O}}(\sigma_{A})$ to the masses measured assuming GR as the fiducial gravitational theory. 

One may wonder whether one has to account for corrections of ${\cal{O}}(\sigma_{A})$ in the masses, as well as 1PN order corrections to the conservative sector, when placing constraints on Einstein-\AE ther and khronometric theory with observations of the orbital decay rate. This is not the case  because in these theories $\dot{P}_b /P_b$ differs from
the GR prediction at \textit{leading} PN order due to corrections to
the dissipative sector.
More precisely, deviations away from GR appear at -1PN order for NS-white dwarf
systems, and at 0PN when the -1PN terms are subdominant (such as in the case of
binary NSs; see discussion after Eq.~\eqref{A-AE} for more details).
Thus, corrections of $\mathcal{O}(\sigma_A)$ and those of 1PN to the conservative
sector are always smaller than observational errors.  

Once one has determined the inferred individual masses and measured the orbital period of the system, one can calculate the predicted rate of change of the orbital period and compare this to the observed value. To date, the GR prediction has always agreed with the observed value
\be
\label{obs}
\left(\frac{\dot{P}_{b}}{P_{b}}\right)_{\obs} = \left(\frac{\dot{P}_{b}}{P_{b}}\right)_{\GR} \left(1 \pm \delta_{\obs} \pm \delta_{\inferred}\right),
\ee
up to the observational uncertainty $\delta_{\obs}$ in $\dot{P}_{b}$ and the inferred error $\delta_{\inferred}$ induced in the predicted value of $\dot{P}_{b}/P_{b}$ due to the error that propagates from the individual masses.

In Einstein-\AE ther and in khronometric theory, the predicted orbital decay rate differs from the GR one:
\be
\label{pred}
\left(\frac{\dot{P}_{b}}{P_{b}}\right)_{\AE,kh} = \left(\frac{\dot{P}_{b}}{P_{b}}\right)_{\GR} \left(1 + \delta \dot{P}_{\AE,kh} \right)\,,
\ee
where $\delta \dot{P}_{\AE, kh}$ is the fractional difference in the prediction. In fact, because of the presence of dipolar radiation, $\delta \dot{P}_{\AE,kh}$ is much larger than unity for small velocities, unless the coupling constants are very small. As we will see later in this section, $\delta \dot{P}_{\AE,kh} = {\cal{O}}(c^2/v_{12}^2)$, while $({\dot{P}_{b}}/{P_{b}})_{\GR} = {\cal{O}}(v_{12}^{10}/c^{10})$, with $v_{12}$ the magnitude of the binary's orbital velocity. 

Comparing Eqs.~\eqref{obs} to~\eqref{pred}, one then finds that the most conservative constraint comes from demanding 
\be
\left| \delta \dot{P}_{\AE,kh} \right| \leq \delta_{\obs} + \delta_{\inferred} = {\cal{O}}(10 \%)\,.
\ee
In what follows, we will instead use the equivalent relation
\be
\label{eq-used}
\delta \dot{P}_{\AE,kh}(m_{A}+\delta m_{A}, P_{b} +\delta P_{b}, \kappa^{i}_{\AE,kh},{\rm{EoS}}) \leq \delta_{\obs}\,,
\ee
to place constraints on the coupling constants $\kappa^{i}_{\AE,kh}$, where $\kappa^{i}_{\AE} = (c_{+},c_{-})$ in Einstein-\AE ther theory and $\kappa^{i}_{kh} = (\lambda,\beta)$ in khronometric theory. 
Here $m_{A}$ and $P_{b}$ are the best-fit parameter for the (inferred) individual masses and the orbital period, while $\delta m_{A}$ and $\delta P_{b}$ are their associated statistical errors.  Recall that, in these theories, the non-GR correction to the prediction of the orbital decay rate not only depends on the individual masses and the orbital period, but also on the EoS through the sensitivities. Thus, we will search for the region in the two-dimensional coupling parameter space where Eq.~\eqref{eq-used} is satisfied, with $\delta \dot{P}_{\AE,kh}$ evaluated within $m_{A} \pm \delta m_{A}$, $P_{b} \pm \delta P_{b}$ and various values of the sensitivities, determined by the EoS and the individual masses. Let us emphasize that Eq.~\eqref{eq-used} does not lead to a line in the two-dimensional coupling parameter space, but rather a two-dimensional region.

\subsection{Binary and Isolated Pulsars \\ As Laboratories for Lorentz-Violation}

To date, all robust pulsar constraints on Lorentz-violation come from studies of its effects on the conservative dynamics of the system that then propagate to the observed pulse sequence. Such effects are modeled with the parameterized post-Newtonian (PPN) formalism~\cite{nordvedt,Nordtvedt:1968qs,1972ApJ...177..757W,1972ApJ...177..775N,1971ApJ...163..611W,1973ApJ...185...31W}, where the point-particle Lagrangian is modified to include preferred-frame effects. The latter are proportional to certain PPN parameters (the relevant ones here are $\alpha_{1}$ and $\alpha_{2}$), as well as contractions of preferred-frame-related velocity vectors and PN vector potentials~\cite{TEGP,will-living}. Such modifications to the Lagrangian induce effects on pulsar observables, which can then be used to place constraints on $\alpha_{1}$ and $\alpha_{2}$. Let us classify these in terms of the sources used: isolated pulsars and binary pulsars. 

Let us first consider the isolated pulsar case. In Lorentz-violating theories of gravity, the presence of a preferred frame
induces precession of the pulsar's spin axis relative to this frame, pushing the pulsar beam in and out of the line of sight~\cite{1987ApJ...320..871N,Damour:1992ah}.  This effect has not been observed in isolated pulsars, which then allows for constraints~\cite{Shao:2013wga} on the strong-field generalization ($\hat{\alpha}_{1},\hat{\alpha}_{2})$~\cite{Damour:1992ah} of the PPN parameters $(\alpha_{1},\alpha_{2})$. The angular precession velocity of the spin angular momentum vector due to preferred frame effects, and thus, the rate of change  of the angle between the spin axis and the line of sight, is proportional to $\hat{\alpha}_{2}$~\cite{1987ApJ...320..871N,Damour:1992ah}, but also depends on the magnitude and direction of the relative 3-velocity of the isolated pulsar with respect to the preferred frame. Using data from PSR B1937+21~\cite{Rutledge:2003xk} and PSR J1744--1134~\cite{1997ApJ...481..386B} that shows no such precession, Shao, {\textit{et al}}~\cite{Shao:2013wga} placed the constraint $\hat{
\alpha}_{2} < 1.6 \times 10^{-9}$ at $95\%$ confidence. Here, they associate the preferred-frame 3-velocity with  our Solar System barycenter velocity relative to the CMB, randomizing over the pulsar's radial velocity with respect to us and the spin orientation. 

Let us now consider the binary pulsar case. For systems with small eccentricity, preferred-frame effects parametrized by $\hat{\alpha}_{1}$ and $\hat{\alpha}_{2}$ decouple. The former induce a correction to the rate of change of the eccentricity, which in turn affects the precession of the pericenter. The latter induce precession of the orbital angular momentum around the center of mass' 3-velocity vector in the  preferred frame, leading to a change in the inclination angle with respect to the line of sight, and thus, precession of the projected semi-major axis~\cite{stairs}.  The orbital plane and pericenter precession produces noticeable effects in the pulse profiles, all of which depend on the magnitude and direction of the 3-velocity of the binary system in the preferred frame.

All current pulsar observations match GR predictions without these extra precession effects, which then allows for constraints on $\hat{\alpha}_{1}$ and $\hat{\alpha}_{2}$. Using data from pulsar-white dwarf binaries with known three-dimensional velocity, PSRs J1012+5307~\cite{Tauris:1994id} and J1738+0333~\cite{freire}, Shao and Wex~\cite{Shao:2012eg} placed the constraints $\hat{\alpha}_{2} < 1.8 \times 10^{-4}$ and $\hat{\alpha}_{1} \lesssim 10^{-5}$ at $95\%$ confidence. We see that the constraint on $\hat{\alpha}_{2}$ is not as strong as that from isolated pulsars. These constraints  depend on the magnitude and direction of the 3-velocity of the binary system in the preferred frame, which Shao and Wex ~\cite{Shao:2012eg} took to be its velocity relative to the CMB.

We will show later that the isolated pulsar constraint is not suitable to bound Einstein-\AE ther or khronometric theory. This is for one main reason. PSR B1937+21~\cite{Rutledge:2003xk} and PSR J1744--1134~\cite{1997ApJ...481..386B} are isolated, millisecond pulsars, and thus their masses have not been measured, precisely because they are isolated. As we will see later, $\hat{\alpha}_{1}$ and $\hat{\alpha}_{2}$ in Einstein-\AE ther and khronometric theory depend on the sensitivities, which cannot be determined without knowing the mass, even if the EoS were known. Without {\textit{a priori}} knowledge of the mass, one cannot map the isolated pulsar constraint on $\hat{\alpha}_{1}$ and $\hat{\alpha}_{2}$ to a constraint on $(c_{+},c_{-})$ or $(\lambda,\beta)$. Note that this will be the case for any Lorentz-violating theory of gravity.

\subsection{Constraining Lorentz-Violating Gravity with Binary Pulsars}
\label{alpha-hat-sec}

Let us first concentrate on observations of the orbital decay rate of PSR J1141-6545~\cite{bhat}, PSR J0348+0432~\cite{2.01NS} and PSR J0737-3039~\cite{kramer-double-pulsar}. The first two are binary systems composed of a NS and white dwarf, in a $0.17$ eccentricity, $4.74$-hour orbit and in a ${\cal O}(10^{-6})$ eccentricity, $2.46$-hour orbit respectively. Therefore, these systems have an orbital velocity of $v_{12}/c = {\cal{O}}(10^{-3})$. The last system is a double binary pulsar with $0.088$ eccentricity and $2.45$-hour orbit. In all cases, the eccentricity is small and will thus be neglected. 
Our constraints are robust to this assumption, because eccentricity corrections to $\delta \dot{P}_{\AE,kh}$ are of ${\cal{O}}(e^{2})$, which is negligible in Eq.~\eqref{eq-used} relative to the uncertainties due to the dependence on the individual masses, orbital period, and EoS. Of course, one would not be able to neglect eccentricity if considering certain binary pulsar systems, such as the very well-studied PSR 1913+16~\cite{Hulse:1974eb,Taylor:1982zz,Taylor:1989sw}, which has an eccentricity of roughly $0.6$.

All pulsars we consider have spin periods significantly larger than $1 \; {\rm{ms}}$. PSR J1141-6545~\cite{bhat} has a spin period of roughly $390 \; {\rm{ms}}$, PSR J0348+0432~\cite{2.01NS} a period of $39 \; {\rm{ms}}$, and PSR J0737-3039A and B~\cite{kramer-double-pulsar} a period of $22 \; {\rm{ms}}$ and $2700 \; {\rm{ms}}$ respectively. Let us convert these spin periods into a dimensionless spin parameter $\bar{a} \equiv J_{*}/(G_N M_{\obs}^{2})$, where $J_{*} = I_{*} \omega_{*}$ is the spin angular momentum, with $I_{*}$ the moment of inertia and $\omega_{*}$ the star's angular velocity. Approximating the moment of inertia with that of a solid sphere, $I_{*} \approx I_{\rm{sphere}} = (2/5) M_{\obs} R_*^{2}$, with radius $11 \; {\rm{km}}$, one finds that $\bar{a} \approx 1.4 \times 10^{-3}$ for PSR J1141-6545~\cite{bhat}, $\bar{a} \approx 8.8 \times 10^{-3}$ for PSR J0348+0432~\cite{2.01NS} and $\bar{a} \approx 2.25 \times 10^{-2}$ and $\bar{a} \approx 1.98 \times 10^{-4}$ for PSR J0737-3039A and B~\
cite{kramer-double-pulsar}. 

One can neglect the spin of the pulsar when modeling the orbital dynamics of binary systems to test GR for the following reason. In the small-spin or {\textit{slow-rotation approximation}}~\cite{Hartle:1967ha,Hartle:1969ht,Hartle:1970ha,Hartle:1973ha}, all observables can be expanded in a series in $\bar{a} \ll 1$. To zeroth-order in this approximation, one recovers the non-spinning results used in this paper. To first-order in this approximation, only observables that depend on the gravitomagnetic sector of the metric, i.e.~the ``cross'' time-space components, are modified. One of these observables is precisely the orbital decay rate, which acquires a spin-orbit coupling correction. Such a correction, however, appears at 1.5PN order), i.e. the leading-order spin correction to the orbital decay rate is actually of ${\cal{O}}(\bar{a} \; v_{12}^{3}/c^3)$ smaller than the leading-order terms, which makes this effect negligible for our purposes. Similarly, there will be spin corrections to observables associated 
with the conservative dynamics, i.e.~spin corrections to the Hamiltonian. As in the orbital decay rate, however, they are suppressed in the PN sense, and thus, can be neglected. Given this, one can model the pulsars in the binary systems that we consider as if they were non-spinning. This allows us to use the results of the previous sections, which do not account for NS spins.

With all of this at hand, consider then a binary system with non-spinning components in a circular orbit in either Einstein-\AE ther or khronometric theory. Because of the circularity condition, all terms proportional to $\dot{r}_{12}$ vanish in Eq.~\eqref{Pdot-AE} to leading PN order. 
After orbit-averaging, the orbital decay rate to leading PN order reduces to
\be
\frac{\dot{P}_{b}}{P_{b}} = - \frac{192 \pi}{5} \left(\frac{2 \pi G_{N} m}{P_{b}}\right)^{5/3} \left(\frac{\m}{m}\right) \frac{1}{P_{b}} \left<\mathcal{A}\right>\,,
\label{flux-AE}
\ee
where we recall one more time that $m_{A}$ are the active masses, $m = m_{1} + m_{2}$ is the total active mass, $P_{b}$ is the orbital period and we have defined
\begin{align}
\label{A-AE}
\left<\mathcal{A}\right> \equiv &
\frac{5(1-c_{14}/2)}{32} \left(s_1 - s_2\right)^2  \left( \frac{P_b}{2 \pi G_N m} \right)^{2/3} \mathcal{C} 
\nn \\
& \times \left[ 1 + {\cal{O}} \left( \frac{v^{2}}{c^{2}}, \frac{V_{CM}^{2}}{c^{2}},(s_{1}-s_{2})^{-1} \frac{V_{CM} v}{c^{2}} \right) \right]
\nn \\
&+  \left( 1-\frac{c_{14}}{2} \right)\left[\left(1 - s_{1}\right) \left(1 - s_{2}\right)\right]^{2/3}  
\nn \\
&\times
\left( \mathcal{A}_1 + \mathcal{S} \mathcal{A}_2 + \mathcal{S}^{2} \mathcal{A}_3  \right) \left[ 1 + {\cal{O}} \left( \frac{v^{2}}{c^{2}} \right) \right]\,.
\end{align}

There are many interesting features in Eqs.~\eqref{flux-AE} and~\eqref{A-AE} that deserve further discussion. 
First, notice that Eq.~\eqref{flux-AE} reduces to the GR prediction when $\mathcal{A}=1$, which is the case when the constants $c_{i} \to 0$ [recall that ${\cal{C}}$ also depends on the $c_{i}$ via Eq.~\eqref{C-func}]. 
Second, notice that the first term in Eq.~\eqref{A-AE} (the one proportional to ${\cal{C}}$) leads to dipolar radiation, which scales as $(v/c)^{-2}$ relative to the GR quadrupole term. 
Third, notice that in Eq.~\eqref{A-AE} we have included the order of uncontrolled remainders explicitly, which can be read off from Eq.~\eqref{Pdot-AE}. In particular, the dipole term is corrected by standard 1PN terms (proportional to $v^{2}$), as well as terms proportional to $V_{CM}^{2}$; the latter  can be safely  assumed to be small because the preferred frame is to be identified with that in which the CMB is isotropic. The dipole term also possesses uncontrolled remainders proportional to the $v V_{CM}$ product, which scale linearly with the difference in sensitivities, as opposed to quadratically as the leading-order dipole term thus. 
  
The rate of orbital decay is dominated by the term with the least powers of velocity, or equivalently the least powers of $G m/P_{b}$, provided the binary is widely separated. Indeed, this is the case for all observed binary pulsars, with $G m/P_{b} = {\cal{O}}(10^{-10})$ a typical value for a binary with a 1-hour orbital period. Clearly then, dipolar radiation dominates the orbital decay rate, unless $s_{1} - s_{2} \approx 0$. On the other hand, the sensitivities of a binary system will be similar to each other if their EoSs and their masses are similar. This is the case for the double pulsar PSR J0737-3039~\cite{kramer-double-pulsar}, for which the quadrupole term [the second term in Eq.~\eqref{A-AE}] becomes comparable to the dipole (the first term in this equation). Of course, the uncontrolled remainders  proportional to $V_{CM}$ that correct  the dipole term at 1PN order  are irrelevant here, as they are also multiplied by the difference in the sensitivities. As we will see, the inclusion of this system 
allows for better constraints on Lorentz-violating theories. 

Let us now concentrate on mapping the constraints on $\hat{\alpha}_{1}$ and $\hat{\alpha}_{2}$ to constraints on $(c_{+},c_{-})$ and $(\lambda,\beta)$ from observations of PSRs J1012+5307~\cite{Tauris:1994id} and J1738+0333~\cite{freire}. The mapping can be obtained by replacing $\alpha_{1} \mapsto \hat{\alpha}_{1}$, $\alpha_{2} \mapsto \hat{\alpha}_{2}$, $s_{1} \mapsto 0$ and $s_{2} \mapsto 0$ in Eq.~\eqref{B-AE}~\cite{damour_esposito_farese,Damour:1992ah}. By equating the latter to Eq.~\eqref{B-AE} without these replacements, we can read off the relation between the strong-field PPN coefficients and the standard PPN coefficients and sensitivities. Notice that this is done here for the first time, since previous work (e.g.~\cite{Bell:1995jz,Damour:1992ah,Shao:2012eg,Shao:2013wga}) did not calculate the sensitivities. This procedure is completely generic and does not depend on the parameters of the orbit. Doing so in Einstein-\AE ther theory, we find
\begin{align}
\label{SF-alpha1}
\hat{\alpha}_{1}^{\AE} &= \alpha_1^{\AE} + \frac{c_{-} (8+\alpha_1^{\AE}) \sigma_A^{\AE}}{2 c_1}\,,
\\ 
\label{SF-alpha2}
\hat{\alpha}_{2}^{\AE} &= \alpha_2^{\AE} + \frac{c_{-} (8 + \alpha_1^{\AE} ) \sigma_A^{\AE}}{4 c_{1}} - \frac{  (c_{14} -2)  (\alpha_1^{\AE} - 2 \alpha_2^{\AE}) \sigma_A^{\AE}}{2  (c_{14} - 2 c_{+})}\,, 
\end{align}
while in khronometric theory we find
\begin{align}
\label{SF-alpha1-HL}
\hat{\alpha}_{1}^{\kh} &= \alpha_1^{\kh} + (8 + \alpha_1^{\kh}) \sigma_A^{\kh}\,,
\\
\label{SF-alpha2-HL}
\hat{\alpha}_{2}^{\kh} &= \alpha_2^{\kh} + \frac{(\alpha_1^{\kh} -  2 \alpha_2^{\kh} + \alpha (4 + \alpha_2^{\kh} ) - (8 + \alpha_1^{\kh} ) \beta ) \sigma_A^{\kh} } {\alpha - 2 \beta}\,,
\end{align}
where we recall that the $c_{i}$ and $(\alpha, \beta)$ are coupling constants in Einstein-\AE ther and khronometric theory.  Observe that both of these quantities depend not only on the weak-field values of the PPN parameters $(\alpha_{1},\alpha_{2})$, but also on the sensitivities of the NSs, which depend on the masses; this is why we use observations of PSRs J1012+5307~\cite{Tauris:1994id} and J1738+0333~\cite{freire}, since we know the NS masses for these systems. Saturating the Solar System constraints on the weak-field parameters $(\alpha_{1},\alpha_{2})$, and using the strong-field constraints on $(\hat{\alpha}_{1},\hat{\alpha}_{2})$, one can place constraints on $(c_{-},c_{+})$ and $(\lambda,\beta)$. 

\subsubsection{Einstein-\AE ther Theory}

Let us now concentrate on Einstein-\AE ther theory. The predicted rate of orbital decay is given by Eq.~\eqref{A-AE} with $(\mathcal{C},\mathcal{A}_{1},\mathcal{A}_{2},\mathcal{A}_{3})$ functions of the coupling constants $(c_{+},c_{-})$, given in Eqs.~\eqref{A-funcs-1}-\eqref{A-funcs-3}, with Eqs.~\eqref{Z-def}, \eqref{w0-def}, ~\eqref{w1-def} and~\eqref{w2-def}. The predicted strong-field generalization of the PPN parameters $(\hat{\alpha}_{1}^{\AE},\hat{\alpha}_{2}^{\AE})$ are given by Eqs.~\eqref{SF-alpha1} and~\eqref{SF-alpha2} in terms of their weak-field versions in Eqs.~\eqref{alpha1} and~\eqref{alpha2}, as well as the sensitivities. Comparing the prediction of the orbital decay rate to observations of PSR J1141-6545~\cite{bhat}, PSR J0348+0432~\cite{2.01NS} and PSR J0737-3039~\cite{kramer-double-pulsar}, we can then place constraints on the $(c_{+},c_{-})$ coupling parameter space of this theory. Comparing the prediction of $(\hat{\alpha}_{1}^{\AE},\hat{\alpha}_{2}^{\AE})$ to the constraint on these 
quantities derived from observations of PSR J1738+0333~\cite{freire}, we can also place constraints on this parameter space. We do not consider constraints derived from the orbital decay rate of PSR J1738+0333~\cite{freire} and from the $(\hat{\alpha}_{1}^{\AE},\hat{\alpha}_{2}^{\AE})$ constraint of PSR J1012+5307~\cite{Tauris:1994id} because they are weaker than constraints derived with the systems described above.

\begin{figure*}[htb]
\begin{center}
\includegraphics[width=8.5cm,clip=true]{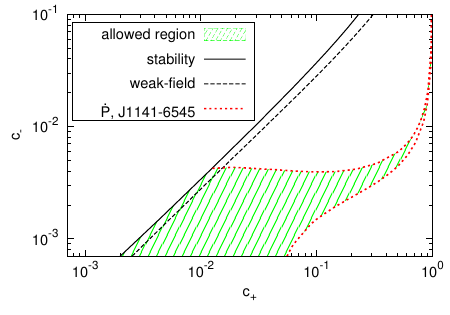}
\includegraphics[width=8.5cm,clip=true]{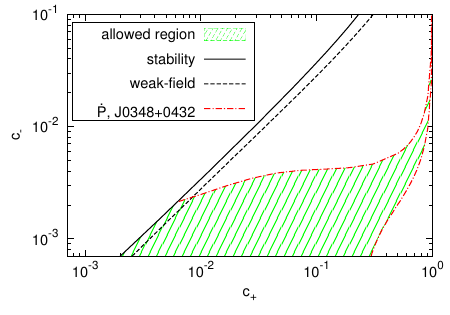}
\includegraphics[width=8.5cm,clip=true]{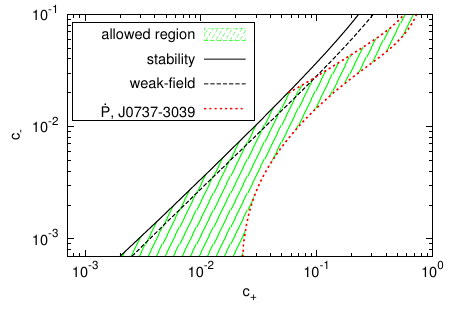}
\includegraphics[width=8.5cm,clip=true]{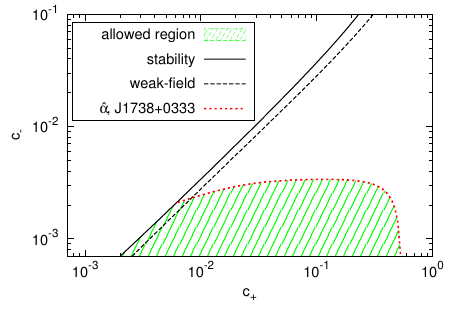}
\caption{\label{fig:Pdot-EA} (Color online) Constraints on Einstein-\AE ther theory from binary pulsar observations. The green-shaded regions correspond to those allowed after imposing constraints from the observation of PSR J1141-6545~\cite{bhat} (top, left panel), PSR J0348+0432~\cite{2.01NS} (top, right panel), PSR J0737-3039~\cite{kramer-double-pulsar} (bottom left panel) and PSR J1738+0333~\cite{freire} (bottom right panel). The first three panels use observations of the orbital decay rate through Eq.~\eqref{eq-used}, allowing for inferred error in the masses, observational error in the orbital period, the allowed range of the PPN parameters $(\alpha_{1}^{\AE},\alpha_{2}^{\AE})$ given Solar System constraints, and different sensitivities given different EoSs. The last panel uses constraints on the strong-field PPN parameters and Eqs.~\eqref{SF-alpha1} and~\eqref{SF-alpha2}. For comparison, we also plot the allowed $(c_{+},c_{-})$ region after imposing Cherenkov and stability constraints~\cite{Foster:2005dk} (below and to the right of the solid black line), as well as the values of $(c_{+},c_{-})$ required for the orbital decay rate to agree exactly with the GR prediction in the zero-sensitivity limit. Observe that the new, strong-field constraints are much more stringent than the Cherenkov/stability bounds.}
\end{center}
\end{figure*}

Figure~\ref{fig:Pdot-EA} shows these constraints. The area below and to the right of the solid black line is the allowed $(c_{+},c_{-})$ region after imposing Cherenkov and stability constraints~\cite{Foster:2005dk}. The green-shaded area with red dotted borders is the allowed $(c_{+},c_{-})$ region after imposing constraints from observations of PSR J1141-6545~\cite{bhat} (top, left panel), PSR J0348+0432~\cite{2.01NS} (top, right panel),  PSR J0737-3039~\cite{kramer-double-pulsar} (bottom left panel) and PSR J1738+0333~\cite{freire} (bottom right panel). As discussed earlier, Eq.~\eqref{eq-used} must be evaluated varying over $(c_{+},c_{-})$, but also allowing for the inferred error in the masses, the observational error in the orbital decay rate, the allowed range on $(\alpha_{1}^{\AE},\alpha_{2}^{\AE})$ given Solar System constraints, and various NS sensitivities associated with different EoSs. Similarly,  Eqs.~\eqref{SF-alpha1} and~\eqref{SF-alpha2} must be evaluated by trying different values of $\alpha_{1}^{\AE}$ and $\alpha_{2}^{\AE}$ within the Solar System constraints, as well as different values of the sensitivities for different EoSs. The result, thus, is not a line in $(c_{+},c_{-})$ space, but rather a two-dimensional region, as shown in the figures.

Observe that the allowed $(c_{+},c_{-})$ regions in Fig.~\ref{fig:Pdot-EA} associated with PSR J1141-6545~\cite{bhat} and PSR J0348+0432~\cite{2.01NS} are rather similar, while they are both quite different from that associated with PSR J0737-3039~\cite{2.01NS}. The first two systems are NS-white dwarf binaries, 
and since white dwarfs have much smaller binding energies than NSs, $s_{\NSS} - s_{\WD} \sim s_{\NSS} \neq 0$, which implies that the orbital decay rate is dominated by dipolar radiation, i.e.~by the first term in Eq.~\eqref{A-AE}. On the other hand, PSR J0737-3039~\cite{kramer-double-pulsar} is a double pulsar binary with similar masses and thus the difference between the sensitivities is small, making the dipolar and quadrupolar terms in the orbital decay rate comparable, i.e.~making the first and second terms in Eq.~\eqref{A-AE} of the same order. Observe also that constraints on $(c_{+},c_{-})$ derived from bounds on $(\hat{\alpha}_{1}^{\AE},\hat{\alpha}_{2}^{\AE})$ in the bottom right panel of Fig.~\ref{fig:Pdot-EA} also possess a characteristically different shape. This is because these constraints derive from comparing the effect of the conservative sector of the theory to observations, i.e.~the Hamiltonian, instead of the effect of the dissipative sector, i.e.~the radiation-reaction force.

For comparison, Fig.~\ref{fig:Pdot-EA} also shows the values of $(c_{+},c_{-})$ (dashed black curve) that would be required for the orbital decay rate to be identical to that of GR in the limit of zero sensitivity~\cite{Foster:2006az}.
Weakly-gravitating objects have small sensitivities, as the latter scales with the binding energy, which is why we label this curve ``weak-field''. This curve was obtained in Ref.~\cite{Foster:2006az} by requiring that $\alpha_{1}^{\AE}$ and $\alpha_{2}^{\AE}$ be identically zero, which results in vanishing sensitivities in the weak-field limit [c.f. Eq.~\eqref{eq:weak-field-AE-s}]. Observe that the strong-field binary pulsar constraints derived here overlap with this weak-field constraint only in a small $(c_{+},c_{-})$ region. Observe also that, although the slope of this weak-field curve is similar to the average slope of the allowed region with PSR J0737-3039~\cite{kramer-double-pulsar} (bottom left panel), it disagrees with the slope of the other allowed regions. This is because setting the sensitivities to zero automatically discards the dipolar term in the orbital decay rate, which in fact dominates over the quadrupolar one for mixed binary pulsar systems.

Reference~\cite{Foster:2006az} argues that the width of the allowed region given binary pulsar constraints should be approximately ${\cal{O}}(0.1\%)$, but this is not the case for the systems studied here. The main difference between our work and that of Ref.~\cite{Foster:2006az} is that the latter only considered constraints from PSR 1913+16~\cite{Hulse:1974eb,Taylor:1982zz,Taylor:1989sw}, which consists of two NSs, in the weak-field/vanishing sensitivity limit, while we here considered other more relativistic double pulsar and mixed pulsar binaries in the strong-field/non-vanishing sensitivity limit. Moreover, the width of the allowed region is not controlled just by the observational error in the orbital decay rate, but also by the inferred error in the individual masses, the observational error in the orbital decay rate, the allowed range of $(\alpha_{1}^{\AE},\alpha_{2}^{\AE})$ given Solar System constraints, and also the different NS sensitivities that depend on the EoS, all of which depend strongly on 
the system that is observed. As we discussed earlier, one must vary over all such parameters to obtain the correct allowed region, which as shown in Fig.~\ref{fig:Pdot-EA} is between ${\cal{O}}(0.1\%)$ and ${\cal{O}}(1 \%)$. 

Reference~\cite{Foster:2007gr} argues that current tests will be satisfied if the weak field conditions are imposed and the coupling constants satisfy $(c_{+},c_{-}) \leq {\cal{O}}(10^{-2})$. For any given system, we find that the allowed region is not a box with sides of ${\cal{O}}(10^{-2})$ but rather a band with width ${\cal{O}}(10^{-3})$. However, when we combine all of these allowed regions in the left panel of Fig.~\ref{fig:Pdot-constraint-summary}, their intersection is indeed almost a box with width of ${\cal{O}}(10^{-3})$ in the $c_{-}$ direction and ${\cal{O}}(10^{-2})$ in the $c_{+}$ direction. The change in shape of the combined allowed region is because different observations lead to allowed bands with different average slopes, which in turn is because for some systems dipolar radiation dominates, while for others it does not. The width of the box is determined by the smallest width of the bands, which is of ${\cal{O}}(10^{-3})$. All together, these constraints constitute the first, accurate 
strong-field test of Einstein-\AE ther theory.

\subsubsection{Khronometric Theory}

Let us now concentrate on khronometric theory. The predicted rate of orbital decay is given by Eq.~\eqref{A-AE} with $(\mathcal{C}, \mathcal{A}_{1},\mathcal{A}_{2},\mathcal{A}_{3})$ functions of the coupling constants $(\lambda,\beta)$, given in Eqs.~\eqref{A1-def-HL}-\eqref{A3-def-HL}. The predicted strong-field generalization of the PPN parameters $(\hat{\alpha}_{1}^{\kh},\hat{\alpha}_{2}^{\kh})$ is given by Eqs.~\eqref{SF-alpha1-HL} and~\eqref{SF-alpha2-HL} in terms of their weak-field versions in Eqs.~\eqref{alpha1-HL} and~\eqref{alpha2-HL}, as well as in terms of the sensitivities. Comparing the prediction of the orbital decay rate to observations of PSR J1141-6545~\cite{bhat}, PSR J0348+0432~\cite{2.01NS} and PSR J0737-3039~\cite{kramer-double-pulsar}, we can then place constraints on the $(\lambda,\beta)$ coupling parameter space of this theory. Comparing the prediction of $(\hat{\alpha}_{1}^{\kh},\hat{\alpha}_{2}^{\kh})$ to the constraint on these quantities derived from observations of PSR 
J1738+0333~\cite{freire}, we also place constraints on this parameter space.

\begin{figure*}[htb]
\begin{center}
\includegraphics[width=8.5cm,clip=true]{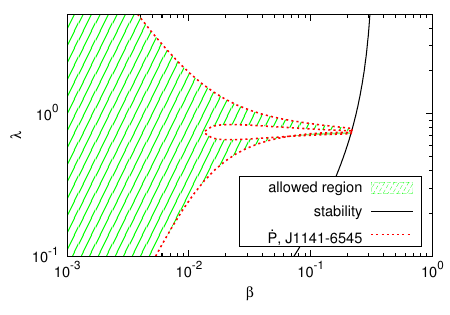}
\includegraphics[width=8.5cm,clip=true]{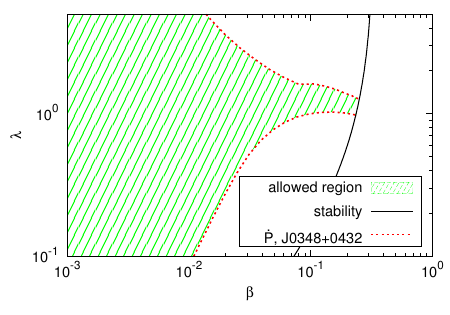}
\includegraphics[width=8.5cm,clip=true]{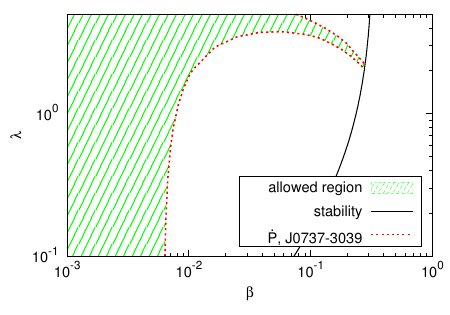}
\includegraphics[width=8.5cm,clip=true]{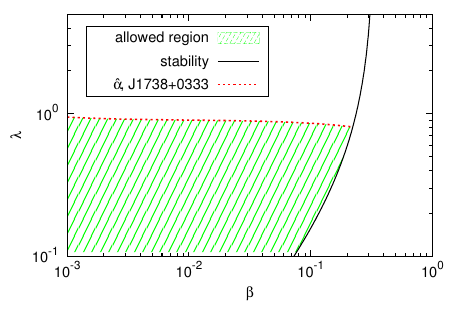}
\caption{\label{fig:Pdot-HL} (Color online) Constraints on khronometric theory from binary pulsar observations. The green-shaded regions correspond to those allowed after imposing constraints from the observation of PSR J1141-6545~\cite{bhat} (top, left panel), PSR J0348+0432~\cite{2.01NS} (top, right panel), PSR J0737-3039~\cite{kramer-double-pulsar} (bottom left panel) and PSR J1738+0333~\cite{freire} (bottom right panel). The first three panels use observations of the orbital decay rate through Eq.~\eqref{eq-used}, allowing for inferred error in the masses, observational error in the orbital period, the allowed range of the PPN parameters $(\alpha_{1}^{\kh},\alpha_{2}^{\kh})$ given Solar System constraints, and different sensitivities given different EoSs. The last panel uses constraints on the strong-field PPN parameters and Eqs.~\eqref{SF-alpha1} and~\eqref{SF-alpha2}. For comparison, we also plot the allowed $(\lambda,\beta)$ region after imposing Cherenkov and stability constraints (above and to the 
left of the solid black line). Observe how stringent the new, strong-field constraints are.}
\end{center}
\end{figure*}

Figure~\ref{fig:Pdot-HL} shows these constraints. The area above and to the left of the solid black line is the allowed $(\lambda,\beta)$ region after imposing Cherenkov and stability constraints. The green-shaded area with red dotted borders is the allowed $(\lambda,\beta)$ region after imposing constraints from observations of PSR J1141-6545~\cite{bhat} (top, left panel), PSR J0348+0432~\cite{2.01NS} (top, right panel),  PSR J0737-3039~\cite{kramer-double-pulsar} (bottom left panel) and PSR J1738+0333~\cite{freire} (bottom right panel). As discussed earlier, Eq.~\eqref{eq-used} must be evaluated varying over $(\lambda,\beta)$, but also allowing for the inferred error in the masses, the observational error in the orbital decay rate, the allowed range on $(\alpha_{1}^{\kh},\alpha_{2}^{\kh})$ given Solar System constraints, and various NS sensitivities associated with different EoSs. Similarly,  Eqs.~\eqref{SF-alpha1-HL} and~\eqref{SF-alpha2-HL} must be evaluated by choosing different values of $\alpha_{1}^{\
kh}$ and $\alpha_{2}^{\kh}$ within the Solar System constraints, as well as different values of the sensitivities for different EoSs. As in the Einstein-\AE ther case, the result is not a line in the $(\lambda,\beta)$ plane, but rather a two-dimensional region.

Observe that, as in the Einstein-\AE ther case, the allowed $(\lambda,\beta)$ regions associated with PSR J1141-6545~\cite{bhat} and PSR J0348+0432~\cite{2.01NS} are similar to each other, but different from that associated with PSR J0737-3039~\cite{2.01NS}. As before, this is because the first two systems are mixed binaries (NS-white dwarf), and thus, $s_{\NSS} - s_{\WD} \sim s_{\NSS} \neq 0$, since $s_{\WD} \ll s_{\NSS}$. In turn, this implies that the orbital decay rate is dominated by dipolar radiation, i.e.~by the first term in Eq.~\eqref{A-AE}. PSR J0737-3039~\cite{kramer-double-pulsar} is a double pulsar binary, i.e.~two NSs with similar masses, and thus the difference in sensitivities is small. This makes dipolar and quadrupolar terms in the orbital decay rate comparable, and thus, changes the shape of the allowed coupling region. The allowed region in the bottom right panel of this figure is also different from the other three, because the former derives from constraints on $(\hat{\alpha}_{1}^{\kh},\
hat{\alpha}_{2}^{\kh})$, and thus from the conservative sector of the theory, instead of the dissipative one. Observe also that the  first three binary systems leave $\lambda$ unconstrained in the $\beta\to 0$ limit. This is to be expected since in this limit the constant dominating the dipole radiation, $\cal C$, vanishes when $\alpha^{kh}_1= 0 = \alpha^{kh}_2$. Furthermore,  the constants in the quadrupole contribution 
also approach their GR value, ${\cal A}_1\to 1$, ${\cal A}_2\to 0$, ${\cal A}_3\to 0$ in this limit, and the only effect comes from the sensitivities which are too small to produce a strong constraint. When we combine all the panels of Fig.~\ref{fig:Pdot-HL}, one obtains the right panel of Fig.~\ref{fig:Pdot-constraint-summary}. Observe that binary pulsar observations further constrain the parameter region previously allowed by stability and BBN considerations.

\section{Conclusions and Future Directions}
\label{sec:Conclusions}

We have investigated Lorentz-violating gravity (focusing in particular on violations of boost invariance) in the light of binary pulsar observations. 
We focused on Einstein-\AE ther theory and khronometric theory,
which are generic theories that break boost invariance at low energies,
and showed that they predict emission of dipolar radiation from binary systems, as first reported in Ref.~\cite{Foster:2006az,Foster:2007gr,Blas:2011zd}. This 
greatly modifies the orbital evolution of these systems, and in particular the decay rate of their orbital period, as compared to the GR prediction. Furthermore, the emission of quadrupolar radiation is also modified.
All these modifications, however, depend on the NS sensitivities, which measure how the binding energy of 
a NS responds to its motion relative to the additional fields of the theory (the \AE ther and khronon).
Thus, constraints on Lorentz-violating gravity can only be placed once the 
sensitivities have been computed.  

Here, we showed in two independent ways that the sensitivities can be extracted, without loss
of generality, from solutions describing slowly-moving NSs to first order in velocity, and we then found these solutions numerically for the first time.
We began with a generic ansatz
for the metric and for the \AE ther and khronon fields. We then expanded
the modified field equations in tensor spherical harmonics,
and succeeded at reducing the field equations to a system of ordinary differential equations, which
we solved numerically.
These slowly-moving solutions then allowed us to extract the NS sensitivities,
for which we also present fits in terms of the coupling constants and NS compactness. 
We checked that the sensitivities approach the weak-field expression derived by Foster~\cite{Foster:2007gr} as one decreases the NS compactness, and showed that the two differ rather significantly in the large compactness regime.

With these solutions at hand, we then computed the predictions of the
orbital decay rate in these theories, compared them to
observations of PSR J1141-6545~\cite{bhat}, PSR J0348+0432~\cite{2.01NS}, 
and PSR J0737-3039~\cite{kramer-double-pulsar}, and obtained very
stringent bounds on the coupling constants of Einstein-\AE ther and khronometric theory. We also
complemented these constraints by computing the bounds on the coupling parameters
induced by constraints on the precession of  PSR J1738+0333~\cite{freire}.
All together, the constraints derived here on the full set of parameters that control gravitational Lorentz violation are much stronger than previously-obtained bounds.

Future work could extend the results obtained here in various
directions. One possibility would be to verify the results of this
paper by carrying a Bayesian hypothesis-testing or model-selection
analysis~\cite{Cornish:2007if,vigeland-vallisneri}. 
Such an analysis would take into account covariances between the system 
parameters and the coupling constants of the theories, which were here neglected. 
However, since the modifications to the orbital decay
rate enter at lower PN order than the leading-order GR prediction, we expect such covariances to have a small effect.

Another possible avenue for future work is to study other binary
pulsar observables, such as rate of advance of perihelion~\cite{stairs}.
In order to compute this observable, one would have to first compute the
acceleration of each component of the binary system, which depends not
only on the sensitivities, but also on their first derivative~\cite{Foster:2007gr}. To calculate the latter, one would have to obtain
moving NS solutions to higher order in the slow-velocity expansion. The
framework we developed here should suffice to obtain such
solutions. For example, the modified field equations should still
separate if one carries out a tensor spherical harmonic decomposition.
(This can be seen e.g.~by extending the procedure of Appendix \ref{app:separability} to next order in the velocity.)

Yet another possibility would be to investigate the constraints that
future GW observations could place on Einstein-\AE
ther and khronometric theory, given a future
detection. GW observations should be particularly
sensitive to the frequency-evolution of the GW phase,
which will be modified in these theories due to dipolar emission. 
Moreover, additional polarizations would be
excited in binary NS inspirals, beyond the two transverse-traceless
ones of GR, which could lead to even stronger constraints if a signal
is detected by multiple interferometers~\cite{Chatziioannou:2012rf}. 

Finally, it would be interesting to compute the sensitivities in Lorentz-violating
gravity for isolated black holes~\cite{Eling:2006ec,Blas:2011ni,Barausse:2011pu,Barausse:2013nwa,Barausse:2012ny,Barausse:2012qh,Barausse:2013nwa}. Once this is accomplished, one could perform an analysis similar to the one carried out here to place constraints on Lorentz violation in gravity from the orbital decay rates of low-mass X-ray binaries~\cite{johannsen1,gonzalez,Hernandez:2013haa}. Indeed, the orbital decay rates of such systems have already been used to place constraints on certain modified gravity theories~\cite{zaglauer,psaltisBD,johannsen1,johannsen2,kent-LMXB}. Moreover, black hole binaries are expected to be a primary
target for future GW detectors. Constraints on the coupling constants 
of Einstein-\AE ther and khronometric theory derived from such observations would 
require knowledge of the sensitivity parameters. The latter could be obtained
by following the analysis carried out here, but generalized to black hole solutions.

\acknowledgments
 We would like to thank Quentin Bailey, Brendan Foster, Samuel Gralla, Ted Jacobson, 
Takahiro Tanaka, Lijing Shao, and Leo Stein for giving us valuable comments. NY
acknowledges support from NSF grant PHY-1114374 and the NSF CAREER
Award PHY-1250636, as well as support provided by the National
Aeronautics and Space Administration from grant NNX11AI49G, under
sub-award 00001944. NY and KY would like to thank, respectively, the Institute
d'Astrophysique de Paris and the Yukawa Institute for Theoretical
Physics for their hospitality, while some of this work was being
carried out. EB acknowledges support from the European Union's Seventh
Framework Programme (FP7/PEOPLE-2011-CIG) through the Marie Curie
Career Integration Grant GALFORMBHS PCIG11-GA-2012-321608.  Some calculations used the
computer algebra-systems MAPLE, in combination with the GRTENSORII
package~\cite{grtensor}. 

\appendix

\section{Separability of Field Equations}
\label{app:separability}

In Secs. \ref{sec:NS-Sols} and \ref{sec:NS-SolsHL}, we have shown explicitly that the field
equations of Einstein-\AE ther and khronometric theory for an isolated non-spinning star moving slowly relative to the \AE ther reduce to a system
of ordinary differential equations (ODEs), if the ``perturbation
potentials'' [i.e.~the functions of $r$ and $\theta$ that describe, at ${\cal
  O}(v)$, the effect of the motion relative to the \AE ther on the system] 
are decomposed in Legendre polynomials.

This is non-trivial and does \textit{not} simply follow from the
fact that the Legendre polynomials are a basis  for the space of the
regular functions of the polar angle $\theta$. In principle, one may
indeed imagine that the equations for the various multipole moments
may not decouple, leaving one with a hierarchy of infinite coupled
ODEs.  In fact, the reason why such a situation does not happen and
the equations for the various multipole moments decouple lies in the
symmetries of the ``unperturbed'' [i.e.~${\cal O}(v^0)$] solution,
which we assumed to be static and spherically symmetric.  In GR, 
it has long been known that whenever the background has
those symmetries, the equations that govern (linear) perturbations 
decouple and therefore reduce to a (finite) system of
ODEs~\cite{Regge:1957rw,Zerilli:1970fj}.  As far as we are aware, a
general proof of this fact is not yet available in Einstein-\AE ther or khronometric theory, 
but we will present another route to this result in the
particular case considered here.  Relative to the
``brute-force'' decomposition of Secs. \ref{sec:NS-Sols} and \ref{sec:NS-SolsHL}, this approach has
the advantage  of showing more explicitly the role of the symmetries
of the background.

We begin by noting that the spacetime outside a spherically symmetric
static star can be described, in cylindrical isotropic coordinates, by
the following metric
\be ds^2 = f(r) dt^2 - h(r) (d\rho^2 + \rho^2 d\theta^2+dz^2)\,,  \ee
with $r=\sqrt{\rho^2+z^2}$. Note that the use of isotropic coordinates
(rather than the areal coordinates used elsewhere in this paper) has
the advantage that the spatial part of the metric is conformally
flat. The \AE ther field is instead simply
\be U_\mu = \delta_\mu^t \sqrt{f(r)} \,.  \ee
Consider now a star slowly moving in the $z$-direction relative to the
\AE ther. Based on the transformation properties of the metric and
\AE ther under a coordinate change $z\mapsto-z$ and $t\mapsto-t$,
it is clear that only
the $g_{tz}$ and $g_{t\rho}$ components of the metric and the $U_t$,
$U_z$ and $U_{\rho}$ components of the \AE ther can be affected. The
normalization condition for the \AE ther immediately requires that the
perturbation to $U_t$ be zero, because $\delta (U^\mu U_\mu) = 2
g^{\mu\nu} U_\mu \delta U_\nu= 2 \delta U_t/\sqrt{f(r)}=0$. One is
then left with the perturbations $\delta g_{tz}$, $\delta g_{t\rho}$,
$\delta U_z$ and $\delta U_{\rho}$.

We only have two vectors at our disposal to construct the
perturbations to these fields, namely $\boldsymbol{n}=(\rho,z)/r$
(clearly, $|\boldsymbol{n}|=1$) and $\boldsymbol{v}=(0,v)$, so
 we can write 
\begin{gather}
\label{ansatz1}
\binom {\delta g_{t\rho}} {\delta g_{tz}} = S(r) \boldsymbol{v}+ V(r)
(\boldsymbol{v}\cdot \boldsymbol{n}) \boldsymbol{n}\,,\\ \binom {\delta
  U_{\rho}} {\delta U_{z}} = \sqrt{f(r)} \left[Q(r) \boldsymbol{v}+
  W(r) (\boldsymbol{v}\cdot \boldsymbol{n})
  \boldsymbol{n}\right]\,.\label{ansatz2}
\end{gather}
Clearly, the functions $S,V,Q,W$ can only depend on $r$, because in
order to introduce a dependence on $\rho$ and $z$ singularly one would
have to use the vector $\boldsymbol{v}$ one more time, i.e.~such a
dependence only appears at ${\cal O} (v^2)$.

The most generic ansatz describing the system at linear order in the
velocity is therefore
\begin{multline}
ds^2 = f(r) dt^2 - h(r) (d\rho^2 + \rho^2 d\theta^2+dz^2)\\+ 2 v
\left[S(r)+V(r) \frac{z^2}{r^2}\right]dz dt+ 2v  V(r)\frac{z\rho}{r^2}
d\rho dt +{\cal O}(v)^2\,, 
\end{multline}
while that for the \AE ther is 
\begin{gather}
\label{U-ansatz}
U_\mu = \sqrt{f(r)} \left\{ \delta_\mu^t +
v\left[\tilde{Q}(\rho,z)\delta_\mu^z+ \tilde{W}(\rho,z)
  \delta_\mu^\rho\right]\right\} +{\cal
  O}(v)^2\,,\\ \tilde{Q}(\rho,z)=Q(r)+W(r)\frac{z^2}{r^2},\quad \tilde{W}(\rho,z)=
W(r) \frac{z\rho}{r^2}.
\end{gather}
Because this decomposition is completely general and simply based on
the symmetries of the problem, and because the metric and \AE ther now
only depend on potentials $S,V,Q,W$ that are functions of $r$ only,
the field equations \textit{must} reduce to a system of differential
equations for $S,V,Q,W$.  (Clearly, if they did not, it would mean
that the above general decomposition is inconsistent with the field
equations, which would in turn mean that slowly moving stars do not
exist in the theory under consideration).

In fact, inserting this ansatz into the field equations for
 Einstein-\AE ther theory, one finds that the non-trivial equations  follow from the
$r$ and $\theta$ components of the \AE ther equation  and from the $(t,r)$
and $(t,\theta)$ components of the Einstein equations,  and have the
following structure
\begin{gather}
e_1\equiv\sum_{i=1}^{12}A_i(r)
v_i=0\label{e1}\,,\\ e_2\equiv\sum_{i=1}^{10}B_i(r)
\tilde{v}_i=0\label{e2}\,,\\ e_3\equiv\sum_{i=1}^{6}C_i(r)
w_i=0\label{e3}\,,\\ e_4\equiv\sum_{i=1}^{10}D_i(r)
\tilde{v}_i=0\label{e4}\,,
\end{gather}
where 
\begin{align}
&\boldsymbol{v}=(S(r),S'(r),S''(r),V(r),V'(r),V''(r),\nonumber\\&
  \qquad
  Q(r),Q'(r),Q''(r),W(r),W'(r),W''(r));\\
   &\boldsymbol{\tilde{v}}=(S(r),S'(r),S''(r),V(r),V'(r),\nonumber\\ &
  \qquad Q(r),Q'(r),Q''(r),W(r),W'(r));\\
   &\boldsymbol{w}=(S(r),S'(r),
  V(r), Q(r),Q'(r),W(r));
\end{align}
and where the functions $A_i(r),B_i(r),C_i(r), D_i(r)$ depend on the
coupling constants as well as on the ${\cal O} (v^0)$ potentials
$f(r)$ and $h(r)$. Their explicit form is not particularly
illuminating and we will not write it down here, but for our purposes
it is sufficient to mention that they are functions of $r$ only, so
the system of Eqs.~\eqref{e1}--\eqref{e4} does indeed become a system of
differential equations in the radial coordinate, as expected.

To show that these differential equations are indeed
\textit{ordinary}, i.e.~that they can be put in the form
$\mbox{d}\boldsymbol{y}/\mbox{d}r=\boldsymbol{F}(\boldsymbol{y})$,
where $\boldsymbol{y}$ is a suitable array of variables, a more
detailed understanding of the gauge degrees of freedom is required.
As already shown in Sec.~\ref{subsec:ansatz-AE}, it is clear that one can set one of the potentials
$S,V,Q,W$ exactly to zero in Einstein-\AE ther theory with a gauge choice. [This can
be checked explicitly by looking at the transformation properties of
the metric and \AE ther under a change of coordinates $t'=t+vH(r)  z$.]
Therefore, one of the four equations in Eqs.~\eqref{e1}--\eqref{e4} 
must be algebraically related to the other three, i.e.~one
should have just three independent equations for the three physical
degrees of freedom.  In fact, it is possible to show that the  radial
derivative $d e_3/dr$ can be expressed as a linear combination of
$e_1$, $e_2$, $e_3$ and $e_4$, once the ${\cal O}(v^0)$ equations are
satisfied. This fact can be checked explicitly, but follows elegantly
from the existence  of a generalized Bianchi identity in
 Einstein-\AE ther theory~\cite{Jacobson:2011cc} (see also
Refs.~\cite{Barausse:2011pu,Barausse:2013nwa}), which  relates the
Einstein equations to the \AE ther equations [Eq.~\eqref{AE_def}] and is given by
\be
\label{identity}
 \nabla_\mu \left(
T_{\AE}^{\mu\nu}-G^{\mu\nu} +
U^\mu\AE^\nu\right)= -\AE_\mu\nabla^\nu U^\mu. 
\ee 
 As a result, one
can set e.g.~$W=0$ with a gauge transformation, and rewrite the system
$e_1=e_2=e_4=0$ as an ODE system
\begin{gather}
S''(r)=S_2(S,S',Q,Q',V,V')\,,\label{ev1}\\ Q''(r)=Q_2(S,S',Q,Q',V,V')\,,\label{ev2}\\ V''(r)=V_2(S,S',Q,Q',V,V')\,,\label{ev3}
\end{gather}
which can be evolved in the radial coordinate (starting e.g.~from an
initial radius $r_\epsilon$ outwards). The additional equation $e_3=0$ is an
initial value constraint for this evolution system, i.e.~$e_3=0$ only
involves initial data for the system \eqref{ev1}-\eqref{ev3} [as can
  be seen from the derivative structure of Eq.~\eqref{e3}], and by
virtue of the generalized Bianchi identity \eqref{identity}, it is
satisfied everywhere  if imposed at the initial radius $r_\epsilon$
(c.f. Refs.~\cite{Barausse:2011pu,Barausse:2013nwa} for more details).

Similarly, in the case of khronometric theory, the only non-trivial
equations are the  $(t,r)$ and $(t,\theta)$ components of the Einstein
equations (the equation following from the variation of the khronon
field is actually implied by the Einstein equations, and does not need
to be imposed explicitly~\cite{Jacobson:2010mx}). These equations take
the form
\begin{gather}
e_1\equiv\label{etq}\sum_{i=1}^{12}F_i(r)
v_i=0\,,\\ e_2\equiv\label{erq}\sum_{i=1}^{10}H_i(r) \tilde{v}_i=0\,,
\end{gather}
where $F_i$ and $H_i$ are again functions of $r$ only [depending on
  the couplings and on the background potentials $f(r)$ and
  $h(r)$]. Also, the requirement that the \AE ther be hypersurface
orthogonal simply imposes a relation   $W(r)=rQ'(r)$ between the
\AE ther potentials at ${\cal O}(v)$. Therefore, the field equations
reduce to a system of differential equations  in the radial isotropic
coordinate $r$, exactly as expected.  More precisely, as we have shown
in Sec.~\ref{sec:NS-SolsHL}, a gauge transformation allows to set both
\AE ther perturbations $Q$ and $W$ to zero. (This can be checked explicitly
by considering a gauge transformation $t'=t+v H(r) z$.) By doing so,
the system takes the form
\begin{gather}
S''(r)=S_2(S,S',V,V')\,,\\ V''(r)=V_2(S,S',V,V')\,,
\end{gather}
which is indeed a set of ODEs.

\section{Mapping between the PN Metric and the Hartle-Thorne Metric}
\label{app:mapping}

Let us here spell out the mapping between the PN metric of Eqs.~\eqref{g00PN}--\eqref{gijPN},
and the asymptotic solution given by Eqs.~\eqref{asymp-v} and \eqref{asymp-s}
for Einstein-\AE ther theory, or by Eqs.~\eqref{asymp-v-HL} and \eqref{asymp-s-HL} for khronometric theory.

Without loss of generality, one can take the $z'$-direction to be the
direction of motion.
Let us then transform the metric of Eqs.~\eqref{g00PN}--\eqref{gijPN},
which is written in the standard PN gauge, to comoving PN coordinates
$(t'',x'',y'',z'')$ via the Lorenz boost
\ba t' &=& t''+vz''+\frac{M_* x''^i v^i}{2r''}\,, \\ x' &=& x''\,,
\\ y' &=& y''\,, \\ z' &=& z''+vt''\,. \ea  
With this coordinate transformation, the line element becomes 
 \ba ds^2 &=& \left( 1-\frac{2M_*}{r''} \right) dt''^2 - \left(
1+\frac{2M_*}{r''} \right) \delta_{ij}d(x'')^i d(x'')^j \nn \\ & &   - v
(2B^{+}+1) \frac{M_* z''}{r''^3} (x''dt''dx'' + y'' dt''dy'') \nn\\&&\hspace{-.2cm}- v \frac{M_*}{r''} \left[ (2B^{-}+7) +
  (2B^{+}+1) \frac{z''^2}{r''^2} \right]dt''dz''\,, 
\ea
where $r''$ is the distance from the origin to the field point in the
comoving standard PN gauge. Note that in the GR limit, $B^{+} \to
-1/2$ and $B^{-} \to -7/2$, and thus, $g_{t''i}$ vanishes. Next, let us transform this metric
to comoving spherical coordinates
$(t,r,\theta,\varphi)$ by
\ba t'' &=& t\,, \\ x'' &=& (r-M_*) \sin\theta \cos\varphi\,, \\ y''
&=& (r-M_*) \sin\theta \sin\varphi\,, \\ z'' &=& (r-M_*) \cos\theta\,,
\ea
where the subtraction by $M_{*}$ is needed to remove the harmonic
factors. This transformation gives 
\ba
\label{metric-Foster-sph0}
ds^2 &=& \left( 1-\frac{2M_*}{r} \right) dt^2 - \left( 1 +
\frac{2M_*}{r} \right) dr^2\nn\\ && - r^2 d\theta^2 - r^2 \sin^2 \theta
d\varphi^2 \nn \\ & &   -2 v (B^{-} + B^{+}+4) \frac{M_*}{r} \cos \theta
dt dr \nn\\ && + v (7+2B^{-}) M_*  \sin\theta dt d\theta\,. 
\ea
Now, let us perform the coordinate transformation given by
Eq.~\eqref{t-transform}, where we choose $H(r,\theta)$ to be 
\begin{align} 
H(r,\theta) &= - \left( 1 -  \frac{1+2 B^{-}+2 C^{-}}{2}
\frac{M_\NS}{r} \right) r \cos\theta 
\nn \\ 
&+ \mathcal{O} \left(\frac{M_\NS^2}{r^2} \right)\,, 
\end{align}
which yields
\ba
\label{metric-Foster-sph}
ds^2 &=& \left( 1-\frac{2M_*}{r} \right) dt^2 - \left( 1 +
\frac{2M_*}{r} \right) dr^2 \nn\\&&- r^2 d\theta^2 - r^2 \sin^2 \theta
d\varphi^2 \nn \\ & &   -2 v \left( 1+ (B^{-} + B^{+}+2) \frac{M_*}{r}
\right) \cos \theta dt dr \nn\\&&+ 2v r \left( 1- (C^{-}-1) \frac{M_*}{r}
\right) \sin\theta dt d\theta\,,
\ea
to the appropriate order. 

Comparing Eq.~\eqref{metric-Foster-sph} to the metric ansatz in
Eq.~\eqref{metric-ansatz}, one can read off the asymptotic behavior of
$k_1(r)$ and $s_1(r)$:
\ba
\label{asymp-v2}
k_1(r) &=& - \left( 1+ (B^{-} + B^{+}+2) \frac{M_*}{r} \right) +
\mathcal{O}\left( \frac{1}{r^2} \right)\,, \nn \\ \\
\label{asymp-s2}
s_1(r) &=& -\left( 1- (C^{-}-1) \frac{M_*}{r} \right) +
\mathcal{O}\left( \frac{1}{r^2} \right)\,.
\ea
Finally, comparing these asymptotic relations to the asymptotic
expansion of $k_1(r)$ and $s_1(r)$ in Eqs.~\eqref{asymp-v}
and~\eqref{asymp-s} for Einstein-\AE ther theory, or in Eqs.~\eqref{asymp-v-HL} and \eqref{asymp-s-HL} for khronometric theory, 
we can read off the integration constant $A$, given by Eq.~\eqref{Eq:A-text}.

\section{High-Order Fit of the Neutron Star Sensitivity in Einstein-\AE ther Theory }
\label{app:fit}

The function Eq.~\eqref{s-fit} used to fit the numerical results for the
NS sensitivities in Einstein-\AE ther theory contained only $27$ terms, which thus led to an $r^{2}$
value of approximately $0.96$. In this Appendix, we repeat the fit but with a function 
that contains $120$ terms, thus yielding an $r^{2}$ value closer to $0.999$. The function that
we fit is
\be 
\label{s-fit2}
s_{\AE} = \sum_{\ell=0}^{5} \sum_{m=0}^{3} \sum_{n=0}^{4} c_{\ell, m, n} c_{+}^{\ell} c_{-}^{m} C_{*}^{n}\,.
\ee
\\
We experimented with different fitting functions (e.g.~polynomials of different orders in $c_{+}$, $c_{-}$ and $C_{*}$) and found that Eq.~\eqref{s-fit2} leads 
to the highest $r^{2}$ value with less than $130$ coefficients. After the fit, the latter are
\bw
\[
\begin{array}{cccc}
 
\begin{array}{c}
 c_{0,0,0}=7.5614\times 10^{-5} \\
 c_{0,0,1}=-1.9843\times 10^{-3} \\
 c_{0,0,2}=1.7522\times 10^{-2} \\
 c_{0,0,3}=-7.1593\times 10^{-2} \\
 c_{0,0,4}=1.0796\times 10^{-1} \\
\end{array}
 & 
\begin{array}{c}
 c_{0,1,0}=-4.1773\times 10^{-1} \\
 c_{0,1,1}=1.0465\times 10^1 \\
 c_{0,1,2}=-9.5518\times 10^1 \\
 c_{0,1,3}=3.8615\times 10^2 \\
 c_{0,1,4}=-5.7411\times 10^2 \\
\end{array}
 & 
\begin{array}{c}
 c_{0,2,0}=2.4191\times 10^2 \\
 c_{0,2,1}=-6.0529\times 10^3 \\
 c_{0,2,2}=5.5053\times 10^4 \\
 c_{0,2,3}=-2.2216\times 10^5 \\
 c_{0,2,4}=3.2636\times 10^5 \\
\end{array}
 & 
\begin{array}{c}
 c_{0,3,0}=-3.7821\times 10^4 \\
 c_{0,3,1}=9.4537\times 10^5 \\
 c_{0,3,2}=-8.569\times 10^6 \\
 c_{0,3,3}=3.4574\times 10^7 \\
 c_{0,3,4}=-5.0288\times 10^7 \\
\end{array}
 \\
 
\begin{array}{c}
 c_{1,0,0}=-1.4808\times 10^{-1} \\
 c_{1,0,1}=3.7377 \\
 c_{1,0,2}=-3.4865\times 10^1 \\
 c_{1,0,3}=1.4124\times 10^2 \\
 c_{1,0,4}=-2.2149\times 10^2 \\
\end{array}
 & 
\begin{array}{c}
 c_{1,1,0}=1.2786\times 10^3 \\
 c_{1,1,1}=-3.2087\times 10^4 \\
 c_{1,1,2}=2.9613\times 10^5 \\
 c_{1,1,3}=-1.1938\times 10^6 \\
 c_{1,1,4}=1.8188\times 10^6 \\
\end{array}
 & 
\begin{array}{c}
 c_{1,2,0}=-8.6676\times 10^5 \\
 c_{1,2,1}=2.1741\times 10^7 \\
 c_{1,2,2}=-2.0015\times 10^8 \\
 c_{1,2,3}=8.0668\times 10^8 \\
 c_{1,2,4}=-1.2206\times 10^9 \\
\end{array}
 & 
\begin{array}{c}
 c_{1,3,0}=1.5257\times 10^8 \\
 c_{1,3,1}=-3.8261\times 10^9 \\
 c_{1,3,2}=3.5182\times 10^{10} \\
 c_{1,3,3}=-1.4181\times 10^{11} \\
 c_{1,3,4}=2.1391\times 10^{11} \\
\end{array}
 \\
 
\begin{array}{c}
 c_{2,0,0}=2.3829\times 10^1 \\
 c_{2,0,1}=-6.0241\times 10^2 \\
 c_{2,0,2}=5.6602\times 10^3 \\
 c_{2,0,3}=-2.2968\times 10^4 \\
 c_{2,0,4}=3.6916\times 10^4 \\
\end{array}
 & 
\begin{array}{c}
 c_{2,1,0}=-2.8516\times 10^5 \\
 c_{2,1,1}=7.1666\times 10^6 \\
 c_{2,1,2}=-6.6286\times 10^7 \\
 c_{2,1,3}=2.6753\times 10^8 \\
 c_{2,1,4}=-4.1104\times 10^8 \\
\end{array}
 & 
\begin{array}{c}
 c_{2,2,0}=2.474\times 10^8 \\
 c_{2,2,1}=-6.2074\times 10^9 \\
 c_{2,2,2}=5.7222\times 10^{10} \\
 c_{2,2,3}=-2.3067\times 10^{11} \\
 c_{2,2,4}=3.5102\times 10^{11} \\
\end{array}
 & 
\begin{array}{c}
 c_{2,3,0}=-4.7786\times 10^{10} \\
 c_{2,3,1}=1.1984\times 10^{12} \\
 c_{2,3,2}=-1.1032\times 10^{13} \\
 c_{2,3,3}=4.446\times 10^{13} \\
 c_{2,3,4}=-6.7418\times 10^{13} \\
\end{array}
 \end{array}
\]
\ew
\bw
\[
\begin{array}{cccc}
\begin{array}{c}
 c_{3,0,0}=-1.6988\times 10^3 \\
 c_{3,0,1}=4.306\times 10^4 \\
 c_{3,0,2}=-4.0809\times 10^5 \\
 c_{3,0,3}=1.6593\times 10^6 \\
 c_{3,0,4}=-2.7415\times 10^6 \\
\end{array}
 & 
\begin{array}{c}
 c_{3,1,0}=2.6497\times 10^7 \\
 c_{3,1,1}=-6.6618\times 10^8 \\
 c_{3,1,2}=6.1694\times 10^9 \\
 c_{3,1,3}=-2.4913\times 10^{10} \\
 c_{3,1,4}=3.8467\times 10^{10} \\
\end{array}
 & 
\begin{array}{c}
 c_{3,2,0}=-2.5671\times 10^{10} \\
 c_{3,2,1}=6.4414\times 10^{11} \\
 c_{3,2,2}=-5.9414\times 10^{12} \\
 c_{3,2,3}=2.3951\times 10^{13} \\
 c_{3,2,4}=-3.6549\times 10^{13} \\
\end{array}
 & 
\begin{array}{c}
 c_{3,3,0}=5.1948\times 10^{12} \\
 c_{3,3,1}=-1.3027\times 10^{14} \\
 c_{3,3,2}=1.1998\times 10^{15} \\
 c_{3,3,3}=-4.8345\times 10^{15} \\
 c_{3,3,4}=7.3477\times 10^{15} \\
\end{array}
 \\
 
\begin{array}{c}
 c_{4,0,0}=5.5764\times 10^4 \\
 c_{4,0,1}=-1.4174\times 10^6 \\
 c_{4,0,2}=1.3554\times 10^7 \\
 c_{4,0,3}=-5.5226\times 10^7 \\
 c_{4,0,4}=9.3758\times 10^7 \\
\end{array}
 & 
\begin{array}{c}
 c_{4,1,0}=-1.0772\times 10^9 \\
 c_{4,1,1}=2.7088\times 10^{10} \\
 c_{4,1,2}=-2.5106\times 10^{11} \\
 c_{4,1,3}=1.0141\times 10^{12} \\
 c_{4,1,4}=-1.5706\times 10^{12} \\
\end{array}
 & 
\begin{array}{c}
 c_{4,2,0}=1.1071\times 10^{12} \\
 c_{4,2,1}=-2.778\times 10^{13} \\
 c_{4,2,2}=2.5632\times 10^{14} \\
 c_{4,2,3}=-1.0333\times 10^{15} \\
 c_{4,2,4}=1.5792\times 10^{15} \\
\end{array}
 & 
\begin{array}{c}
 c_{4,3,0}=-2.2974\times 10^{14} \\
 c_{4,3,1}=5.7611\times 10^{15} \\
 c_{4,3,2}=-5.3069\times 10^{16} \\
 c_{4,3,3}=2.1382\times 10^{17} \\
 c_{4,3,4}=-3.2537\times 10^{17} \\
\end{array}
 \\
 
\begin{array}{c}
 c_{5,0,0}=-6.8247\times 10^5 \\
 c_{5,0,1}=1.7392\times 10^7 \\
 c_{5,0,2}=-1.6767\times 10^8 \\
 c_{5,0,3}=6.8445\times 10^8 \\
 c_{5,0,4}=-1.1899\times 10^9 \\
\end{array}
 & 
\begin{array}{c}
 c_{5,1,0}=1.55\times 10^{10} \\
 c_{5,1,1}=-3.8986\times 10^{11} \\
 c_{5,1,2}=3.615\times 10^{12} \\
 c_{5,1,3}=-1.4604\times 10^{13} \\
 c_{5,1,4}=2.2663\times 10^{13} \\
\end{array}
 & 
\begin{array}{c}
 c_{5,2,0}=-1.6476\times 10^{13} \\
 c_{5,2,1}=4.1344\times 10^{14} \\
 c_{5,2,2}=-3.8153\times 10^{15} \\
 c_{5,2,3}=1.538\times 10^{16} \\
 c_{5,2,4}=-2.3528\times 10^{16} \\
\end{array}
 & 
\begin{array}{c}
 c_{5,3,0}=3.4679\times 10^{15} \\
 c_{5,3,1}=-8.6962\times 10^{16} \\
 c_{5,3,2}=8.0115\times 10^{17} \\
 c_{5,3,3}=-3.2277\times 10^{18} \\
 c_{5,3,4}=4.9149\times 10^{18} \,.
\end{array}
 \\
\end{array}
\]
\ew
%

\bibliography{master}
\end{document}